\documentclass[reqno,12pt,oneside]{report} 

\usepackage{rac}         
\usepackage{tgpagella}

\usepackage[intlimits]{amsmath} 
\usepackage{amsxtra}     
\usepackage{amsthm}
\usepackage{amssymb}
\usepackage{amsfonts}
\usepackage{graphicx}    
\usepackage{epstopdf} 
\usepackage{rotating}
\usepackage{color}
\usepackage{epsfig}
\usepackage{verbatim}
\usepackage[numbers]{natbib}   
\usepackage[printonlyused]{acronym} 
\usepackage{setspace}    
\usepackage{siunitx}
\usepackage{multirow} 
\usepackage{amsmath}
\usepackage{graphicx}
\usepackage{xcolor}
\long\def\red#1{\bgroup\color{red}#1\egroup}
\usepackage{threeparttable}
\newcommand{\fref}[1] {Fig.~\ref{#1}\xspace}
\newcommand{\tref}[1]{Table~\ref{#1}}

\usepackage{amsmath,amssymb,amsfonts}
\usepackage{algorithm}
\usepackage{algorithmic}
\usepackage{graphicx}
\usepackage{textcomp}
\usepackage{multirow}
\usepackage{makecell}
\usepackage{url}
\usepackage{siunitx}
\usepackage{float}
\usepackage{xspace}

\usepackage{bm}
\newcommand{\blmath}[1]{\bm{#1}}

\def\argmin#1{\mathop{\mathrm{arg\,min}}_{#1}}
\newcommand{\M}{\sum_{m=1}^{M}}
\newcommand{\N}{3}

\newcommand{\paren}[1] {\left( #1 \right)}
\newcommand{\of}[1] {\!\left( #1 \right)}
\newcommand{\Pm} {\xmath{\mathcal{P}_m}}
\newcommand{\Pmi}[1] {\mathcal{P}_{mi}\of{#1}}
\newcommand{\xmath}[1] {\ensuremath{#1}\xspace}
\newcommand{\x} {\blmath{x}}
\newcommand{\y} {\blmath{y}}
\newcommand{\w} {\blmath{w}}
\newcommand{\Q} {\mathrm{Q}}
\newcommand{\K} {\mathrm{K}}
\newcommand{\V} {\mathrm{V}}
\newcommand{\X} {\blmath{X}}
\renewcommand{\Xi} {\xmath{\blmath{X}_i}}

\newcommand{\Xikk} {\xmath{\blmath{X}_i^{k+1}}}
\newcommand{\A} {\xmath{\mathcal{A}}}
\newcommand{\cB} {\xmath{\mathcal{B}}}
\newcommand{\cL} {\xmath{\mathcal{L}}}
\newcommand{\eps} {\xmath{\varepsilon}}
\newcommand{\Shift}[1] {\textsc{Shift}\of{#1}}
\newcommand{\pshift}[1] {\Pm\of{\Shift{#1}}}
\newcommand{\refoldi}[1] {\textsc{Refold}_i\of{#1}}
\newcommand{\svti}[1] {\textsc{SVT}_{\lambda_i/\rho}\of{#1}}
\long\def\comment#1{}

\usepackage{url}
\usepackage{caption}
\usepackage{subcaption}
\usepackage{tikz,psfrag}
\usetikzlibrary{arrows.meta}
\usetikzlibrary{arrows,positioning}
\usetikzlibrary{fit}
\usepackage[usestackEOL]{stackengine}
\tikzstyle{i} = [pin edge={to-,thin,black}]

\def\argmin#1{\underset{#1}{\text{arg min\ }}}
\def\argmax#1{\underset{#1}{\text{arg max\ }}}

\newcommand{\z}{\blmath{z}}

\renewcommand{\Re}{\mathbb{R}}
\newcommand{\Cp}{\mathbb{C}} 
\newcommand{\Bp}{\blmath{\phi}}
\newcommand{\BP}{\blmath{\Phi}}
\newcommand{\bv}{\blmath{v}}
\newcommand{\ba}{\blmath{\theta}}

\newcommand{\norm}[1]{\left\lVert#1\right\rVert}
\newcommand{\normr}[1]{\lVert#1\rVert} 
\newcommand{\abs}[1]{\left|#1\right|}

\providecommand{\icomplex}       {\xmath{\imath}}
\providecommand{\eexp}           {\xmath{\mathinner{\mathrm{e}}}}
\newcommand{\Expni}[1] {\xmath{\eexp^{-\icomplex #1}}}
\providecommand{\Expn}[1]       {\xmath{\eexp^{-#1}}}

\providecommand{\der}[1]        {\xmath{\mathop{\mathrm{d}#1}\nolimits}}
\newcommand{\df}{\xmath{\der f}}

\providecommand{\desa}[1]       {\begin{equation}%
    \begin{aligned}#1\end{aligned}\end{equation}} 

\usepackage{hyperref}
\usepackage{tikz-cd}

\usepackage[capitalize]{cleveref}
\crefname{section}{Sec.}{Secs.}
\Crefname{section}{Section}{Sections}
\Crefname{table}{Table}{Tables}
\crefname{table}{Tab.}{Tabs.}

\theoremstyle{plain}

\theoremstyle{definition}

\theoremstyle{remark}

\numberwithin{theorem}{chapter}     

\makeatletter
\def\cleardoublepage{\clearpage\if@twoside \ifodd\c@page\else
\hbox{}
\thispagestyle{empty}
\newpage
\if@twocolumn\hbox{}\newpage\fi\fi\fi}
\makeatother 

\begin{document}

\titlepage{Novel Models for High-Dimensional Imaging: High-Resolution fMRI Acceleration and Quantification}{Shouchang Guo}{Doctor of Philosophy}{Electrical and Computer Engineering}{2022}
{
 Professor Jeffrey A. Fessler, Co-Chair\\
 Professor Douglas C. Noll, Co-Chair\\
 Associate Professor Laura K. Balzano\\
 Assistant Professor David F. Fouhey\\
}

\initializefrontsections

\frontispiece{
\[
\begin{tikzcd}[row sep=16em,column sep=6.5em]
 & \textrm{SNR} \arrow[dr,dash] \\
\textrm{Resolution} \arrow[ur,dash] \arrow[rr,dash] && \textrm{Scan Time}
\end{tikzcd}
\] 
}

\copyrightpage{Shouchang Guo}{2022}{shoucguo@umich.edu}{\href{https://orcid.org/0000-0002-0329-1638}{0000-0002-0329-1638}}

\makeatletter
\if@twoside \setcounter{page}{4} \else \setcounter{page}{1} \fi
\makeatother
 
\dedicationpage{\emph{To mom and dad, grandparents, water and wind.}}

\startacknowledgementspage
As a student who knew nothing about medical imaging when started, I am eternally grateful to my advisors Prof. Doug Noll and Prof. Jeff Fessler for their trust and patience, and for everything I have learned to become who I am today. Their genuine zest and devotion for science and teaching, and genuine care and support for our growth have always been the source of courage and enthusiasm for me. Prof. Noll’s incredible insights and intuition, and profound knowledge of physics and MRI have guided me through darkness and uncertainty at every stage of research. It is magnificent and magical to see how Prof. Fessler tackles sophisticated problems with elegant mathematical modeling and inspiring flows of equations. I have been extremely lucky to learn from both of them about how to connect and innovate, how to solve problems with intuition and skills, and how to pursue my goals with passion and perseverance. 

I am deeply grateful to my committee members Prof. Laura Balzano and Prof. David Fouhey for their guidance and encouragement, all the invaluable ideas and advice, and their amazing research that intrigues me to think above and beyond. 

I want to express my deep appreciation to all the senior members in the group, Drs. Scott Peltier, Jon-Frederik Nielsen, Luis Hernandez-Garcia, Kathleen Ropella, Sydney Williams, Tianrui Luo, Anish Lahiri, Amos Cao, Melissa Haskell, Sai Ravishankar, David Hong, Hao Sun, Jiabei Zheng, and my roommate Dr. Shriya Sethuraman for their generous support and all the important insights and discussions that help me grow.

I have been very fortunate to have spent innumerable happy hours with my friends and labmates Michelle Karker, Claire Lin, Dinank Gupta, Guanhua Wang, Mingjie Gao, Haowei Xiang, Mariama Salifu, Caroline Crockett, Cameron Blocker, Naveen Murthy, Steven Whitaker, Zongyu Li, Eric Cheek, Tao Hong, Xijia Quan, David Frey, and others.

Special thanks to NIH Grants U01EB026977 and R01EB023618. I also wish to thank all the participants in MRI studies and staff members Kristen Thornton, Theresa Russ, Erik Keup, Kim Sharma, and others for being so kind and supportive. 
Ann Arbor is such a beautiful place with all the great people.

Finally, I want to thank my parents, grandparents, my brother, and my all-time best friends, for their love and trust, and their sometimes unwarranted confidence in me.

``What does a fish know about the water in which he swims all his life?" It has been a wonderful journey with infinite dimensions of things that I do not know, and I hope that I have infinite courage and enthusiasm to learn and do something meaningful.

\label{Acknowledgements}


\tableofcontents     
\listoffigures       
\listoftables        

\startabstractpage
{Novel Models for High-Resolution fMRI Acceleration and Quantification}{Shouchang Guo}{Co-Chairs:}
The goals of functional Magnetic Resonance Imaging (fMRI) include high spatial and temporal resolutions with a high signal-to-noise ratio (SNR). We introduce a novel method for fMRI named \emph{Oscillating Steady-State Imaging (OSSI)}. OSSI can provide 2 to 3 times higher SNR than the standard method. 
However, the SNR improvement comes at a cost of spatial-temporal resolution.

To simultaneously improve spatial and temporal resolutions and maintain the high SNR advantage of OSSI,
we present novel pipelines for fast acquisition and high-resolution fMRI reconstruction and physics parameter quantification.
We design a sparse sampling pattern to accelerate scan time.
Because OSSI images are high-dimensional, we propose a \emph{patch-tensor low-rank model} to exploit the local spatial-temporal low-rankness of the images. The proposed method enables high-resolution 3D fMRI
with a factor 10 acceleration and 1.3 mm spatial resolution,
and yields 2 times higher SNR than the standard fMRI methods with 2 times more brain activation.

To accurately model the nonlinearity of OSSI oscillation pattern, instead of applying subspace models that might not be perfectly suited for the data,
we propose a \emph{physics-based manifold model} that builds the MR physics for OSSI signal generation as a regularizer for the undersampled reconstruction.
The proposed manifold model reconstructs high-resolution fMRI images with high SNR and a factor of 12 acceleration. Furthermore, the model enables dynamic tracking of important physics parameters for more accurate brain activity monitoring with a 150 ms temporal resolution. 

To exploit learning-based approaches for dynamic MRI with better temporal modeling and richer representations, 
we propose a \emph{voxel-wise attention network} that combines MR physics with the attention mechanism for temporal learning and mapping.
We also develop a two-stage learning scheme to resolve the training data limitation. 
The proposed network reconstructs dynamic MRI sequences with a factor of 12 undersampling and provides high-quality functional maps with 4 times faster reconstruction than model-based approaches.

With novel models for acquisition and reconstruction, we demonstrate that we can improve SNR and resolution simultaneously without compromising scan time. All the proposed models outperform other comparison approaches with higher resolution and more functional information. 


\label{Abstract}

\startthechapters 

\chapter{Introduction}
\label{chap:0intro}
\section{Motivation}
Three factors, simple but overwhelmingly important, have governed the quality of magnetic resonance imaging (MRI): the signal-to-noise ratio (SNR), the resolution, and the scan time. The trade-offs among SNR, resolution, and scan time, have been sources of inspiration for MRI research.

In functional MRI (fMRI), a time series of MRI images are acquired to track brain activity.
Because signal changes for brain activation are very small,
we need high SNR to distinguish brain signals from noise sources.
Because functional units of the brain are on the order of 1 mm or smaller,
we need fine spatial resolution to precisely locate functional signals. 
As SNR is proportional to voxel size,
high SNR is essential for high spatial resolution.
To achieve high spatial resolution or high SNR,
traditional methods must
increase scan time for each image in the fMRI time series.
The increased scan time per image or decreased temporal resolution
would diminish the temporal accuracy of fMRI signals.

\section{Background}
\subsection{High SNR Functional MRI}
Functional MRI acquires a time series of MRI images to track brain activity. 
The SNR of an fMRI image \cite{macovski1996noise,lai1998three,noll1995methodologic} is determined by 
\begin{equation}
\text{SNR} \propto B C V \sqrt{T},
\end{equation}
where $B$ represents magnetic field strength, $C$ is a head coil dependent term, $V$ is the voxel volume for the brain image, and $T$ is proportional to the scan time for each image. The scan time for collecting each of the images in the time series is also referred to as the temporal resolution of fMRI. 

Brain activity related signal changes are small and can be easily buried in noise. As SNR is proportional to voxel volume and functional units of the brain are on the order of 1 mm or smaller, high SNR is critical for high-quality and high-resolution fMRI. However, 
current methods for SNR improvement are limited. 

Improvements in $B$ and $C$ require a new set of hardware. Increasing $B$ with higher field strength systems is a costly investment and leads to severe distortion issues in the images. 
Increasing the number of coils for $C$ in a head array suffers from diminishing returns as coil elements get smaller, particularly for deep brain structures. 

For software related factors, $V$ and $T$ correspond to spatial resolution and scan time of an image, respectively.  
There is a triangle trade-off between SNR, spatial resolution, and scan time in MRI. To increase the SNR, one would need to increase scan time, or compromise resolution; to improve resolution, SNR would be sacrificed or the scan time would need to increase; to reduce scan time, the SNR and/or the resolution of the image would decrease.
It is very hard to improve all three factors at the same time, and the main goal of this thesis is to improve SNR, resolution, and scan time simultaneously without costly equipment.

\subsection{High-Resolution Image Reconstruction}

In MRI, the data collected via scanning are in ``$k$-space", and a Fourier transform relationship holds between the object image and the acquired $k$-space data \cite{liang2000principles,nishimura2010principles}. Therefore, the $k$-space is basically the Fourier domain of an MRI image, and the simplest way to reconstruct the image is to take the inverse Fourier transform of the $k$-space data. Collecting data with a larger $k$-space extent can increase spatial resolution at the expense of scan time. 

To improve resolution without compromising scanning time, compressed sensing \cite{donoho2006compressed,lustig2008compressed} and model-based reconstruction \cite{fessler2010model,lustig2007sparse} propose random sparse sampling (dramatically reduced sampling rates compared to the Nyquist sampling criteria) in $k$-space. Furthermore, prior information on images is imposed to solve the undetermined problem with a limited amount of measurements. The image reconstruction problem is formulated as 
\begin{equation}
\argmin{\X} 
\frac{1}{2}\lVert\mathcal{A}(\X)-\y \rVert_2^2 + 
\alpha \mathcal{R}(\X),
\end{equation}
where $\X$ are the images to be reconstructed, and $\y$ denotes the small number of $k$-space measurements. $\mathcal{A}$ is a linear operator representing the MR physics, and $\mathcal{A}$ represents the Fourier transform for single-coil MRI with Cartesian sampling. 
$\mathcal{R}(\cdot)$ regularizes the images with prior information and assumptions. $\alpha$ is the regularization parameter.

Typical priors used as constraints on the images include total variation \cite{fessler2010model,lustig2007sparse}, low-rank and/or sparse \cite{Otazo2015Low-rankComponents,lustig2007sparse,ravishankar2017low}, and learned dictionary \cite{ravishankar2010mr}. Recent works \cite{schlemper2017deep,hammernik2018learning,aggarwal2018modl} use neural networks as regularizers for undersampled reconstruction. In this work, we propose novel models and techniques for high SNR and high-resolution fMRI image reconstruction.

\section{Outline and Contributions}

This thesis is organized as follows: 

\autoref{chap:1ossi},
published in \cite{ossi2019},
describes a new Oscillating Steady-State Imaging (OSSI) method for high SNR fMRI.
OSSI establishes a new steady state by combining balanced gradients in balanced steady-state free precession \cite{bieri2013fundamentals} and quadratic RF phase progression in RF-spoiled GRE \cite{zur1991spoiling}. 
The resulting oscillating steady-state signal combines high SNR of the balanced steady state and the $T_2^*$ contrast of gradient echo (GRE) imaging for fMRI. OSSI provides at least 2 times higher SNR than standard GRE fMRI without costly equipment investments. However, the SNR advantage of OSSI comes at a price of spatial and temporal resolutions. 

\autoref{chap:2tensor},
published in \cite{tensor2020},
describes a novel pipeline for fast acquisition and high-resolution and high-dimensional fMRI. 
As the unique oscillation pattern of OSSI images makes it well suited for high-dimensional modeling, we propose a patch-tensor low-rank model to exploit the inherent high-dimensional structures and local spatial-temporal low-rankness of the images. 
We also develop a practical
sparse sampling scheme with improved sampling incoherence. 
With an alternating direction method
of multipliers based algorithm, we improve OSSI
spatial and temporal resolutions with a factor of 12 acquisition acceleration and 1.3 mm isotropic spatial resolution
in prospectively undersampled experiments. 
Compared to the standard GRE imaging at the same spatial-temporal resolution, the proposed model demonstrates
2 times higher SNR with 2 times more functional
activation.

\autoref{chap:3manifold}, under revision and
available in \cite{2021manifold},
describes a new manifold model for high-resolution fMRI joint quantification and reconstruction.
Because OSSI signals exhibit a nonlinear oscillation pattern and to accurately model the nonlinearity, instead of using subspace models that might not be perfectly suited for the data,
we build the MR physics for OSSI signal generation
as a regularizer for the undersampled reconstruction.
Our proposed physics-based manifold model
turns the disadvantages of OSSI acquisition into advantages. 
OSSI manifold model (OSSIMM) outperforms subspace models
and reconstructs high-resolution fMRI images
with a factor of 12 acceleration and without spatial-temporal smoothing. 
Furthermore, OSSIMM can dynamically quantify and track important physics parameters for more accurate brain activity monitoring
with a 150 mm temporal resolution.

\autoref{chap:4attnet} 
describes a novel learning-based approach and training scheme for dynamic MRI acceleration and reconstruction.
Because learning-based temporal modeling in dynamic MRI is an open question and often requires large amounts of training data, we propose a voxel-wise attention network that incorporates an attention mechanism for temporal learning and mapping. The proposed network combines MR physics with a data fidelity layer for end-to-end inference. 
We also develop a two-stage learning scheme that pretrains the network with voxel-wise simulated data, and then fine-tunes with human data to resolve the lack of training data.
Our proposed model reconstructs dynamic MRI images with a factor of 12 undersampling, and provides high-quality images and functional maps. The proposed voxel-wise, attention-based model can potentially be used for MR fingering reconstruction and other dynamic reconstruction applications.

\autoref{chap:5future} proposes future work
on other novel spatial-temporal models for MRI image sequence reconstruction and acceleration.

\chapter{Oscillating Steady‐State Imaging (OSSI): A Novel Method for Functional MRI}
\label{chap:1ossi}
Signal-to-noise ratio (SNR) is crucial for high-resolution fMRI, however, current methods for SNR improvement are limited. A new approach, called Oscillating Steady-State Imaging (OSSI), produces a signal that is large and $T_2^*$-weighted, and is demonstrated to produce improved SNR compared to gradient echo (GRE) imaging with matched TE and spatial-temporal acquisition characteristics for high-resolution fMRI.
Quadratic phase sequences were combined with balanced gradients to produce a large, oscillating steady-state signal. The quadratic phase progression was periodic over short intervals such as 10 TRs, inducing a frequency-dependent phase dispersal. Images over one period were combined to produce a single image with effectively $T_2^*$-weighting.  The
OSSI parameters were explored through simulation and phantom data, and 2D and 3D human fMRI data were collected using OSSI and GRE imaging.
Phantom and human OSSI data showed highly reproducible signal oscillations with greater signal strength than GRE.  Compared to single slice GRE with matched TE and spatial-temporal resolution, OSSI yielded more activation in visual cortex by a factor of 1.84 and an improvement in temporal SNR by a factor of 1.83.  Voxelwise percentage change comparisons between OSSI and GRE demonstrate a similar $T_2^*$-weighted contrast mechanism with additional $T_2'$-weighting of about 15 ms immediately after the RF pulse.
OSSI is a new acquisition method that exploits a large, oscillating signal that is $T_2^*$-weighted and suitable for fMRI. The steady-state signal from balanced gradients creates higher signal strength than single slice GRE at varying TEs, enabling greater volumes of functional activity and higher SNR for high-resolution fMRI. \footnote{
This chapter was published in \cite{ossi2019,ossi,ossi2}.
}

\section{Introduction}
Because the signal-to-noise ratio (SNR) in MRI is proportional to voxel volume, and the functional units of the brain are on the order of 1 mm, high SNR is required for functional MRI (fMRI) of these small brain structures.  Many common methods for improving SNR have already been well-used, but now face limitations.  For example, extending readouts increases sensitivity to off-resonance distortions, and increasing the number of coils in a head array suffers from diminishing returns as coil elements get smaller, particularly for deep brain structures.  One can also enhance SNR by going to higher field systems, but this requires a costly investment.  Thus, there is a compelling need for alternative approaches to improving the SNR in fMRI.

Functional MRI using the blood oxygenation (BOLD) effect has been based on $T_2^*$-weighted gradient echo (GRE) imaging from its inception and has commonly been implemented using single-shot fast imaging methods like echo-planar imaging (EPI) or spiral imaging. There has also been some work on acquisition using steady-state methods.  These include $T_2^*$-weighted, 3D GRE acquisitions of several variations \cite{VanGelderen1995,Rettenmeier2019} and short TR, fast recovery (STFR) sequences that preserve magnetization through principles of driven equilibrium \cite{Sun2015}.  There are also variants of balanced steady-state methods like balanced steady-state free precession (bSSFP, also known as True FISP, FIESTA, bFFE), such as transition-band bSSFP \cite{Scheffler2001,Miller2003}, have exploited shifts in resonant frequency associated with changes in blood oxygenation.  At the same time, blood oxygenation changes have also led to observable signal changes using passband bSSFP resulting from changes in $T_2$ directly and from diffusive effects around small vessels \cite{Zhong2007,Miller2008,Scheffler2016}.

Standard implementations of bSSFP use constant excitation phase or a linear phase sequence for RF pulses.  In this work we use a similar balanced-gradient pulse sequence, but with quadratic phase sequences, which is equivalent to a linearly sweeping frequency.  Since the frequency response is periodic in the frequency domain, a frequency sweep will lead to periodic signal oscillation.  We note that if the gradients are not balanced (e.g. gradient spoiled), the quadratic phase sequence will lead to a RF-spoiled gradient echo acquisition, provided that the sweep rate is sufficiently fast.  Also, if the quadratic phase sequence is sufficiently slowly evolving, then the balanced-gradient acquisition leads to contrast that is very similar to the standard bSSFP contrast though the response slowly shifts over time.  In this work, we explore a novel domain using balanced-gradients but with a quadratic phase sequence that is rapid, having a period on the order of 10 TRs, which leads to an oscillatory signal.  We refer to this approach as Oscillating Steady-State Imaging (OSSI).  We show that the OSSI signal is large compared to Ernst angle GRE imaging and further show that the OSSI signal is sensitive to changes in $T_2^*$, making it suitable for high-resolution fMRI. We distinguish our approach using quadratic phase sequences from other oscillatory steady states resulting from sequences of alternating patterns of phase \cite{Scheffler2006,Overall2002}, which have a different contrast. 

In this work, we demonstrate a novel fMRI acquisition method that has the potential to improve the SNR over GRE with  matched TE and spatial-temporal acquisition characteristics. It focuses on a unique oscillating steady-state source of signal that is large and $T_2^*$-weighted, and we explore its signal properties in both simulation and experimental studies. 

\begin{figure}
\centering
\includegraphics[width=0.9\textwidth]{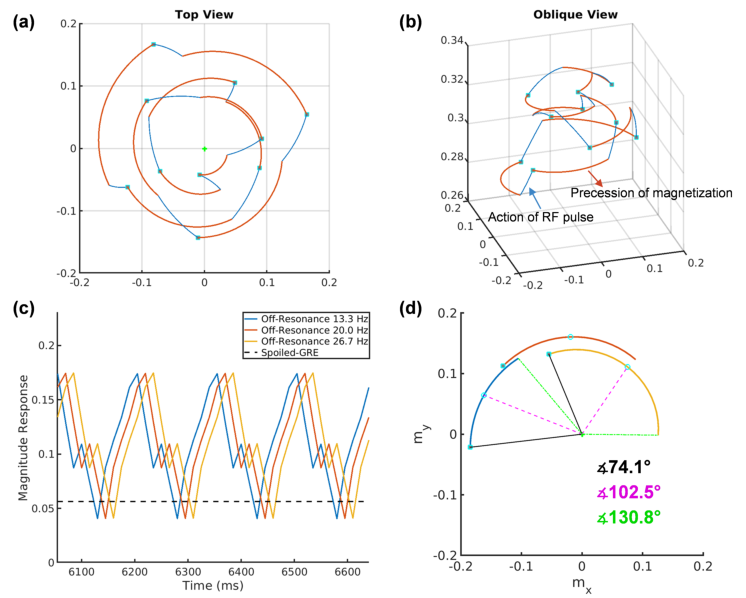}
\caption{Simulation of OSSI spin behavior and signals.\\
(a) and (b) Periodic motion of magnetization through RF pulses (the filled squares are at the end of the RF pulse) and free precession for a gray matter spin at -20 Hz off-resonance frequency, $T_1$ = 1433.2 ms, $T_2$ = 92.6 ms, TR = 15 ms, $n_c$ = 10, and FA = 10$^\circ$ from two different views.
(c) Magnitude signal variation of different isochromats (6.67 Hz apart) for the magnetization in (a) and (b) just after the RF pulse, the black dashed line is the Ernst angle signal for spoiled-GRE.
(d) Spin positions during free precession for different isochromats (same isocromats as in (c)) leading to phase dispersion and $T_2^*$-weighting. The cyan circles mark the center of the precession interval.
}
\label{of1}
\end{figure}

\section{Theory}
\subsection{Oscillating Steady-State Imaging}
Quadratic phase sequences in conjunction with a constant gradient dephasing is a well-recognized approach for establishing a spoiled steady state. The sequence is typically applied using the RF phase increment \cite{Zur1991}
\begin{equation}\label{e1}
\phi(n) - \phi(n-1) = \psi_{A} n + \psi_B,
\end{equation}
where $\psi_A$ is commonly chosen to provide full cancellation of the transverse magnetization prior to the next RF pulse; typical values for spoiling are $\psi_A = 117^\circ,50^\circ,150^\circ$, etc.  The constant term, $\psi_B$, represents a constant frequency shift and is not important in most of these analyses.  The linear phase increment is equivalent to a quadratic phase sequence, for example, $\phi (n) = \psi_A n^2/2$ is the same as in \eqref{e1} for the case of $\psi_B = -\psi_A/2$.  In this work, we examine such quadratic phase sequences with balanced gradients, which maintains the steady-state components leading to stronger signals.  This approach was proposed by Foxall \cite{Foxall2002} to implement bSSFP with $T_2$-like weighting whereby the frequency-dependent bands in image intensity slowly shifted over the acquisition.  Foxall argued that bSSFP-like contrast would be preserved if the phase increment is kept small ($\psi_A < 3^\circ$).  We have observed that larger phase increments also leads to steady-state signals, however that the contrast is no longer similar to bSSFP contrast, but instead has contrast that is both $T_2$- and $T_2'$-weighted, and thus effectively $T_2^*$-weighted.  With appropriate selection of $\psi_A$, the phase sequence can be made to be periodic with cycle length $n_c$ by setting
\begin{equation}
\psi_{A} = \frac{2\pi}{n_c}.
\end{equation}
This periodic sequence leads to oscillations in the steady-state signal with period $\text{T}_\text{OSSI} = n_c\text{TR}$.  Maintenance of transverse components via a steady state tends to make the resultant signals $T_2$-weighted, while the different phases of different isochromats lead to $T_2'$-weighting.  We note that Wang et al. \cite{Wang2019} have similarly observed that quadratic phase RF pulses lead to frequency dependent phase variations and $T_2^*$-weighting. 

The OSSI signals have a variety of interesting properties.  Like bSSFP, the OSSI response is frequency dependent and the spectral properties are periodic with 1/TR in the frequency domain. Further, it can be shown that shifts in frequency will lead to signal being shifted in time.  Specifically, a frequency shift of 1/T\textsubscript{OSSI} will lead to the phase sequence being shifted by exactly one TR, which leads to the OSSI response being similarly shifted in time by one TR.  Note that a frequency shift of 1/T\textsubscript{OSSI} is equivalent to $\Delta\psi_B = 2\pi/\text{T}_\text{OSSI}$.  Frequency shifts that are not integer multiples of 1/T\textsubscript{OSSI} will also have oscillatory behavior, but with slightly different temporal signal responses.  Thus, different isochromats within an image will have unique time courses, each of which is periodic with T\textsubscript{OSSI}, and depending on the frequency, these time courses will be shifted in time and/or have slightly different shape.  The shifts in time for different isochromats induces a frequency dependent phase dispersal, effectively leading to $T_2^*$-like contrast.  In order to produce a stable and usable time course for fMRI analyses, we commonly combine the $n_c$ images for one period of the OSSI signal by some method, for example using root mean square (RMS) or 2-norm combination.  

\begin{figure}
\centering
\includegraphics[width=10cm]{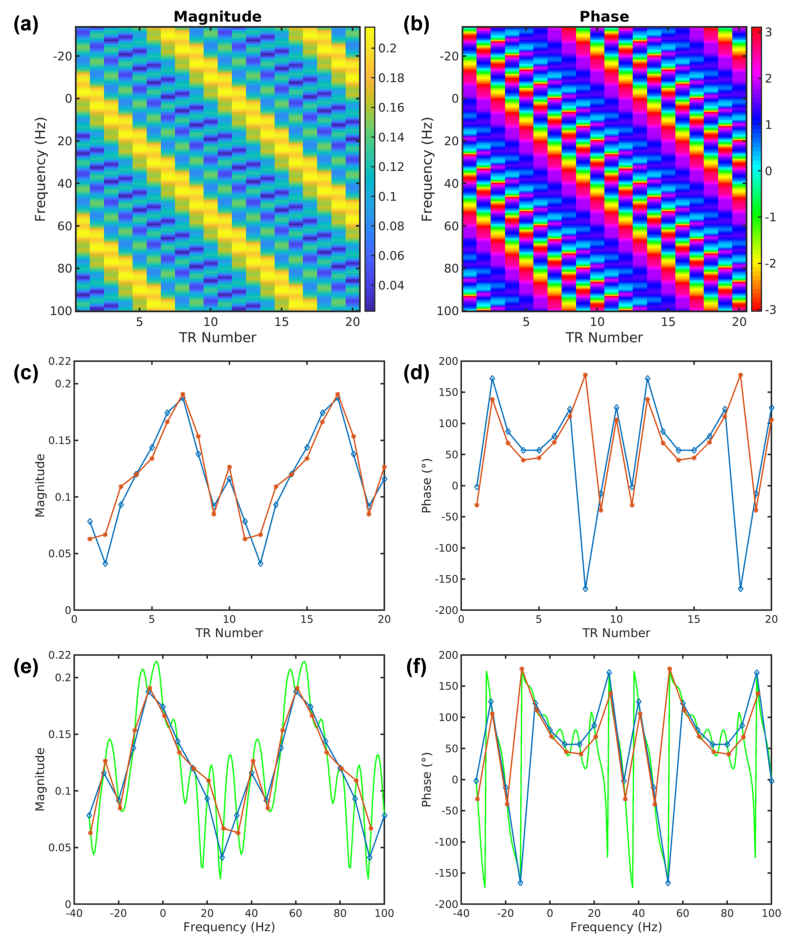} 
\caption{Simulation for signal properties just after the RF pulse, where the pulse duration was adjusted to minimize off-resonance phase accumulation during the RF pulses (TE $<$ 0.02 ms). The left and right panels show simulated OSSI signal magnitude and phase, respectively.  (a) and (b) show magnitude and phase responses as a function of off-resonance frequency and time (TR number), observe the periodicity in time ($\text{T}_\text{OSSI} = n_c\text{TR}$) and frequency (1/TR = 66.67 Hz).  (c) and (e) are magnitude response of the signal vs. time and frequency, respectively, and (d) and (f) are the phase responses showing phases after correction for the excitation RF phase. The blue and red lines in temporal plots (c) and (d) correspond to two isochromats at off-resonance -33.33 Hz and -32.67 Hz, respectively. It can be seen that an off-resonance amount of less than 1/T\textsubscript{OSSI} lead to some modest changes in the shape of the response. The green curve in (e) and (f) are the magnitude and phase of the frequency response, respectively, and indicate the manifold on which the steady-state response exists.  The blue and red lines connect 6.67 Hz apart samples of the manifold and start from off-resonance -33.33 Hz and -32.67 Hz respectively. Particularly, by comparing (c) and (e), (d) and (f), it is clearly shown that the time and samples of frequency responses have exactly the same shape, only flipped.}
\label{of2}
\end{figure}

\subsection{OSSI Spin Behavior and Signal Simulation}

OSSI spin behavior was examined using a Bloch equation simulator for spins having relaxation parameters similar to gray matter using the average of reported values \cite{Parameters} $T_1$ = 1433.2 ms and $T_2$ = 92.6 ms at 3T, and pulse sequence parameters TR = 15 ms with an excitation pulse length of 3.2 ms, number in phase cycle ($n_c$) = 10, and flip angle (FA) = 10$^\circ$. The phase progression with $n_c$ = 10 is equivalent to a spoiling seed of 36$^\circ$ for spoiled-GRE. An example of magnetization progression at steady state is shown for a spin with off-resonance -20 Hz in \fref{of1}, with the pattern repeating every $n_c$ TRs.  The signal intensity varies as magnetization moves towards and away from the center in (a), and the spin moves up and down in $m_z$ (b). From \fref{of1} (c), we can see that the magnitude of the OSSI signal right after the excitation at TE = 1.6 ms has a periodicity of $n_c$TR and is substantially larger than the spoiled GRE signal for the same parameters.  Observe that off-resonance shifts of multiples of $1/\text{T}_\text{OSSI} = 1/(n_c\text{TR})$ = 6.67 Hz lead to exactly the same temporal waveform with a shift of 1 TR in time. The isochromats in \fref{of1} (c) and (d) cover a frequency range of 13.3 Hz and result in a 74.1$^\circ$ phase spread for the time point right after the RF pulse. Note that the phase between isochromats increases during the readout, which indicates increased $T_2'$-weighting, and there is no spin-echo signal formed at the center of readout, demonstrating a very different contrast mechanism compared to bSSFP.  The observed phase accumulation is equivalent to a $T_2'$-weighting with an effective TE of 15.4 ms at beginning and 27.2 ms at the end of the readout interval, respectively.
 
\fref{of2} (a) and (b) shows the magnitude and phase responses, respectively, as a function of time and frequency.  In \fref{of2} (c) and (d), one can see that frequency shifts that are not multiples of 1/T\textsubscript{OSSI} lead to slightly different time courses in magnitude and phase. The duality between time and frequency is shown in \fref{of2} (e) and (f).  Here one can see that samples in frequency spaced at integer multiples of 1/T\textsubscript{OSSI} will give exactly the same waveform as the time courses in \fref{of2} (c) and (d), but reversed.  More specifically, the OSSI signal $M_T$ and the frequency response $M_F$ have the following relationship 
\begin{equation}
M_T(k\ \mathrm{mod} \ n_c;f_0) = M_F\left(f_0 + \frac{1-k \ \mathrm{mod} \ n_c}{n_c\text{TR}}\right),
\end{equation}
where $k$ is the TR number, and $f_0$ denotes the off-resonance frequency.  From this expression, one can clearly see that the steady-state frequency response is the manifold on which the time-course signals are found.

\begin{figure}
\centering
\includegraphics[width=11cm]{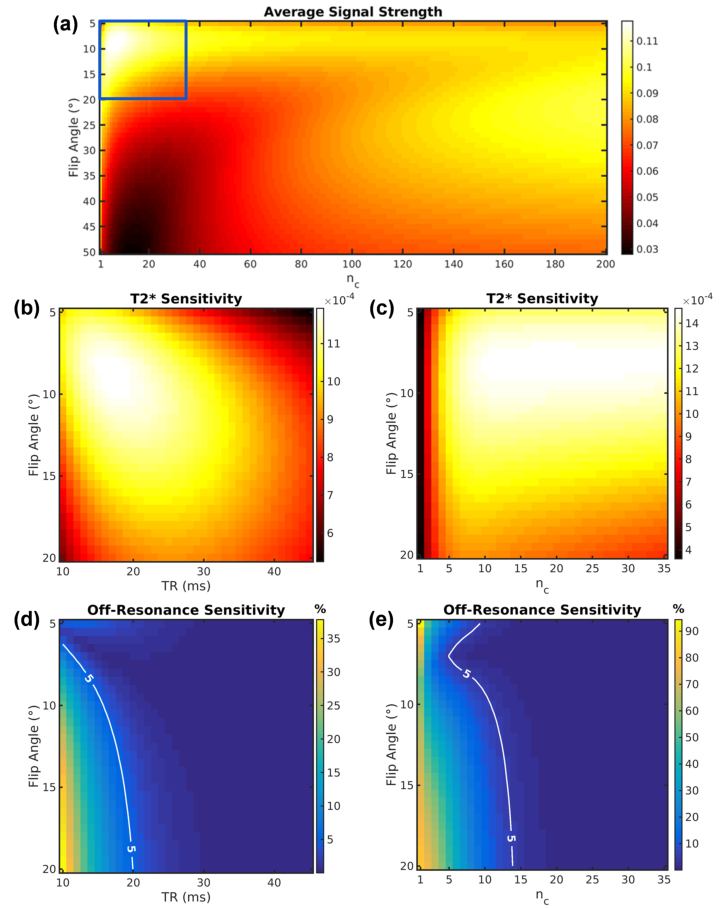}
\caption{Simulation of acquisition parameters for spiral-out readouts (TE = 1.6 ms). (a) to (c) are $T_2^*$ sensitivity defined as S\textsubscript{activated} – S\textsubscript{rest} in units of $M_0$ = 1.\\
(a) shows the RMS combined magnitude signal as a function of $n_c$ and flip angle for a fixed TR of 15 ms. Notice the bright spot around $n_c$ = 10 and flip angle = 10$^\circ$. We focus on the region denoted by the blue square for OSSI fMRI acquisition parameter optimization, and the results are in (b) to (e). 
(b) shows how $T_2^*$ sensitivity varies with TR and flip angle for a fixed $n_c$ = 10. The signal is normalized by $\sqrt{(\text{TR}-c)/\text{TR}} \approx \sqrt{\text{T}_\text{A/D}}$ with $c$ = 5 ms for SNR efficiency.
(c) shows how $T_2^*$ sensitivity varies with $n_c$ and flip angle for TR = 15 ms.
(d) gives off-resonance sensitivity at different TR and flip angles for $n_c$ = 10.
(e) gives off-resonance sensitivity at different $n_c$ and flip angles for TR = 15 ms. 
}
\label{of3}
\end{figure}

\subsection{Acquisition Parameter Optimization}

In seeking to optimize the OSSI signal, there are a variety of measures of goodness.  First, since we are interested in applying this method to functional MRI, we wish to maximize sensitivity to changes in the signal resulting from changes in $T_2'$, normalized by the square root of imaging time.  It is desirable to have smaller $n_c$ as fewer TRs are needed to complete a single image, while longer TRs are preferred because they allow a longer time for acquisition.  We also desire to maximize uniformity of the RMS combined OSSI signal as a function of frequency shifts smaller than 1/T\textsubscript{OSSI}.

To understand the impact of pulse sequence parameters on the OSSI signal, additional Bloch simulations were carried out.  \fref{of3} (a) shows the RMS combined signal intensity for OSSI as a function of $n_c$ and flip angle for TR = 15 ms.  Note that $n_c = 1$ corresponds to bSSFP, and for $n_c > 120$, the OSSI signal behaves similarly to bSSFP with a flat phase response (spin-echo-like contrast) over some range of off-resonance frequencies \cite{Foxall2002}.  However, the bright signals in the upper left corner, bounded by the box, were somewhat unexpected and are the focus of this chapter.  Here, we examine a range of parameters with respect to sensitivity for fMRI studies and to undesired sources of signal variation.  
Deoxygenation of blood at 3T primarily affects $T_2'$ in tissue \cite{Boxerman1995} and causes an approximately exponential decay $\exp (-t/T_2')$ of the BOLD signal. This effect can be modeled by averaging complex signals from a large number of spins with different off-resonance frequencies. When the number of spins is sufficiently large, there exists a Fourier relationship between $\exp (-|t|/T_2')$ and the probability density function of off-resonance frequency $f$, yielding the Cauchy distribution $G(f) = \gamma/(\pi(\gamma^2+f^2))$, where $\gamma$ is the scale parameter of the distribution and $T_2' = 1/(2 \pi \gamma)$.

Therefore, to simulate the $T_2^*$-weighted signal of a voxel in the static dephasing regime, we generated complex OSSI signals from 2000 spins with off-resonance frequencies uniformly ranging from -150 Hz to 150 Hz, and calculated weighted sum of the complex signals. The weighting function is the Cauchy distribution $G(f)$ centered at a specific off-resonance frequency and using $T_2'$ = 148.3 ms and 135.5 ms,  corresponding to $T_2^*$ of 57 ms and 55 ms given an underlying $T_2$ = 92.6 ms for gray matter, which were selected to model baseline and active conditions, respectively. The $T_2^*$ difference represents a typical 1.9\% signal change for a $T_2^*$-weighted GRE image with TE = 30 ms. The OSSI baseline and active signals were obtained by applying RMS combination to every $n_c$ = 10 consecutive and non-overlapping time points of the $T_2^*$-weighted signals. 

The OSSI signal of each spin was simulated using a range of parameters for TR, FA, and $n_c$. We varied two parameters while fixing the third parameter, and performed the simulation for at least 5 $T_1$s to ensure the signal was in steady state.  The $T_2^*$ sensitivity is defined by the difference of the active ($T_2^*$ = 55 ms) and baseline ($T_2^*$ = 57 ms) signals in units of $M_0$ either just after the RF pulse for spiral out acquisitions or just before the subsequent RF pulse for spiral-in acquisition. \fref{of3} (b) and (c) gives the $T_2^*$ sensitivity for a spiral-out acquisition (TE = 1.6 ms) as a function of different TRs and flip angles for $n_c$ = 10, and different $n_c$ and flip angles at a fixed TR of 15 ms, respectively.  Supporting Information \fref{osf1} (b) and (c) presents the same relationship for a spiral-in acquisition (TE = TR - 1.6 ms). 
 
As noted above and shown in \fref{of2}, the OSSI pulse sequence is very frequency sensitive but for the use in fMRI an important question is the sensitivity of the combined (RMS over $n_c$ points) signal vs. frequency. An example of this effect is the small difference between RMS combined signal of blue and red lines in \fref{of2} (c). The combined signal is periodic in the frequency domain with $1/\text{T}_\text{OSSI} = 1/(n_c\text{TR})$, so we varied the central frequency offset over this range to obtain the signal variability due to field inhomogeneity. The variability was calculated by taking the maximum difference of the combined signals at different central frequencies. \fref{of3} (d) and (e) give the frequency-dependent signal variability for the spiral-out acquisition as a function of different TRs and flip angles for $n_c$ = 10, and different $n_c$ and flip angles at a fixed TR of 15 ms, respectively. Supporting Information \fref{osf1} (d) and (e) shows the same relationship for a spiral-in acquisition.  Note that the small central frequency dependent variations were averaged across $1/\text{T}_\text{OSSI}$ for \fref{of3} (a)-(c). To assess the $T_2^*$-weighting of OSSI in comparison to GRE using as long of a TE as possible (equivalent to a spiral-in acquisition), we plot the maximal $T_2^*$-weighted signal change vs. TR in the Supporting Information, \fref{osf2}.

\section{Methods}

All the studies were performed on a 3T GE MR750 scanner (GE Healthcare, Waukesha, WI) with a 32-channel head coil (Nova Medical, Wilmington, MA). We implemented the OSSI pulse sequence using the vendor's standard pulse programming language, EPIC, and collected data with matched spatial and temporal resolutions using both OSS and GRE approaches.

\subsection{Phantom Experiments}

\begin{figure}
\centering
\includegraphics[width=15cm]{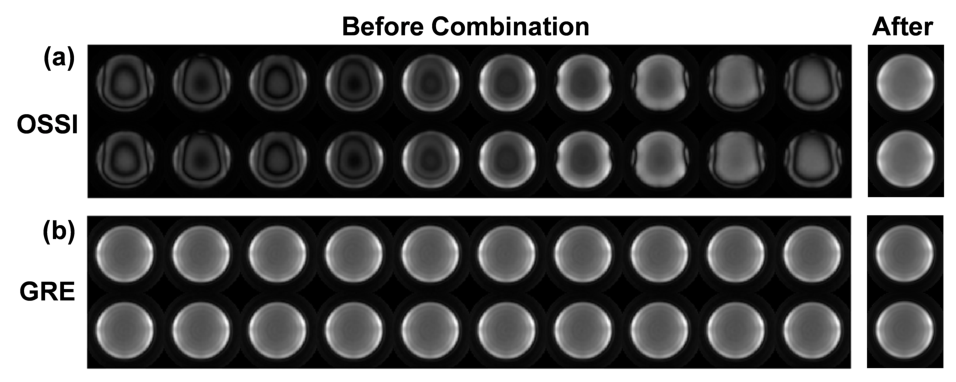}
\caption{Images of steady state with quadratic phase progression ($n_c$ = 10) with (a) balance gradients (OSSI) and (b) spoiling gradients (GRE). Each panel has 10 images across the periodic phase pattern and is shown twice to demonstrate the reproducibility. The 2-norm combined images are given on the right. The OSSI and GRE images are not on the same intensity scale.}
\label{of4}
\end{figure}

To demonstrate the principles of OSSI, we collected images of the FBIRN phantom \cite{Glover2012} (approximate $T_1$/$T_2$ = 530/60 ms) using both balanced and spoiled gradients. An oblique slice with FOV = 220 mm and slice thickness = 2.5 mm was acquired, and the voxel size = 6.29 $\times$ 6.29 $\times$ 2.5 mm$^3$.  For OSSI, we chose TR = 15 ms, $n_c$ = 10, FA = 10$^\circ$, and a fully sampled single-shot spiral-out trajectory. The spoiled-GRE data were acquired with the same parameters, except for the addition of spoiling gradients and the use of a spiral-in readout to make the effective TEs of the two acquisitions more similar. Specifically, the OSSI spiral-out TE = 2.7 ms, which corresponds to an effective TE of 17.5 ms. To bring GRE TE closer to OSSI effective TE and to increase GRE $T_2^*$-sensitivity with the limited TR = 15 ms, we used GRE spiral-in TE = 11.2 ms. The number of time points = 100 with 10 s discarded acquisition prior to collecting data. Every $n_c$ = 10 images (1 period of the oscillations) were combined pixel-wise using the 2-norm. 

\begin{figure}
\centering
\includegraphics[width=11cm]{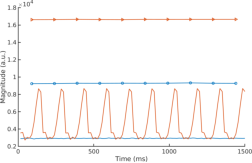}
\caption{Time courses for a 4-voxel ROI in the phantom for OSSI (red) and GRE (blue). Both before and after 2-norm combination, OSSI shows signal strengths roughly two times larger than the spoiled GRE signal.}
\label{of5}
\end{figure}

\subsection{Human Experiments}

\subsubsection{2D Human Studies}

Human functional imaging studies were performed on 5 subjects using both OSSI and GRE methods with informed consent and IRB approval. The functional task was a left vs. right reversing-checkerboard visual stimulus (with 5 cycles of 20 s L/20 s R).  The 2D sampling pattern for both GRE and OSSI was multi-shot (number of interleaves $n_i$ = 8) fully sampled variable-density spirals with a densely sampled core (300 k-space points). A single oblique slice through visual cortex was selected with FOV = 220 mm and 2.5 mm slice thickness. The voxel size was 1.77 $\times$ 1.77 $\times$ 2.5 mm$^3$ (matrix size 124 $\times$ 124). All the 2D images were reconstructed as 128 $\times$ 128 matrices. The experiments include 4 spiral-out acquisitions of 4 subjects and 4 spiral-in acquisitions of 4 subjects.

For the OSSI method, we chose TR = 15 ms, $n_c$ = 10, and nominal FA = 10$^\circ$. The OSSI effective TR for each spiral = 150 ms ($\text{TR} \cdot n_c$) and the volume TR = 1.2 s ($\text{TR} \cdot n_c \cdot n_i$). For the GRE acquisition, we carefully matched spatial-temporal resolution of OSSI, each interleave was acquired with GRE TR = 150 ms, volume TR = 1.2 s ($\text{TR} \cdot n_i$), and the Ernst flip angle FA = 27$^\circ$ to optimize SNR. The number of time points for OSSI was 1670 or 167 combined images, and the number of time points for GRE was 167 with no combination necessary, corresponding to the 200 s of the functional task. To establish the steady state, no data were collected for the first 10 s for both acquisitions.  OSSI actual TE was set to minimum (TE = 2.7 ms) for spiral-out imaging and TE = 11.6 ms for the spiral-in case.  Recognizing that the OSSI acquisition has some inherent $T_2^*$-weighting with spiral-out effective TE = 17.5 ms and spiral-in effective TE = 27.5 ms according to the simulations, we used slightly varying GRE TEs for different experiments to get an robust real data estimation of OSSI effective TE. For the 4 spiral-out experiments, we selected GRE TE = 17.5, 20, 20, and 23 ms, and for the 4 spiral-in experiments, we selected GRE TE = 27.5, 30, 30, and 33 ms.
 
Additionally, a $T_1$-weighted image was acquired for each subject and used to create a mask for the brain regions using the Brain Extraction Tool \cite{Smith2002}.

\subsubsection{3D Human Studies}

As an anecdotal demonstration, we acquired a 3D data set for a single human subject.  The functional study was the same visual stimulus as in 2D studies (5 cycles of 20s on/20s off). An oblique 12-slice 3D volume was acquired using a stack of single-shot spirals with spiral-out readouts.  The matrix size = 64 $\times$ 64 $\times$ 12, and the voxel size = 3.44 $\times$ 3.44 $\times$ 3 mm$^3$. For 3D OSSI imaging, TR = 15 ms, $n_c$ = 10, FA = 10$^\circ$, TE = 2.2 ms for each slice, and the volume TR = 1.8 s ($\text{TR} \cdot n_c \cdot n_z$). The spiral sampling trajectory in the $k_x$-$k_y$ plane was a variable-density spiral with a linearly decreasing sampling density, leading to a factor of 3 undersampling. Along the $k_z$ direction, the spirals were rotated 45$^\circ$ for each spiral platter to reduce undersampling artifacts. For the GRE imaging, the 12 slices were collected using a 2D spiral-out sequence with fully sampled uniform-density spirals, GRE TR = 1.8 s, TE = 23 ms to approximately match OSSI effective TE, and FA = 75$^\circ$.  The number of volumes = 112 for both OSSI (after 2-norm combination) and GRE, for a total about 200 s of acquisition, which followed 10 s of discarded acquisition used to establish the steady state.

We also acquired 2D multi-slice images using a standard spin-warp acquisition for generating SENSE maps. The 32-channel coil images were compressed to 28 virtual coils, and the SENSE maps were generated using ESPIRiT \cite{Uecker2014,Bart}. The 3D OSSI images were reconstructed from the undersampled measurements using the conjugate gradient SENSE \cite{Pruessmann2001,Sutton2003}, with an edge-preserving regularizer implemented through \cite{fesslermirt}. The fully sampled GRE data were reconstructed using the gridding method.     

\subsection{Data Analysis}

As mentioned above, every $n_c$ = 10 consecutive and non-overlapping OSSI images were combined by taking the 2-norm. Functional imaging performance was evaluated for both OSSI and GRE BOLD by evaluating activation maps and the temporal SNR (tSNR). The data from the first cycle (40 s) of the task were discarded to avoid the modeling error in the initial rest period. To reduce the effects of scanner drift, detrending was applied using lower order discrete cosine transform bases. The correlation coefficients were determined by correlation with a reference waveform, and the activated regions were defined by the magnitude of the correlation coefficients larger than a 0.5 threshold. The reference waveform was generated by convolving the canonical hemodynamic response function \cite{spm} with the task. The number of activated voxels were counted at the bottom third of the brain, where the primary visual cortex is located.  The tSNR maps were calculated by dividing the mean of the time course by the standard deviation of the time course residual (after removing the mean and the task) for each voxel. We calculated the average tSNR over the whole brain over an ROI limited to the brain region and excluding the skull and scalp. 
 
To determine the effective TE of OSSI, we generated scatter plots based on the percent signal change for voxels that were active in both GRE and OSSI acquisitions. In GRE, the percent signal change is approximately equal to $\Delta R_2' \cdot \text{TE}$ \cite{Jin2006}.  By establishing the relationship between OSSI and GRE percent change and under the assumption that activation change ($\Delta R_2'$) is the same in both cases, we can estimate the effective TE for OSSI using $\text{TE}_\text{eff} = b  \cdot  \text{TE}_\text{GRE}$ when the percent changes of OSSI and GRE are highly correlated, where $b$ is the slope of the OSSI-GRE percent change relationship.  Due to variability in both data sets, we performed a model II fit with standardized major axis (SMA) regression and 0 intercept to estimate the slope of the relationship for each experiment.  Voxels with a percent change greater than 4\% in either method, which likely represent vascular signals, were found to be highly variable and were excluded from the regression. In addition, the linearity of the relationship between OSSI and GRE percent signal changes was assessed using Pearson's correlation coefficient.

\section{Results}

\begin{figure}
\centering
\includegraphics[width=16cm]{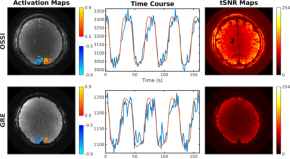}
\caption{OSSI and GRE functional results from multi-shot spiral-out acquisition with OSSI TE = 2.7 ms and GRE TE = 23 ms. At left, the activation map uses a threshold of 0.5 for the correlation with a reference waveform, and the background is the mean image of the OSSI combined or GRE images. The time course for a 4-voxel ROI is shown for each method together with the reference waveform (intensity units are arbitrary signal units). At right, the temporal SNR maps are also shown for both methods.}
\label{of6}
\end{figure}

The phantom images in \fref{of4} (a) present the evolution of the oscillation pattern for OSSI over the $n_c$ = 10 phase cycles and show the highly reproducible nature of the oscillations. Note that magnetic field inhomogeneity leads to an inhomogeneous spatial pattern in the OSSI data, and that different isochromats have different temporal patterns.  \fref{of4} (b) shows the same slice with spoiled gradients. Although the spoiled steady-state images are free of oscillations, their magnitudes are much lower.  The 2-norm combination of every non-overlapping $n_c$ = 10 OSSI images produces spatially and temporally uniform signals. The time courses in \fref{of5} show oscillating steady-state signal and the stable signal after the 2-norm combination.  In comparison to spoiled GRE with matched resolutions and TE, the OSSI signal strength was roughly 2 times larger than the spoiled signal, though the exact relationship is highly dependent on phantom/tissue $T_1$s and $T_2$s.

\begin{figure}
\centering
\includegraphics[width=16cm]{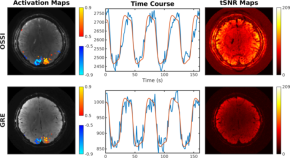}
\caption{OSSI and GRE functional results from multi-shot spiral-in acquisition with OSSI TE = 11.6 ms and GRE TE = 33 ms. The activation map uses a threshold of 0.5 for the correlation with a reference waveform, and the time course for a 4-voxel ROI is shown with the reference waveform for each method (intensity units are arbitrary signal units). The temporal SNR maps are also shown for both OSSI and GRE acquisitions. Compared to the spiral-out results in \fref{of6}, we can see that spiral-in gives more activations, but relatively lower signal strength and temporal SNR for both OSSI and GRE.}
\label{of7}
\end{figure}

\fref{of6} gives 2D human spiral-out activation maps, time courses, and tSNR maps for visual stimulation. 
Compared to GRE with matched acquisition characteristics and TE, the OSSI result shows more activations according to the activation maps, a larger task-related signal change as in the time course, and higher tSNR presented by the tSNR maps. The OSSI signals appear localized more in parenchyma with less signal from the sulci and vascular regions near the sagittal sinus.
 
\fref{of7} shows 2D human spiral-in functional results and tSNR maps.  The OSSI acquisition results in larger activation regions and much higher tSNR in comparison to GRE. Though anecdotal, the time course of OSSI appears to be less noisy. The spiral-in scheme uses a closer to optimal TE for fMRI (at 3T, a common choice is TE = 30 ms), thereby leading to more activations for both OSSI and GRE compared to the spiral-out results in \fref{of6} in spite of the overall lower signal intensity and tSNR. 

\begin{figure}
\centering
\includegraphics[width=16cm]{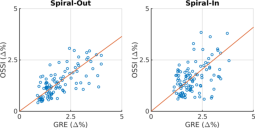}
\caption{Percent signal change of OSSI vs. GRE for active voxels in \fref{of6} and \fref{of7} where the percentage signal change was below 4\% in both methods (spiral out TEs: OSSI = 2.7 ms, GRE = 23 ms; spiral in: OSSI = 11.6 ms, GRE = 33 ms). These figures demonstrate a high correlation between the methods, indicating the potential utility of OSSI as an alternative to GRE fMRI.  The slope of the line was fit via Model II regression.}
\label{of8}
\end{figure}

\fref{of8} shows OSSI and GRE percent signal change scatter plots for spiral-out data of \fref{of6} and spiral-in data of \fref{of7}, and the slope of the scatter plots depicts the relationships between OSSI effective TE and GRE TE. The slope resulted from the SMA regression is 0.73 for spiral-out and 0.82 for spiral-in. As described in Methods, we can calculate the effective TE for OSSI as spiral-out OSSI TE\textsubscript{eff} = 16.7 ms, and spiral-in OSSI TE\textsubscript{eff} = 27.1 ms for this subject.  Scatter plots for the other subjects can be found in the Supporting Information, \fref{osf7}.  The mean OSSI TE effective across all the subjects is 17.8 ms for spiral-out, and is 27.1 ms for spiral-in, given the actual TE’s of 2.7 ms and 11.6 ms, respectively, which correspond to an effective $T_2'$-weighting of about 15 ms at the time of the excitation pulse for both spiral-out and spiral-in cases. The average correlation coefficient between OSSI and GRE across all subjects was 0.5, and linearity of the relationship was found to be significant ($p < 0.05$) for all data sets.
The high correlation of OSSI and GRE percent signals in the common activated regions is consistent with a similar contrast mechanism for OSSI and GRE acquisitions.

Quantitative measurements for all visual fMRI experiments including number of activated voxels (at the bottom third of the brain) and average tSNR of the whole brain are given in \tref{ot1}. OSSI shows a 1.84 ratio (s.d. = 0.5) of number of activation voxels in comparison to GRE with matched spatial-temporal resolutions and similar effective TEs. The tSNR ratio of OSSI to GRE has a mean of 1.83 (s.d. = 0.19). tSNR values were compared using a paired t-test, and OSSI was found to be significantly higher ($p < 0.05$).  The columns in \tref{ot1} directly corresponds to the columns in Supporting Information \fref{osf4}, which presents activation maps and tSNR maps for the 5 subjects. For each subject and GRE TE ranging from 17.5 ms to 33 ms, the OSSI acquisition provides larger activation regions and higher tSNR than GRE. Subject 2 demonstrated motion artifacts, which led to lower tSNR ratios, artifacts in the tSNR maps, and some false positive activations (near the edge of the brain). The circular spatial variation in tSNR maps are believe to result from pulsatile flow at ventricles and vessels in combination with the multi-shot (8-shot) acquisition. When averaging tSNR over an ROI that is away from artifacts, the tSNR ratio of OSSI to GRE is generally greater than 2.
 
\begin{figure}
\centering
\includegraphics[width=12cm]{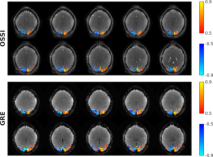}
\caption{Functional MRI of 10 slices drawn from volumetric 3D OSSI acquisition (volume TR = 1.8 s, TE = 2.2 ms, matrix size = 64 through undersampling in-plane) and 2D multi-slice GRE (TR = 1.8 s, TE = 23 ms, matrix size = 64) showing similar activation patterns in visual cortex.}
\label{of9}
\end{figure}

\fref{of9} is a preliminary demonstration of 3D activation results in visual cortex. OSSI and GRE acquisitions give comparable activation maps even through the OSSI data were undersampled. For OSSI, the number of activated voxels = 705 and the average tSNR = 57.2. For GRE, the number of activated voxels = 883 and the average tSNR = 62.4.

\begin{table}
\caption{Quantitative results including number of activated voxels and average tSNR.}
\label{ot1}
\begin{threeparttable}
\begin{tabular}{|c|c|c|c|c|c|c|c|c|c|c|} \hline
\multicolumn{2}{|c|}{} & \multicolumn{4}{c|}{\textbf{Spiral-Out}} & \multicolumn{4}{c|}{\textbf{Spiral-In}} & \\ \hline
\multicolumn{2}{|c|}{Subject ID} & 1 & 2 & 3 & 4 & 1 & 2 & 3 & 5 & Mean (SD)\\ \hline
& OSSI & 215 & 159 & 210 & 84 & 264 & 165 & 236 & 123 & 182 \\ \cline{2-11} 
\multirow{-1.5}{*}
{\begin{tabular}[c]{@{}c@{}}\# Activated\\ Voxels\end{tabular}}
& GRE & 133 & 113 & 116 & 55 & 151 & 84 & 144 & 41 & 105 \\ \cline{2-11} 
& Ratio & 1.62 & 1.41 & 1.81 & 1.53 & 1.75 & 1.96 & 1.64 & 3.0 & 1.84 (0.5) \\ \hline
\multirow{3}{*}
{\begin{tabular}[c]{@{}c@{}}Average\\ tSNR\end{tabular}}
& OSSI & 85.1 & 55.1 & 74.7 & 68.2 & 71.4 & 47.2 & 60.6 & 47.9 & 63.8 \\ \cline{2-11} 
& GRE & 40.9 & 34.9 & 41.9 & 37.8 & 34.6 & 29.5 & 34.1 & 24.7 & 34.8 \\ \cline{2-11} 
& Ratio & 2.08 & 1.58 & 1.78 & 1.80 & 2.06 & 1.60 & 1.78 & 1.94 & 1.83 (0.19) \\ \hline
\end{tabular}
\begin{tablenotes}
\item OSSI, oscillating steady-state imaging; GRE, gradient echo imaging; tSNR, temporal signal-to-noise ratio.
\end{tablenotes}
\end{threeparttable}
\end{table}

\section{Discussion}

This chapter describes a fundamentally new approach to fMRI acquisition that uses a novel oscillating steady-state source of signal that is very large and also sensitive to the blood oxygenation, thereby offering the potential for high SNR fMRI.  The proposed quadratic phase progression in conjunction with balanced gradients produces this new steady state. As with other steady-state imaging methods, the OSSI method has large signals because it reuses rather than spoils the magnetization.  The oscillating steady-state signals available prior to gradient dephasing contain typically more than twice the average signal amplitude of spoiled signals. We have also noted that this pulse sequence with its quadratic phase sequence is very sensitive to off-resonance.  Indeed, a frequency dependent phase dispersal is important for generating the $T_2'$- or $T_2^*$-contrast that makes it suitable for fMRI.  In our experiments, we found $T_2'$-weighting of approximately 15 ms at the time of excitation pulse. Additional $T_2'$ weighting can be obtained with increased TE as shown in Supporting Information \fref{osf3}. 
 
OSSI signals oscillate with a periodicity of $n_c$TR, however, the oscillations are highly reproducible, and by combining $n_c$ time points generates stable time courses required for fMRI analysis. As demonstrated in the high-resolution visual stimulation fMRI study, the OSSI approach improves tSNR by about 83\% and the number of activated voxels was increased by about 84\%, both relative to GRE imaging at the Ernst angle with the carefully matched spatial-temporal acquisition characteristics and effective TEs. 
The 2D human fMRI experiments used relatively high spatial resolution and thus were closer to thermal noise limit.  The same data were subjected to low spatial resolution reconstructions with the results shown in the Supporting Information, \fref{osf10} and \tref{ost1}.  While the OSSI data still had SNR advantages, the SNR gain of OSSI was reduced relative to the high-resolution and more thermal noise limited cases. Similarly for the anecdotal low-resolution 3D human data that was likely to be physiological noise limited and for the low-resolution phantom data that is systematical noise limited \cite{Kruger2001}, the SNR advantage is compromised.

Because we acquired GRE with a longer TE, there is a possible concern that the longer TR might alter temporal noise characterstics, so we compared the OSSI method to GRE with a 3$\times$ shorter GRE TR. As shown in Supporting Information \fref{osf11} and \tref{ost2}, for the experimental conditions used (high spatial resolution, likely thermal noise limited regime), the shorter TR leads to similar functional results and tSNR values as GRE with longer TR.
We also considered the possible use of other steady state methods, such as FISP/S1 SSFP and found that while it has more signal than GRE for short TRs, it does not have the increased $T_2^*$ sensitivity of OSSI.

We note that further improvements in performance are possible and in fact, likely.  For example, our simulations show that TR = 15 ms, $n_c$ = 10, and FA = 10$^\circ$ is a good combination to get high SNR and functional MRI responses, but it is by no means optimal. The short readouts can limit SNR efficiency, so there are potential advantages to going to longer TRs and longer readouts.  As shown in \fref{of3} and Supporting Information \fref{osf1}, multiple combinations of imaging parameters give a similar $T_2^*$-sensitivity and off-resonance sensitivity. There is a complex interplay between these sensitivity measures and the major imaging parameters including TR, $n_c$, FA, and TE (including TE locations from different readouts, e.g. spiral-out, spiral-in, or EPI with TE in the center). The RF (FA) inhomogeneity in the brain at 3T may influence the actual FA to use when acquiring slices at different parts of the brain, which further complicates the optimization.  We also note that the optimal FA appears to be small in comparison to many bSSFP applicaitons where FAs $>30^\circ$ are common, which would indicate that RF heating is unlikely to be an issue with OSSI.  Curiously, the optimal FAs are often not far from the Ernst angle, e.g. 8.3$^\circ$ for TR = 15 ms.  The final optimization will require practical experience regarding which factors are most important for particular fMRI studies.
 
The sensitivity to frequency as noted above, will lead to substantial physiological noise, and in particular, artifacts and noise from respiration, which is known to lead to oscillations shifts in resonant frequency \cite{Noll}. \fref{osf9} presents residual time courses and spectra of OSSI and GRE at a non-active region, and the OSSI spectrum shows a prominent peak near respiration frequencies.  We investigated the standard physiological noise removal technique RETROICOR \cite{Glover2000} applied to individual temporal phases as well as the combined images, and also k-domain methods (RETROKCOR \cite{Hu1995}) applied to individual temporal phases.  We found only modest improvements in tSNR and activation maps when applying corrections over limited time windows and no improvements over longer windows \cite{ossi}.  This, we believe, is due to the complex and non-linear nature of the interaction between frequency and the temporal signal (see \fref{of2} (c, d), for example).  In addition, the use of a 2D slice for the visual study makes it sensitive to inflow and pulsatility artifacts.  Physiological noise correction is an active area of research \cite{Amos19} and will be the topic of a future manuscript.  As such, no physiological noise corrections for OSSI were applied in the present work, but we believe that after correction, further tSNR and activation improvements close to the increases in signal strength will be possible.
 
Like most steady-state methods, the short TR largely prevents interleaving of slices, when combined with the time needed to reach steady state, dictates that OSSI methods are best suited to 3D acquisitions.  Furthermore, the need to acquire volumetric images for each temporal phase implies that $n_c$ times as many images are required for a study.  Fortunately, the reproducible nature of oscillating signal may allow dramatic reductions in the acquisition time.  For example, the use of sparse sampling in k-space and modeling of the oscillations using patch-tensor low-rank \cite{tensor} or a dictionary based regularizer \cite{dictionary} can fully recover the missing data in the image reconstruction process.  This again, is a topic of active research, and preliminary results suggest that larger than a 13-fold reduction in k-space is possible with minimal performance degradation. 
We have anecdotally demonstrated the ability to acquire 3D images, though without acceleration using the spatiotemporal models described here.  
We note that most 2D acquisitions would include acceleration using 2D simultaneous multi-slice imaging \cite{Moeller2010,Setsompop2012}, but also note that undersampling in 3D exploits roughly the same parallel imaging concepts \cite{Zahneisen2014}. So, we believe that similiar accelerations are possible for OSSI and the use of temporal modeling will help resolve the inefficiency of acquiring $n_c$ images.  
The slow volume TR reduces temporal resolution, but does not reduce SNR due to averaging of signal and noise across the $n_c$ temporal phases.  As pointed out above, the short TR does limit the length of the readout which can reduce the SNR efficiency.

In prior work, bSSFP imaging for fMRI has taken advantage of different phenomena, for example, frequency shifts, changes in $T_2$ associated with changes in blood oxygenation, or changes due to inhomogeneous effects and diffusion around small vessels \cite{Zhong2007,Miller2008}.  In this work, we argue that the OSSI signal changes are due to more traditional, size-scale invariant changes in $T_2^*$ or $T_2'$ of the tissues, again in response to changes in blood oxygenation. We argue that this sensitivity is due to frequency sensitivity of OSSI signal that leads to frequency-dependent phase variations as shown in the simulations of \fref{of1} (d).  The percent signal plots in \fref{of8} and \fref{osf7} clearly show a very linear relationship (average $p$-value for the slope = 0.01 using standard linear regression) of OSSI and GRE percent signal changes, which would be consistent with a similar signal change mechanism between the two methods. We note that further work is necessary to fully elucidate the mechanism, including the effects of diffusion around vessels. The simulation and percent signal change analysis are both consistent with the OSSI signal being inherently $T_2^*$ or $T_2'$ weighted, specifically OSSI leads to an additional $T_2'$-weighting of approximately 15 ms for the parameters used (TR = 15 ms, $n_c$ = 10, FA = 10$^\circ$).

The analysis of percent change signal in OSSI and GRE excluded voxels with a percent change greater than 4\% in at least one of the methods.  Above 4\%, the signal change for OSSI seemed to flatten and the relationship was no-longer suitable for linear regression.  These very high GRE percent changes, which likely represent vascular signals as shown in \fref{osf8}, have had a lower signal change in OSSI perhaps due to flow-related signal changes. If so, this partial suppression of vascular signals could be seen as a desirable feature as it will improve functional localization.

There are a number of unstudied phenomena we wish to address in the future. Long $T_2$-species like cerebrospinal fluid in the ventricles are very bright in OSSI, but when combined with cardiac pulsatility lead to low tSNR as seen in the first and third rows of \fref{osf4}.  Part of the high variability may arise from the in-flow effects associated with the 2D acquisition and may be partially resolved by 3D imaging.  Pulsatile effects and in-flow phenomena with vessels require further investigation.  The short TR makes implementation of fat suppression more challenging, however, the relaxation and spectral characteristics of lipids seem to lead to relatively low signal intensity and limited artifacts in the images.  Still, the signal characteristics of lipids, as well as the possible use of slab-selective spectral spatial pulses, should be investigated. As with most fMRI studies, detection and bulk correction of head motion will be needed.  In this case, we will also need to consider any impact on the steady-state signal due to head motion. 

Another interesting question is what occurs in the presence of large magnetic fields gradients near regions of large susceptibility differences in the brain, for example, the orbitofrontal cortex.  Such gradients might have a similar impact as applying unbalanced gradients, leading to signal spoiling and a reduction of the additional $T_2'$-weighting of 15 ms common to OSSI. The signal may gracefully transition to a spoiled GRE signal with relatively short TE. This phenomenon is closely related to partial spoiling described by Ganter \cite{Ganter2006}, except that very small phase increments with large gradients are used in Ganter’s paper, while here we have large phase increments between RF pulses but partial gradient spoiling.  We are also interested in other possible applications of the OSSI signal.  Frequency sensitivity may be useful in applications where frequency tracking is needed, for example, in tracking temperature-dependent frequency changes in therapeutic ultrasound.

\section{Conclusion}
The OSSI approach departs from traditional acquisition approaches by exploiting a novel $T_2^*$-weighted signal mechanism that produces large steady-state signals, and to our knowledge, has never been used before for fMRI. We show in both simulations and experimental data that the proposed approach has a similar contrast mechanism and percent signal change as GRE and leads to a substantial increase in signal strength and tSNR with matched spatial-temporal resolutions and effective TE, thereby enabling detection of 84\% greater volumes of functional activity. The SNR advantages were shown for a specific case of single slice fMRI using a short TR, and extensions to volumetric acquisition and implementation of physiological noise corrections will be critical for general application.  Still, this approach offers the prospect of high-resolution fMRI without the need for higher magnetic field strength systems.
\newpage

\section{Supporting Information}

The supplemental material presents Oscillating Steady-State Imaging (OSSI) simulation and human data results. 

Simulations consist of acquisition parameter optimization for spiral-in readouts, OSSI to GRE $T_2^*$-sensitivity comparison, and how OSSI $T_2^*$-weighting changes with increased TE. Human data results include activation maps, tSNR maps, z-maps, histograms of all 8 visual experiments, additional percent signal change plots, maps showing vascular activations, residual time courses, low spatial resolution reconstructions, and comparison to GRE TR = 50 ms.

\subsection{Simulations}

\begin{figure}
\centering
\includegraphics[height=15cm]{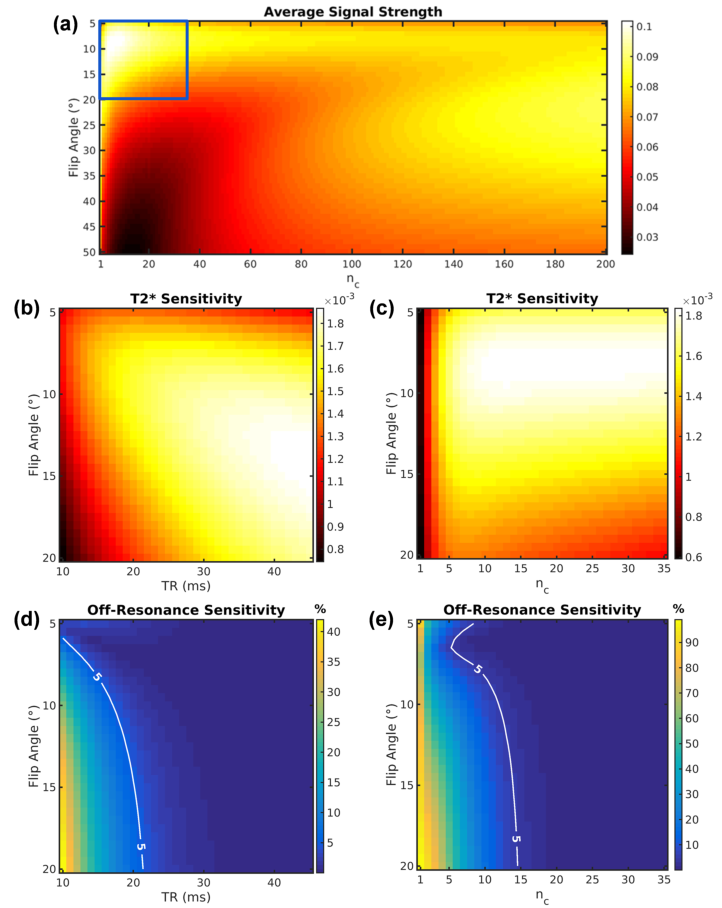}
\caption{Simulation of acquisition parameters for spiral-in readouts (TE = TR - 1.6 ms). (a) to (c) are in units of $M_0$ = 1.\\
(a) shows the RMS combined magnitude signal as a function of $n_c$ and flip angle for a fixed TR of 15 ms. We focus on the region denoted by the blue square for OSSI fMRI acquisition parameter optimization, and the results are given in (b) to (e).
(b) shows how $T_2^*$ sensitivity (S\textsubscript{activated} – S\textsubscript{rest}) varies with TR and flip angle for a fixed $n_c$ = 10. The signal is normalized by $\sqrt{(\text{TR}-c)/\text{TR}} \approx \sqrt{\text{T}_\text{A/D}}$ with $c$ = 5 ms for SNR efficiency.
(c) shows how $T_2^*$ sensitivity varies with $n_c$ and flip angle for TR = 15 ms.
(d) gives off-resonance sensitivity at different TR and flip angles for $n_c$ = 10.
(e) gives off-resonance sensitivity at different $n_c$ and flip angles for TR = 15 ms. 
}
\label{osf1}
\end{figure}

\begin{figure}
\centering
\includegraphics[width=12cm]{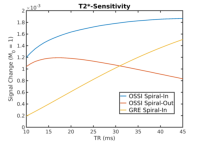}
\caption{$T_2^*$-sensitivity (S\textsubscript{activated} – S\textsubscript{rest} in units of $M_0$) changes with varying TR for GRE spiral-in (TE = TR - 1.6 ms), OSSI spiral-out (TE = 1.6 ms), and OSSI spiral-in (TE = TR - 1.6 ms). The signals are normalized by $\sqrt{(\text{TR}-c)/\text{TR}} \approx \sqrt{\text{T}_\text{A/D}}$ with $c$ = 5 ms for SNR efficiency and are maximized over flip angle for each method.}
\label{osf2}
\end{figure}

\begin{figure}
\centering
\hspace{-2em}
\includegraphics[width=15cm]{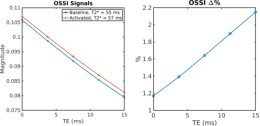}
\caption{Simulated OSSI $T_2^*$-weighting and percent signal increase almost linearly with increased TE for TR = 15 ms, $n_c$ = 10, and flip angle = 10$^\circ$.}
\label{osf3}
\end{figure}

\fref{osf1} presents spiral-in acquisition parameter optimization results. OSSI to GRE $T_2^*$-sensitivity comparison as a function of TR is shown in \fref{osf2}. OSSI $T_2^*$-weighting increases almost linearly with increased TE as given in \fref{osf3}.

\subsection{Human Data}

\subsubsection{All Visual Experiments}

\begin{figure}
\centering
\hspace{-3em}
\includegraphics[width=16cm]{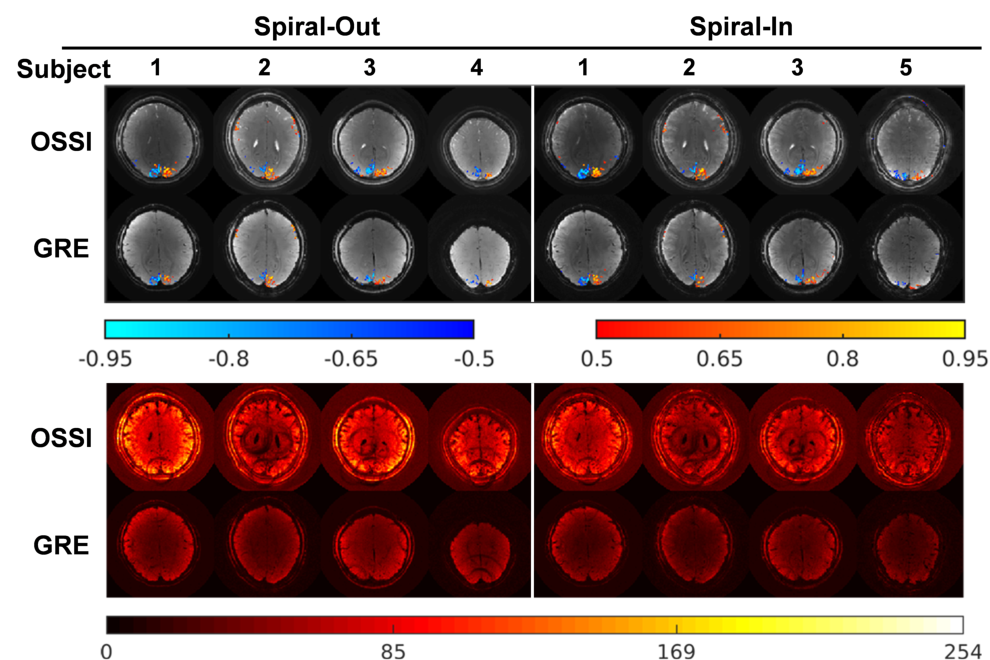}
\caption{Comparison of OSSI and GRE activation maps and tSNR maps for all 5 subjects.}
\label{osf4}
\end{figure}

\begin{figure}
\centering
\hspace{-3em}
\includegraphics[width=16cm]{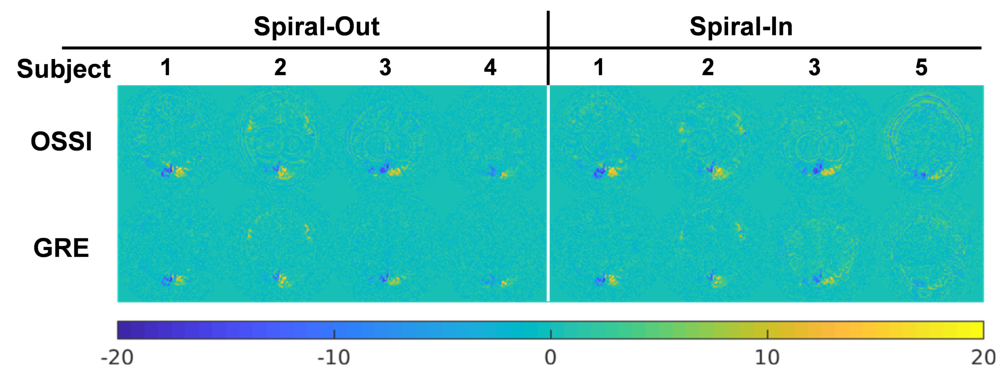}
\caption{Z-maps of all the experiments.}
\label{osf5}
\end{figure}

\begin{figure}
\centering
\includegraphics[width=12cm]{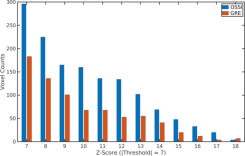}
\caption{Histograms of voxel counts over z-score threshold = $\pm 7$, which corresponds to correlation = $\pm 0.5$ for the GRE TR = 150 ms case.}
\label{osf6}
\end{figure}

\begin{figure}
\centering
\hspace{-3em}
\includegraphics[width=8.8cm]{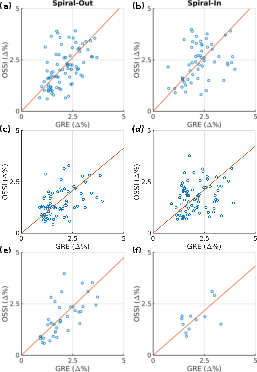}
\caption{Percent signal change of OSSI vs. GRE for active voxels for subjects 2-5 where the percentage signal change was below 4\% in both methods. Actual OSSI TE is 2.7 ms for spiral-out and is 11.6 ms for spiral-in. \\  These figures demonstrate a high correlation between the methods, indicating the potential utility of OSSI as an alternative to GRE fMRI.  The slope of the line was fit via Model II regression.\\
(a) subject 2 spiral-out acquisition, GRE TE = 20 ms, slope = 1.05, and OSSI TE\textsubscript{eff} = 21 ms.\\
(b) subject 2 spiral-in acquisition, GRE TE = 30 ms, slope = 1.06, and OSSI TE\textsubscript{eff} = 31.7 ms.\\
(c) subject 3 spiral-out acquisition, GRE TE = 17.5 ms, slope = 0.83, and OSSI TE\textsubscript{eff} = 14.5 ms.\\
(d) subject 3 spiral-in acquisition, GRE TE = 27.5 ms, slope = 0.85, and OSSI TE\textsubscript{eff} = 23.2 ms.\\
(e) subject 4 spiral-out acquisition, GRE TE = 20 ms, slope = 0.95, and OSSI TE\textsubscript{eff} = 19 ms.\\
(f) subject 5 spiral-in acquisition, GRE TE = 30 ms, slope = 0.87, and OSSI TE\textsubscript{eff} = 26.2 ms.
}
\label{osf7}
\end{figure}

\begin{figure}
\centering
\includegraphics[width=10cm]{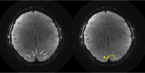}
\caption{GRE background image with vasculature (left) and the activated voxels with $>$ 4\% percent signal changes overlaid to the GRE background image (right).}
\label{osf8}
\end{figure}

\begin{figure}
\centering
\includegraphics[width=12cm]{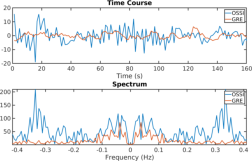}
\caption{The residual time courses and spectra averaged over a larger ROI (20$\times$20 voxels) away from the active regions after mean and drift removal. The large ROI eliminates the effect of thermal noise. The OSSI spectrum has higher physiological noise due, in part, to larger signals, but the presence of a prominent peak near respiration frequencies demonstrates potential greater sensitivity of physiological noise.  The phyisological noise may cause the tSNR improvement to be less than that predicted from signal strength alone.}
\label{osf9}
\end{figure}

\fref{osf4} presents activation maps and tSNR maps for all visual fMRI experiments. For each experiment, the OSSI acquisition provides larger activation regions and higher tSNR compared to the standard GRE approach. \fref{osf5} shows the corresponding z-maps, and \fref{osf6} histograms compare OSSI and GRE voxel counts over a z-threshold. The relationships between OSSI and GRE percent signal changes of subject 2-5 are in \fref{osf7}. \fref{osf8} shows the potential vasculature nature for signals where percent signal changes were large and thus excluded from effective TE analysis. \fref{osf9} contains example residual time courses and spectra averaged over a large ROI away from activations comparing noise patterns of OSSI and GRE. 

\subsubsection{Low Spatial Resolution Reconstructions}

\begin{figure}
\centering
\hspace{-3em}
\includegraphics[width=14cm]{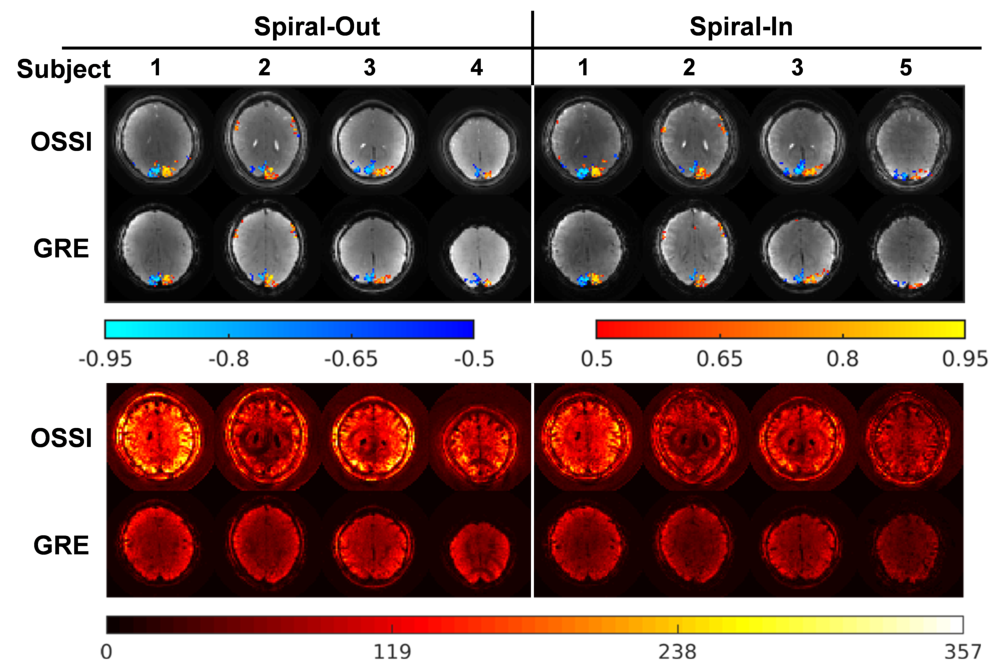}
\caption{Comparison of OSSI and GRE activation maps and tSNR maps for all the experiments reconstructed at a lower spatial resolution.}
\label{osf10}
\end{figure}

Human data shown in \fref{osf4} were reconstructed with a limited k-space (64/FOV) to obtain images with a lower spatial resolution. The activation maps and tSNR maps in \fref{osf10} and quantitative measurements in \tref{ost1} demonstrates a reduced SNR advantage of OSSI to GRE at lower resolutions.  

\begin{table}
\caption{Quantitative results including number of activated voxels and average tSNR from low spatial resolution reconstructions.}\label{ost1}
\begin{threeparttable}
\begin{tabular}{|c|c|c|c|c|c|c|c|c|c|c|} \hline
\multicolumn{2}{|c|}{\textbf{Low-Resolution}} & \multicolumn{4}{c|}{\textbf{Spiral-Out}} & \multicolumn{4}{c|}{\textbf{Spiral-In}} & \\ \hline
\multicolumn{2}{|c|}{Subject ID} & 1 & 2 & 3 & 4 & 1 & 2 & 3 & 5 & Mean (SD)\\ \hline
& OSSI & 124 & 90 & 118 & 51 & 129 & 97 & 134 & 98 & 105 \\ \cline{2-11} 
\multirow{-1.5}{*}
{\begin{tabular}[c]{@{}c@{}}\# Activated\\ Voxels\end{tabular}}
& GRE & 89 & 62 & 88 & 53 & 92 & 72 & 116 & 97 & 84 \\ \cline{2-11} 
& Ratio & 1.39 & 1.45 & 1.34 & 0.96 & 1.40 & 1.35 & 1.16 & 1.01 & 1.26 (0.19) \\ \hline
\multirow{3}{*}
{\begin{tabular}[c]{@{}c@{}}Average\\ tSNR\end{tabular}}
& OSSI & 138.3 & 78.9 & 118.1 & 94.9 & 109.9 & 66.4 & 88.9 & 65.6 & 95.1 \\ \cline{2-11} 
& GRE & 68.7 & 60.6 & 72.6 & 60.0 & 58.0 & 50.8 & 55.9 & 36.2 & 57.8 \\ \cline{2-11} 
& Ratio & 2.01 & 1.3 & 1.63 & 1.58 & 1.89 & 1.31 & 1.59 & 1.81 & 1.64 (0.26) \\ \hline
\end{tabular}
\begin{tablenotes}
\item OSSI, oscillating steady-state imaging; GRE, gradient echo imaging; tSNR, temporal signal-to-noise ratio.
\end{tablenotes}
\end{threeparttable}
\end{table}

\subsubsection{Comparison to GRE TR = 50 ms}

\begin{figure}
\centering
\includegraphics[width=12cm]{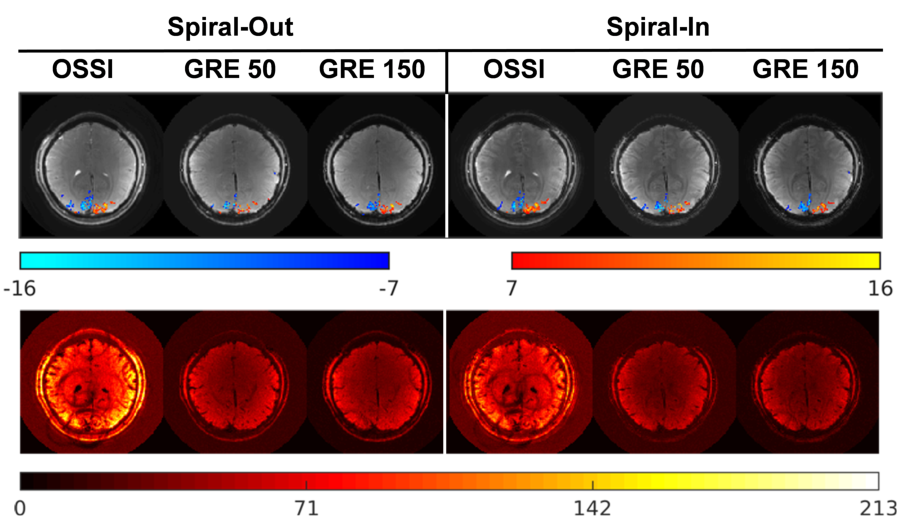}
\caption{Z-maps and tSNR maps of OSSI, GRE TR = 50 ms, and GRE TR = 150 ms for both spiral-out and spiral-in acquisitions. The z-score threshold = $\pm 7$ and corresponds to correlation = $\pm 0.5$ for the GRE TR = 150 ms case. For GRE TR = 50 ms, averaged images of every 3 time points are used for tSNR calculation.}
\label{osf11}
\end{figure}

We further compared OSSI to GRE TR = 50 ms with 3$\times$ more time points. It is shown in \fref{osf11} and \tref{ost2} that GRE TR = 50 ms or 150 ms for each interleave give similar functional results and tSNR values for the thermal noise limited experimental conditions.

\begin{table}
\caption{Quantitative measures including number of voxels beyond a z-score threshold of $\pm 7$ and average tSNR within the brain.}\label{ost2}
\begin{threeparttable}
\begin{tabular}{|c|c|c|c|c|c|c|}
\hline
\multirow{2}{*}{} & \multicolumn{3}{c|}{Spiral-Out} & \multicolumn{3}{c|}{Spiral-In} \\ \hline
& OSSI 
& {\begin{tabular}[c]{@{}c@{}}GRE TR\\ 50 ms\end{tabular}}
& {\begin{tabular}[c]{@{}c@{}}GRE TR\\ 150 ms\end{tabular}}
& OSSI 
& {\begin{tabular}[c]{@{}c@{}}GRE TR\\ 50 ms\end{tabular}}
& {\begin{tabular}[c]{@{}c@{}}GRE TR\\ 150 ms\end{tabular}}
\\ \hline
\begin{tabular}[c]{@{}c@{}}\# Activated\\ Voxels\end{tabular} & 182 & 120 & 125 & 194 & 145 & 128 \\ \hline
{\begin{tabular}[c]{@{}c@{}}Average\\ tSNR\end{tabular}} & 74.4 & 41.1 & 41.9 & 60.5 & 36.3 & 34.0 \\ \hline
\end{tabular}
\begin{tablenotes}
\item OSSI, oscillating steady-state imaging; GRE, gradient echo imaging; tSNR, temporal signal-to-noise ratio.
\end{tablenotes}
\end{threeparttable}
\end{table}
 
\chapter{High-Resolution Oscillating Steady-State fMRI using Patch-Tensor Low-Rank Reconstruction}
\label{chap:2tensor}
The goals of fMRI acquisition include high spatial and temporal resolutions with a high signal to noise ratio (SNR). Oscillating Steady-State Imaging (OSSI) is a new fMRI acquisition method that provides large oscillating signals with the potential for high SNR, but does so at the expense of spatial and temporal resolutions. The unique oscillation pattern of OSSI images makes it well suited for high-dimensional modeling. We propose a patch-tensor low-rank model to exploit the local spatial-temporal low-rankness of OSSI images. We also develop a practical sparse sampling scheme with improved sampling incoherence for OSSI. With an alternating direction method of multipliers (ADMM) based algorithm, we improve OSSI spatial and temporal resolutions with a factor of 12 acquisition acceleration and 1.3 mm isotropic spatial resolution in prospectively undersampled experiments. The proposed model yields high temporal SNR with more activation than other low-rank methods. Compared to the standard gradient echo (GRE) imaging with the same spatial-temporal resolution, 3D OSSI tensor model reconstruction demonstrates 2 times higher temporal SNR with 2 times more functional activation. \footnote{
This chapter was published in \cite{tensor2020,tensor,tensor2}.
}

\section{Introduction}
\label{sec:introduction}
Functional magnetic resonance imaging (fMRI) measures neural activity based on blood-oxygenation-level-dependent (BOLD) contrast and the hemodynamic correlations \cite{Huettel2009FunctionalImaging} by acquiring a time series of $T_2^*$-weighted brain images. BOLD signal change from fMRI images acquired with the standard gradient echo (GRE) imaging is small and can be easily buried in noise. Furthermore, as signal to noise ratio (SNR) is proportional to voxel size and functional units of the brain are on the order of 1 mm or smaller, high SNR is critical for high-resolution and high-quality fMRI. However, current methods for SNR improvements are limited:
multi-coil head arrays suffer from diminished returns for deep brain structures,
and high magnetic field systems are costly.
This chapter focuses on Oscillating Steady-State Imaging (OSSI) \cite{ossi2019},
a new fMRI acquisition method that has the potential to provide 2 times higher SNR
than the standard GRE approach. 

OSSI combines balanced gradients and a quadratic RF phase progression with large phase increments,
and leads to a combination of high SNR of the balanced steady state and $T_2^*$-weighting of GRE imaging.
The quadratic RF phase cycling is $\phi(n) = {\pi n^2}/{n_c}$,
where $n$ is the RF index and $n_c$ is the cycle length.
For $n_c = 1$, $\Delta\phi$ between RF pulses is 180$^\circ$,
which is balanced steady-state free precession (bSSFP).
For $n_c \geq 120$ with very small $\Delta\phi$,
the mechanism leads to bSSFP-like contrast \cite{Foxall2002}.
OSSI acquisitions use $1 < n_c < 120$ that produce large and oscillating signals.
Specifically, by selecting a short repetition time (TR) with $n_c = 10$,
OSSI demonstrates a similar $T_2^*$-weighted contrast mechanism as GRE
with additional $T_2'$-weighting of about 15~ms immediately after the RF pulse.
Details on how the SNR and $T_2^*$-sensitivity vary with $n_c$ and other acquisition parameters
can be found in \cite{ossi2019}.

The OSSI signal oscillates with a periodicity dictated by the quadratic RF phase cycling,
and OSSI images have a periodic oscillation pattern that repeats every $n_c$
images as illustrated in Figs.~\ref{ossi}, \ref{tsf1} and \ref{tsf2}.
Thus, one must acquire and combine $n_c$ as many images
to get images that are free of oscillations and suitable for fMRI analysis.
With standard reconstruction methods, this need would compromise temporal resolution by a factor of $n_c$,
and the short TR requirement necessary for steady-state imaging (e.g., TR = 15~ms)
limits the time for traversing k-space
and thus limits the single-shot spatial resolution.
We aspire to improve the spatial and temporal resolutions
by designing a sparse sampling scheme and an accurate reconstruction method.

Past works on reconstructing fMRI time series use models such as low-rank \cite{Chiew2015K-tConstraints}, low-rank and sparse \cite{Lam2013}, and low-rank plus Fourier domain sparsity \cite{Petrov2017ImprovingReconstruction,Otazo2015Low-rankComponents} that impose low-rankness and/or sparsity on matrices of the vectorized space dimension and time.
We found them insufficient for OSSI,
as the oscillations in OSSI images make them neither low-rank nor sparse along the time dimension.
To simultaneously exploit redundancy in the oscillation pattern of OSSI
and the repeated acquisition for fMRI time courses,
we structure OSSI images to have two time dimensions and develop a patch-based tensor model.

Based on the $n$-rank definition \cite{KoldaTensor} and tensor nuclear norm \cite{liu2012tensor} for tensor competition \cite{liu2012tensor,Gandy2011TensorOptimization}, global tensor low-rank or low-rank plus sparse reconstruction models have been applied to dynamic MRI via space $x\, \times$ space $y\, \times$ time  \cite{Roohi2017Multi-dimensionalMRIb}, cardiac MRI via space $\times$ time $\times$ cardiac phases \cite{Ramb2017Low-rankMRI}, and quantitative cardiovascular magnetic resonance multitasking with multiple time dimensions \cite{Christodoulou2018}.

Instead of tensor nuclear norm, global tensor low-rank models have also been explored
via Tucker decomposition or higher-order SVD (HOSVD) \cite{KoldaTensor,DeLathauwer2000ADecomposition}
for dynamic MRI with sparse core tensors \cite{Yu2014MultidimensionalTransform},
high-dimensional MR imaging with sparsity constraints and tensor subspace
estimated from navigator data \cite{He2016},
multi-dimensional dynamic phosphorus-31 magnetic resonance spectroscopy and imaging \cite{Ma2017},
and electron paramagnetic resonance oxygen imaging \cite{Christodoulou2016}
with specialized sparse sampling strategies.
Furthermore, the CANDECOMP/PARAFAC (CP) decomposition \cite{KoldaTensor}
was exploited for multi-contrast dynamic cardiac MRI denoising \cite{Yaman2019}
and for tensor completion with designed regular sub-Nyquist sampling with applications
for fMRI acceleration \cite{Kanatsoulis2020}.

Previous patch-wise tensor low-rank models impose low-rank constraints on spatial submatrices of the tensor unfoldings \cite{Trzasko2013AReconstruction,Trzasko2013}, select patches with both local and non-local similarities and exploit patch-tensor low-rankness using HOSVD for multi-contrast MRI reconstruction \cite{Bustin2019}, or compare CP and Tucker decompositions for local and global low-rank tensor denoising \cite{Yaman2019}. Because both CP and Tucker decompositions require selection of tensor ranks, our work focuses on tensor nuclear norm minimization that avoids explicit selection of tensor ranks, and structures local patch-tensors
to exploit the local and high-dimensional spatial-temporal low-rankness.
We further design a sparse sampling scheme that
prospectively undersamples the data with a 12-fold acceleration for 2D and a 10-fold acceleration for 3D.
The proposed model provides high-resolution reconstructions with high temporal SNR (tSNR) and more functional activation than global tensor or matrix low-rank models.

Patch-tensor low-rank (patch-tensor LR) reconstruction and the sparse sampling schemes are new for fMRI, and the application to OSSI fMRI data is also new. Compared to standard GRE imaging, the proposed OSSI tensor model demonstrates a factor of 2 tSNR improvement for fMRI with 2 times larger functional activation.

The chapter is organized as follows.
Section~II presents notations and definitions for tensors.
Section~III proposes the patch-tensor model and optimization algorithm.
Section~IV develops the incoherent undersampling
and describes the experimental setup for OSSI fMRI studies.
Section~V demonstrates the improved functional performance using the proposed approach
compared to other reconstruction and acquisition methods.
Section~VI discusses future directions,
and Section~VII concludes the chapter.

\section{Background and Notation}
\label{sec:def}

A tensor is a multidimensional array \cite{KoldaTensor}.
We denote tensors according to their dimensions.
One-dimensional tensors or vectors are denoted by bold lowercase letters, e.g., $\x$,
and tensors of dimension two or higher are denoted by bold capital letters, e.g., $\X$.
Scalars are denoted by italic letters, e.g., $x$.

The inner product of two tensors $\X,\mathbf{Y} \in \mathbb{C}^{I_1 \times I_2 \times \cdots \times I_N}$ is defined as the sum of the element-wise products \cite{DeLathauwer2000ADecomposition},
$$
\langle \X,\mathbf{Y} \rangle = \sum_{i_1 = 1}^{I_1} \sum_{i_2 = 1}^{I_2} \cdots \sum_{i_N = 1}^{I_N} y_{i_1 i_2 \cdots i_N}^* x_{i_1 i_2 \cdots i_N},
$$
where $*$ denotes the complex conjugate.
Naturally, the norm of tensor $\X$ is
$
\lVert \X \rVert = \sqrt{\langle\X,\X\rangle}.
$

\begin{figure}
\centering
\includegraphics[width=0.95\textwidth]{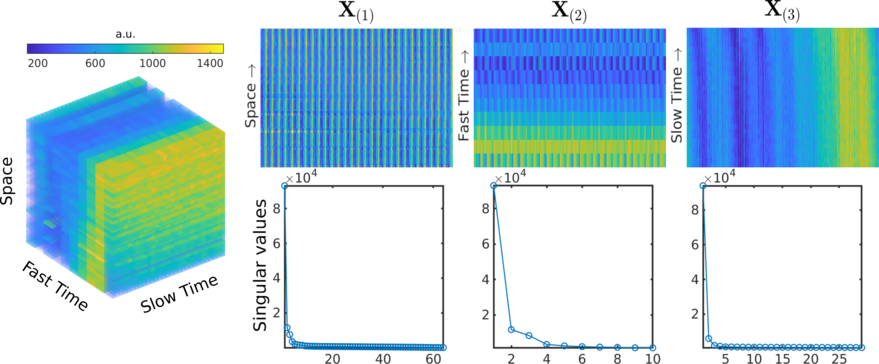}
\caption{A 3D patch-tensor (left), its three matrix unfoldings of different modes (top right), and the singular values of the unfoldings demonstrating the patch-tensor low-rank (bottom right).
}\label{rank}
\end{figure}

The process of reforming a tensor to matrices by reordering the vectors of the tensor
is known as matricization or unfolding.
Each dimension of a tensor is known as a mode,
and the number of modes is known as the tensor's order or number of dimensions.
After unfolding, the tensor becomes matrices of different modes,
and the number of these matrices equals the number of dimensions.
Figure~\ref{rank} illustrates unfolding a three-dimensional tensor to three matrices.
The mode-$n$ unfolding of tensor $\X$
is denoted by $\X_{(n)}$,
accordingly, refolding the mode-$n$ matrix back to $\X$ is
$\textsc{Refold}_n\left(\X_{(n)}\right)$.
As seen in \cite{KoldaTensor} and \cite{DeLathauwer2000ADecomposition},
different papers may use different permutations of the vectors to get the unfoldings;
the specific order is unimportant as long as it is consistent. 

The $n$-rank of $\X$
is the column rank of $\X_{(n)}$ and is denoted by $\text{rank}\left(\X_{(n)}\right) = r_n$.
Therefore, $\X$ is a rank-$(r_1,r_2,\ldots,r_N)$ tensor.

\section{Reconstruction Methods}
\label{sec:recon}

\begin{figure}
\centering
\includegraphics[width=0.95\textwidth]{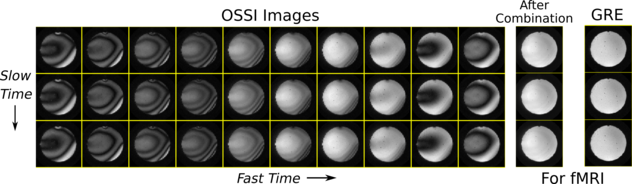}
\caption{OSSI images with periodic oscillation patterns are structured along ``fast time” and ``slow time” dimensions. Every $n_c$ = 10 fast time images can be 2-norm combined to generate fMRI images that are free of oscillations and have $T_2^*$-sensitivity comparable to standard GRE imaging.}\label{ossi}
\end{figure}

This section introduces the patch-tensor LR model based reconstruction problem, the optimization algorithm, important implementation details, and other reconstruction methods for comparison.

\subsection{Tensor Model Problem Formulation}

fMRI involves acquiring a time series of images to track brain activity. 
In OSSI fMRI, the images periodically oscillate with every $n_c$ time points
along with the regular fMRI time course as shown in Fig.~\ref{ossi}.
Typically, we 
combine every $n_c$ consecutive and non-overlapping images
with root sum squared (2-norm) 
to get uniform images for fMRI analysis \cite{ossi2019}.
To simultaneously exploit the redundancy in OSSI oscillatory patterns
and the repetition along fMRI time series,
we structure OSSI fMRI images into two time dimensions.
The fast oscillation dimension is called ``fast time'',
and the fMRI time dimension is called ``slow time''. 

To improve both spatial and temporal resolutions for OSSI fMRI,
and to model the reproducibility in both fast and slow time dimensions,
we propose a tensor low-rank model for the undersampled reconstruction.
The tensor dimensions include vectorized space, fast time = $n_c$, and slow time. 
Since the exact form of the oscillations is resonant frequency dependent
and resonant frequency usually varies slowly across space,
low-rankness involving the fast oscillations is a local feature
(more similarities among neighboring pixels than between non-local pixels or over the whole image).
Furthermore, due to the complexity of functional activity,
imposing low-rankness on temporal blocks instead of the whole fMRI time series
improves the modeling accuracy.
Therefore, we propose a patch-tensor LR model
with limited spatial and temporal extent, 
and impose low-rankness on all the unfoldings of the patch-tensor.

The whole fMRI time series is broken into non-overlapping time blocks.
For each block,
we reshape
3D ($\text{space}\, x \times \text{space}\, y \times \text{time}\, t$)
or 4D ($\text{space}\, x \times \text{space}\, y  \times \text{space}\, z \times \text{time}\, t$)
OSSI images
into 4D ($x \times y \times \text{fast time}\, n_c \times \text{slow time}\, t_s$)
or 5D ($x \times y \times z \times \text{fast time}\, n_c \times \text{slow time}\, t_s$)
tensors.
We partition the 4D or 5D tensors into patches,
and vectorize all the spatial dimensions to form 3D low-rank patch-tensors
($\text{vectorized space}\, s_p \times n_c \times t_s$).
Figure~\ref{rank} visualizes an in vivo 3D patch-tensor,
its three unfoldings,
and the corresponding singular values demonstrating the low-rankness of the unfoldings.
The patch-tensor is from the center of a brain with no activation,
and Fig.~\ref{tsf5}(a) plots the corresponding log-scale singular values.
Figure~\ref{tsf5}(b) presents low-rank unfoldings of a different patch-tensor
in an activated region.

The proposed patch-tensor LR model based reconstruction problem with non-overlapping patches is
\begin{equation}
\argmin{\X} \M \sum_{i=1}^\N \lambda_i \, \mbox{rank} \left( \Pmi{\X} \right)
+ \frac{1}{2}\lVert\A(\X)-\y\rVert_2^2,
\label{primaryf}
\end{equation}
where $\X \in \mathbb{C}^{x \times y \, (\times z) \times t}$
is a complex OSSI fMRI time block to be reconstructed.
Linear operator
$\mathcal{P}(\mathbf{\cdot})$ partitions and reshapes its input
into $M$ locally low-rank patch-tensors with
$\Pm(\X)\in\mathbb{C}^{s_p \times n_c \times t_s}$,
$m = 1,\ldots,M$.
$\Pmi{\X} = \Pm(\X)_{(i)}$ denotes the mode-$i$ unfolding
of the $m$th tensor patch $\Pm(\X)$.
$\lambda_i$ is the regularization parameter for low-rankness of the mode-$i$ unfolding.
Linear operator
\A represents the MRI physics;
it consists of coil sensitivities and the non-uniform Fourier transform (NUFFT) including undersampling. $\y$ denotes sparsely sampled k-space measurements.

We focus on the following convex relaxation of \eqref{primaryf}:
\begin{equation}
\argmin{\X} \M \sum_{i = 1}^{\N}
\lambda_i\lVert \Pmi{\X} \rVert_* + \frac{1}{2}\lVert\A(\X)-\y\rVert_2^2.
\label{costf}
\end{equation}
This formulation encourages low-rankness of all the patch-tensor unfoldings
by minimizing the sum of their singular values.
Meanwhile, the data fidelity term encourages correspondence between the images
and the acquired k-space samples. 

\subsection{Optimization Algorithm}

The regularizers in the unconstrained cost function \eqref{costf} can be handled via the alternating direction method of multipliers (ADMM) \cite{Boyd2010DistributedMultipliers,Gandy2011TensorOptimization}
applied to the equivalent constrained optimization problem:
\begin{equation}
\begin{split}
&\argmin{\mathbf{Z}}\, \min_{\{\Xi\}} \M \sum_{i = 1}^{\N}
\lambda_i\lVert \Pmi{\Xi} \rVert_*
+ \frac{1}{2}\lVert\A\big(\mathbf{Z}\big)-\y\rVert_2^2
\\
&\mbox{subject to}\ \Xi = \mathbf{Z},\ i = 1,2,\N,
\end{split}
\end{equation}
with $\Xi \in \mathbb{C}^{x \times y \, (\times z) \times t},\, i = 1,2,\N$ constrained to be equal to $\mathbf{Z} \in \mathbb{C}^{x \times y \, (\times z) \times t}$.
The scaled form of the corresponding augmented Lagrangian is
\begin{equation}
\begin{split}
& \cL\left(\{\Xi\},\mathbf{Z},\{\mathbf{U}_i\}\right) =
\M \sum_{i = 1}^{\N} \lambda_i\lVert \Pmi{\Xi} \rVert_* 
\\& + \frac{1}{2}\lVert\A(\mathbf{Z})-\y\rVert_2^2
+ \frac{\rho}{2}\sum_{i = 1}^{\N}\lVert \Xi-\mathbf{Z}
+ \mathbf{U}_i\rVert^2
-\frac{\rho}{2}\sum_{i = 1}^{\N}\lVert\mathbf{U}_i\rVert^2.
\end{split}
\end{equation}
We update the variables $\{\Xi\},\,\mathbf{Z}$
and scaled dual variables 
$\{\mathbf{U}_i\}$ sequentially,
holding the other variables fixed.

For non-overlapping patch-tensors,
the update step for each patch of
$\{\Xi\}_{i=1}^{\N}$
is:
\begin{equation}
\begin{split}
\Pm(\Xikk)
&= \argmin{\Pm(\Xi)}\,
\cL_{mi}\left(\Pm(\Xi),\mathbf{Z}^k,\mathbf{U}_i^k\right)
\end{split}
\end{equation}
for $m=1,\ldots,M$
and $i=1,\ldots,\N$
at iteration $k+1$, where 
\begin{equation}
\begin{split}
\cL_{mi}
&= \lambda_i\lVert \Pmi{\Xi} \rVert_* + \frac{\rho}{2}\lVert\Pm(\Xi)
- \Pm(\mathbf{Z}^k - \mathbf{U}_i^k)\rVert^2 \\
&= \lambda_i\lVert \Pmi{\Xi} \rVert_* 
\mbox{} + \frac{\rho}{2}\lVert \Pmi{\Xi} - \Pmi{\mathbf{Z}^k - \mathbf{U}_i^k} \rVert_F^2.
\end{split}
\end{equation}
Because $\Pmi{\Xi}$ and $\Pmi{\mathbf{Z}^k - \mathbf{U}_i^k}$ are matrices,
patch update $\Pmi{\Xikk}$
is easily obtained
with a singular value soft-thresholding operator
$\textsc{SVT}(\mathbf{\cdot})$ with threshold ${\lambda_i}/{\rho}$, 
\begin{equation}
\begin{split}
\Pmi{\Xikk} 
&= \argmin{\Pmi{\Xi}} \, \cL_{mi}
\left( \Pmi{\Xi}, \mathbf{Z}^k, \mathbf{U}_i^k \right) \\
&= \textsc{SVT}_{{\lambda_i}/{\rho}} \left( \Pmi{\mathbf{Z}^k - \mathbf{U}_i^k} \right).
\end{split}
\end{equation}
Therefore, the update for the patches of $\{\Xi\}$ becomes
\begin{equation}
\Pm(\Xikk)
= \textsc{Refold}_i\left(\Pmi{\Xikk}\right).
\end{equation}
We parallelize this step over all the unfoldings and patches. 

The $\mathbf{Z}$ update simplifies to:
\begin{equation}
\begin{split}
\mathbf{Z}^{k+1} 
= & \argmin{\mathbf{Z}}\, \cL\left(\{\Xikk\},\mathbf{Z},\{\mathbf{U}_i^k\}\right)\\
= & \argmin{\mathbf{Z}} \bigg(\frac{1}{2}\lVert\A(\mathbf{Z})-\y\rVert_2^2 \\
& + \frac{\rho}{2}\sum_{i = 1}^{\N}\lVert\mathbf{Z}-\left(\Xikk+\mathbf{U}_i^k\right)\rVert^2\bigg).\\
\end{split}
\label{Zk1}
\end{equation}
We use the conjugate gradient method for this least-squares minimization.

The scaled dual variables $\{\mathbf{U}_i\}_{i=1}^{\N}$ are updated
in the usual ADMM way by
\begin{equation}
\begin{split}
\comment{ NO! THIS argmin IS NOT CORRECT!  see w20 598 course notes.  delete this!
\{\mathbf{U}_i^{k+1}\} &=
\argmin{\{\mathbf{U}_i\}} \, \cL\left(\{\Xikk\},\mathbf{Z}^{k+1},\{\mathbf{U}_i\}\right)
\\
}
\mathbf{U}_i^{k+1}
&= \mathbf{U}_i^{k} + \Xikk - \mathbf{Z}^{k+1}.
\end{split}
\end{equation}

\subsection{Practical Considerations}

\subsubsection{Random Cycle Spinning}

The singular value soft-thresholding operation for non-overlapping patch-tensors
leads to 
blocking artifacts at the boundaries of the patches.
Using overlapping patches would be computationally intensive,
so instead we apply random cycle spinning in every iteration as in \cite{Figueiredo2003AnRestoration,Ong2016BeyondDecomposition}.
We perform a randomly chosen circular shift
along each dimension of the input tensor before partitioning and reshaping, and unshift the tensor back after updating and placing the patch-tensors together. 
Accordingly, the actual update
for the patches of each $\Xi$ is
\begin{equation}
\begin{split}
&\pshift{\Xikk} =
\\
&\hspace*{2em}
\refoldi{\svti{\Pmi{\Shift{\mathbf{Z}^k - \mathbf{U}_i^k}}}}.
\end{split}
\label{Xk1}
\end{equation}

\subsubsection{Overlapping Time Blocks}

We reconstruct each fMRI time block separately to lighten the memory burden,
so random cycle spinning only removes patch boundary artifacts within each block.
To further reduce potential artifacts at the temporal boundaries of the blocks,
we reconstruct overlapping time blocks and discard additional time points near the boundaries for all the methods.
Figure \ref{tsf9} illustrates how the ranges and discarded portions of the time blocks are selected.

\subsubsection{ADMM Implementation Details}

We scale the k-space data to have maximum magnitude of 1
before applying ADMM.
With this normalization,
simply setting the regularization parameters $\lambda_1$ = $\lambda_2$ = 1 works well.
Because $\X_{(3)}$ has lower rank than $\X_{(1)}$ and $\X_{(2)}$
as shown in Fig.~\ref{rank},
we choose $\lambda_3$ = 2 to 
provide more weighting to the low-rankness of $\X_{(3)}$.

For ADMM penalty parameter $\rho$, we investigated a range of $\rho$ values and found $\rho$ = 121 empirically to be a good initialization. Furthermore, for our application, using varying penalty parameter or increasing $\rho$ after a number of inner iterations contributes to a faster convergence. After $T$ inner iterations updating variables $\{\Xi\},\,\mathbf{Z}$, and $\{\mathbf{U}_i\}$, 
the following updates are performed in the outer iteration:
\begin{equation}
\begin{split}
&\rho \mapsto r\rho\\
&\mathbf{U}_i \mapsto {\mathbf{U}_i}/{r}.
\end{split}
\label{stepsize}
\end{equation}
We chose 
rate $r = 3$,
and rescale the scaled dual variable $\mathbf{U}_i$ after updating $\rho$. 
This scheme is adapted from \cite{Boyd2010DistributedMultipliers,Gandy2011TensorOptimization}.
Algorithm \ref{alg1} summarizes the method.

 \begin{algorithm}
 	\caption{Patch-tensor low-rank reconstruction algorithm}  
 	\begin{algorithmic}[1] 
\renewcommand{\algorithmicrequire}{\textbf{Input:}}
\renewcommand{\algorithmicensure}{\textbf{Output:}}
 		\REQUIRE $\A,\,\y,\,\{\lambda_i\} = [1\, 1\, 2],\,\rho = 121,\,r = 3,\,S = 2,\,T = 11$
 		\ENSURE OSSI images $\mathbf{Z}^{k+1}$ 
 		\FOR{$s = 0, \ldots, S-1$}
 		\FOR{$t = 0, \ldots, T-1$}
 		\STATE $k = s*T+t$
 		\STATE Update $\mathbf{Z}^{k+1}$ using \eqref{Zk1} 
 		\FOR{$i = 1, 2, \N$} 
 		\STATE Update $\Xikk$ using \eqref{Xk1} 
 		\STATE $\mathbf{U}_i^{k+1} = \mathbf{U}_i^{k} + \Xikk - \mathbf{Z}^{k+1}$
 		\ENDFOR	
 		\ENDFOR	
 		\STATE Update $\rho$ and each $\mathbf{U}_i$ using \eqref{stepsize} 
 		\ENDFOR	
 		\RETURN $\mathbf{Z}^{k+1}$ 
 	\end{algorithmic}
 	 	\label{alg1}
 \end{algorithm}

\subsection{Other Reconstruction Approaches}

We compare the proposed reconstruction method to matrix local low-rank (MLLR) \cite{Trzasko2011LocalReconstruction}, global tensor low-rank (GTLR), patch-tensor low-rank plus sparse (patch-tensor L$+$S), and conjugate gradient SENSE \cite{Pruessmann2001,Sutton2003} with an edge-preserving regularizer (regularized CG-SENSE). 

MLLR imposes low-rank constraints on space $\times$ time matrices
by vectorizing image patches for the spatial dimension.
The cost function for MLLR is the same as setting $i = 1$ in \eqref{costf}.
GTLR enforces low-rankness on all the unfoldings of the tensor of size
space $xy\, \times \, n_c \times \, t_s$ without taking patches.
The cost function is the same as \eqref{costf} with $M = 1$
and without spatial partitioning. 
GTLR reconstructs fMRI time blocks and is global in spatial sense but not in temporal sense. It is less convenient for computation to impose low-rankness on a temporal global tensor.

The optimization problem for patch-tensor L$+$S is
$$
\argmin{\mathbf{L},\mathbf{S}}\, \frac{1}{2}
\lVert\A(\mathbf{L}+\mathbf{S})-\y\rVert_2^2
+ \M \sum_{i = 1}^{\N}
\lambda_i\lVert \Pmi{\mathbf{L}} \rVert_* + \mu\lVert{\Phi}(\mathbf{S})\rVert_1,
$$
where $\mathbf{L},\mathbf{S} \in \mathbb{C}^{x \times y \times t}$
denote the image components to be reconstructed
and $\Phi$ denotes 2D Fourier transform along both fast and slow time dimensions
to enhance the Fourier domain sparsity of the sparsity component $\mathbf{S}$.
The low-rank component $\mathbf{L}$ has the same regularization as in \eqref{costf},
and $\lambda_i$ and $\mu$ are regularization parameters. 

The optimization problem for regularized CG-SENSE is
$$
\argmin{\X} \frac{1}{2}\lVert\A(\X)-\y\rVert_2^2 + \sum_{j = 1}^{J}\psi\left([\mathbf{C}\X]_j\right),
$$
where $\X \in \mathbb{C}^{xy}$ denotes one vectorized image of the time series,
$\mathbf{C} \in \mathbb{R}^{J \times xy}$ is the 2D spatial finite difference matrix with $J = 2xy$,
and $\psi$ is the Huber potential function. 

We used ADMM to perform the MLLR, GTLR, and patch-tensor L$+$S reconstructions.
The ADMM parameters for patch-tensor L$+$S were the same as \eqref{stepsize} for patch-tensor LR. The CG update in the ADMM inner iterations and the regularized CG-SENSE reconstruction were implemented with the Michigan Image Reconstruction Toolbox \cite{fesslermirt}.

\section{Acquisition Methods}
\label{sec:acq}

Each oscillating state (index $n$) of OSSI was acquired with quadratic RF phases $\phi(n) = {\pi n^2}/{n_c}$, cycle length $n_c$ = 10, TR = 15~ms, and flip angle = 10$^\circ$ for the desired SNR and $T_2^*$-sensitivity \cite{ossi2019}. The short TR of 15~ms limits the readout, and $n_c$ = 10 compromises temporal resolution. Hence, sparse sampling is important for improving OSSI spatial and temporal resolutions.

This section develops practical sparse sampling schemes with increased sampling incoherence for OSSI,
and describes human fMRI studies. 
We collected 2D ``mostly sampled" with retrospective undersampling, 2D prospectively undersampled, and 3D prospectively undersampled data.
With FOV = 220~mm, slice thickness = 2.5~mm, and matrix size = $168 \times 168$,
the spatial resolution = $1.3 \times 1.3 \times 2.5$~mm$^3$ for all experiments.

\subsection{Variable-Density Spiral Sampling Trajectory}

We focus on variable-density (VD) spiral trajectories that travel quickly through k-space.
The sampling density of VD spirals varies at different k-space radii. By dense sampling in the center of k-space where the MR energy concentrates and sparse sampling at outer k-space, VD spirals can reduce imaging time and off-resonance blur \cite{Tsai2000ReducedTrajectories,Santos2006Single3t} compared to uniform-density (UD) spirals. 
We design VD spirals based on \cite{Hargreavesvds,Lee2003FastPerfusionb} with uniform density and over-sampling in the k-space center, and then linearly decrease the sampling density as the spirals approach the outer part of k-space. The trajectory is parameterized by $(n_i,\,a,\,b, \,d)$, where $n_i$ = number of interleaves, $a$ = effective FOV (in mm) at k-space center, $b$ = effective FOV at the edge of k-space, and $d$ denotes the number of central k-space points with uniform sampling density determined by $a$.

We used $(n_i,\,a,\,b,\,d)$ = (9, 310, 110, 300) for the retrospective sampling pattern with spiral-out readouts. The effective FOV for $n_i$ = 9 interleaves was $a = 310$~mm at the center of k-space for the first $d = 300$ sampling points, then decreased linearly to $b = 110$~mm at the edge of k-space.
The readout length for each interleave was 8.3~ms.
The k-space of each image can be mostly covered with all the 9 interleaves.
However, due to the variable-density nature of the trajectories,
the 9-interleave trajectory was still undersampled by approximately a factor of 1.5,
and we refer to this sampling pattern as ``mostly sampled''.
We chose $a = 300$~mm and $b = 80$~mm for prospective undersampling with spiral-in readouts
to increase $T_2^*$-sensitivity, and the readout length was 7.4~ms. 

\begin{figure}
\centering
    \includegraphics[width=0.95\textwidth]{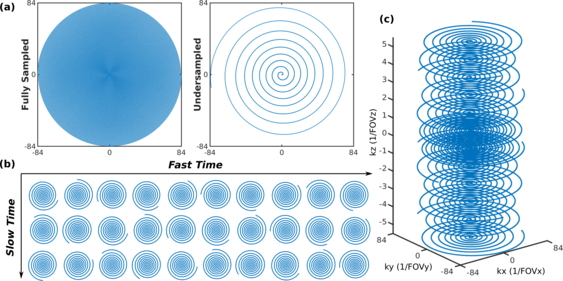}
    \caption{
(a) Compared to the fully sampled trajectory, the designed single-shot variable-density spiral trajectory for each time frame or $k_z$ plane enables a 12-fold acquisition acceleration.
(b) Prospective 2D undersampling pattern with the incoherent rotations between fast time (the oscillation dimension) and slow time (the fMRI time dimension). 
(c) 3D undersampled stack-of-spirals providing a 10-fold acceleration with one spiral for the outer $k_z$ planes, two spirals for the two central $k_z$ planes, and golden-angle rotations between $k_z$ planes.
    }\label{samp}
\end{figure}

We took 1 interleave out of $n_i$ = 9 VD spirals as the undersampled trajectory.
Compared to a UD spiral with the same FOV and matrix size, 
the single-shot undersampled trajectory provided a factor of 12 acceleration in-plane
as presented in Fig.~\ref{samp} (a).
We selected the VD spiral parameters
for a good balance between the undersampling factor and reconstruction performance.

\subsection{Incoherent Sampling for Time Dimensions and 3D}

The proposed spiral trajectory provides aggressive undersampling in-plane and would introduce reconstruction artifacts if used without regularization. 
As we are using a tensor model with two time dimensions for the undetermined reconstruction problem, we prefer the sampling pattern to be incoherently varying along the two dimensions for artifact reduction \cite{Lustig2007SparseImaging}. Therefore, we rotate the VD spiral using a golden-angle based approach for each temporal frame to avoid overlapping trajectories in both fast and slow time dimensions.

We define an interleave index $k= 0, \ldots, K-1$ for a time series of OSSI images with $K$ interleaves in total. For 2D retrospective sampling with multi-shot acquisition, the rotation angle for each interleave $k$ was
\begin{equation}
\begin{aligned}
ga\cdot k + 2\cdot ga\cdot\left\lfloor k/n_c/n_i \right\rfloor,
\end{aligned}
\label{retro}
\end{equation}
where $ga$ = 111.246$^\circ$ is the golden angle, $n_c$ = 10 is the size of fast time dimension, $n_i$ is the number of interleaves, and $\lfloor \cdot \rfloor$ denotes the floor function. The acquisition for the interleaves first looped through OSSI oscillation states 1 to $n_c$, then looped through multi-shot 1 to $n_i$, and after that proceeded to the next slow time point.

For 2D prospective undersampling, only 1 VD interleave was collected for each image, and the rotation angle was
\begin{equation}
\begin{aligned}
ga\cdot k + ga\cdot\left\lfloor k/n_c \right\rfloor.
\end{aligned}
\label{prosp}
\end{equation}
The index $k$ looped through OSSI fast time oscillations for every slow time point. 
Figure~\ref{samp} (b) presents the prospective sampling pattern. The baseline rotation was determined by the golden angle.
The second terms in \eqref{retro} and \eqref{prosp} were designed specifically
to increase sampling incoherence along slow time as shown in Figs.~\ref{tsf6} and \ref{tsf7}
for prospective undersampling and retrospective undersampling, respectively.

For 3D prospective undersampling, we used a stack of VD spirals with 2-shot acquisition at the 2 central $k_z$ planes and single-shot acquisition at other $k_z$ locations as in Fig.~\ref{samp} (c), providing a 10-fold acceleration compared to the fully sampled k-space. Rotations in \eqref{prosp} were applied, where $k$ looped through OSSI oscillations, then $k_z$ planes, and finally the slow time points. 

Because of the increased sampling incoherence in the two time dimensions, the angular dimension of k-space can be mostly covered with sampling trajectories of 9 or 10 consecutive time frames. We used this feature and combined k-space data of every 10 slow time points to compute data-shared initialization for reconstruction, which helped decrease the number of CG iterations and computation time.

\subsection{Human fMRI Studies}

We implemented the OSSI pulse sequence and the proposed sampling scheme using GE’s standard pulse programming environment EPIC. All the data were collected on a 3T GE MR750 scanner (GE Healthcare, Waukesha, WI) with a 32-channel head coil (Nova Medical, Wilmington, MA) using the proposed retrospective and prospective undersampling schemes. 
Prospectively undersampled OSSI studies were further compared to standard GRE fMRI with matched spatial-temporal resolution.

The human studies were carried out under IRB approval. The fMRI task was a left vs. right reversing-checkerboard visual stimulus of 210 s with 10 s rest, 5 cycles of left or right stimulus of 20 s (20 s L/20 s R $\times$ 5 cycles). 

2D OSSI used an oblique slice passing through the visual cortex. 
The 2D mostly sampled data were acquired with multi-shot VD spirals with number of interleaves $n_i$ = 9, volume TR = 1.35 s ($\text{TR} \cdot n_c \cdot n_i$), and spiral-out TE = 2.7~ms. The rotation angles between interleaves and time frames were determined by \eqref{retro}. The number of time frames (both fast time $n_c$ and slow time) was 1490 with 10 s discarded acquisition to ensure the steady state. The retrospectively undersampled data only contained the first VD interleave of every 9 interleaves. 

The 2D prospectively undersampled data were collected with single-shot VD spirals ($n_i$ = 1) with volume TR = 150~ms ($\text{TR} \cdot n_c$) and spiral-in TE = 11.7~ms. The rotation angles of the spirals were selected by \eqref{prosp}. The number of fast time frames was 13340 with 10 s discarded acquisition. As every $n_c$ images were 2-norm combined for fMRI analysis, the temporal resolution for the prospectively undersampled data was 150~ms.
2D GRE fMRI images with the same spatial resolution and temporal resolution of 150~ms as OSSI were also acquired for comparison.
Specifically, GRE imaging used multi-shot spiral acquisition with $n_i$ = 3, TR = 50~ms, Ernst flip angle = 16$^\circ$, and spiral-in TE = 30~ms. Each interleave was VD spiral with $(n_i,\,a,\,b,\,d)$ = (3, 240, 120, 300) and readout length = 21.9~ms.

For 3D imaging, an oblique slab was selected.
Prospectively undersampled OSSI was compared to GRE imaging with matched spatial resolution and matched temporal resolution of 2.1 s. The number of 3D volumes was 96 for the 200 s task. For OSSI, the number of $k_z$ planes $n_z$ = 12, volume TR = 2.1 s ($\text{TR} \cdot n_c \cdot n_z$), and spiral-in TE = 10.3~ms. For GRE, multi-slice TR = 700~ms with 14 slices, multi-shot acquisition with $n_i$ = 3, spiral-in TE = 30~ms, Ernst flip angle = 16$^\circ$, and same VD spiral trajectories for each slice as in 2D GRE imaging were used.

For calculation of coil sensitivity maps, we collected images with a standard spin-warp sequence at TR = 50~ms, TE = 6.3~ms, and Ernst flip angle = 16$^\circ$. The 32-channel coil images were compressed to 16 virtual coils for 2D and 24 virtual coils for 3D via PCA \cite{Huang2008AChannels}, and coil sensitivity maps were calculated using ESPIRiT \cite{Uecker2014,Bart}. 
We also created coil-combined images for extraction of the brain region using the Brain Extraction Tool \cite{Smith2002}.

\subsection{Performance Evaluation}

The reconstruction and functional performances were evaluated with normalized root-mean-square difference (NRMSD) for retrospectively undersampled data, activation maps, and tSNR maps. 

The retrospectively undersampled reconstruction $\hat{\X}$ was compared to $\X_{\mathrm{ref}}$ reconstruction from ``mostly sampled” data by regularized CG-SENSE, using the metric
$
\text{NRMSD} = {\lVert\X_{\mathrm{ref}} - \hat{\X}\rVert}/{\lVert\X_{\mathrm{ref}}\rVert}.
\label{nrmsd}
$

Every $n_c$ = 10 reconstructed images of OSSI were combined via 2-norm for functional analysis. 
The data from the first cycle (40 s) of the task were discarded to avoid the modeling error in the initial rest period.
To reduce scanner drift effects, we detrended the data using the first 4 discrete cosine transform basis functions for both OSSI combined and GRE fMRI images.

The background of the activation map was the mean of reconstructed fMRI images. The activated regions were determined by correlation coefficients above a 0.45 threshold. Correlation coefficients were defined by correlating the fMRI time course for each voxel with the task-related reference waveform, and the reference waveform was given by convolving the task with the canonical hemodynamic response function \cite{spm}. The tSNR maps were generated by dividing the time course mean by the standard deviation of the time course residual (without the mean and the task) for each voxel. 
NMRSD within the brain (excluding the scalp and skull) from reconstructed images, number of activated voxels at the lower third of the brain (where the visual activation concentrates), and averaged tSNR within the brain were calculated for quantitative evaluations. 

\section{Reconstruction and Results}
\label{sec:results}

This section compares OSSI undersampled reconstructions using the proposed tensor model and other low-rank related approaches. 3D OSSI reconstruction is further compared to multi-slice GRE to demonstrate the SNR advantage of OSSI.

\subsection{Regularization Parameter Adjustment}

To ensure that different reconstructions have similar spatial-temporal resolutions,
we compared the local impulse responses \cite{Fessler1996,Olafsson2018FastFMRI}
of the reconstruction methods.
Specifically, we added an impulse perturbation
$\eps \A\left(\delta_{j,t}\right)$ to the undersampled k-space data $\y$
and reconstructed the perturbed data with different models.
We selected $j$ and $t$ to be in the spatial and temporal center
of the time block being reconstructed, respectively,
and we chose $\eps = 1$ (about 10\% of the OSSI signal magnitude). 
Accordingly, the local impulse response is
$
h(j,t) = \paren{\cB\big(\eps\A\left(\delta_{j,t}\right)+\y\big)-\cB(\y)} / \eps
$,
where $\cB(\mathbf{\cdot})$ denotes a reconstruction method.

Profiles of the impulse response along spatial dimension and temporal dimensions
can help assess the spatial-temporal sharpness of the reconstructions for $\cB \neq \A^{-1}$.
As shown in Fig.~\ref{tsf8},
we selected regularization parameters
to ensure that impulse responses of different reconstructions had similar peaks
and were close to the magnitude for the regularized CG-SENSE reconstruction.
Based on the ratios for the $\lambda_i$ values in \eqref{costf},
the final 2D reconstruction parameters were
$\{\lambda_i\} = [1\, 1\, 2]*0.4$ for patch-tensor LR,
$\lambda_3 = 1.6$ for MLLR,
$\{\lambda_i\} = [1\, 1\, 2]*4$ for GTLR,
$\{\lambda_i\} = [1\, 1\, 2]*0.3$ and $\mu$ = 3 for patch-tensor L$+$S.

Furthermore, with carefully adjusted regularization parameters,
reconstructing overlapping time blocks or non-overlapping time blocks for the fMRI time series
led to similar results,
as demonstrated by example time courses and spectra of the patch-tensor LR reconstruction in Fig.~\ref{tsf10}.

\begin{figure}
\centering
\includegraphics[width=0.95\textwidth]{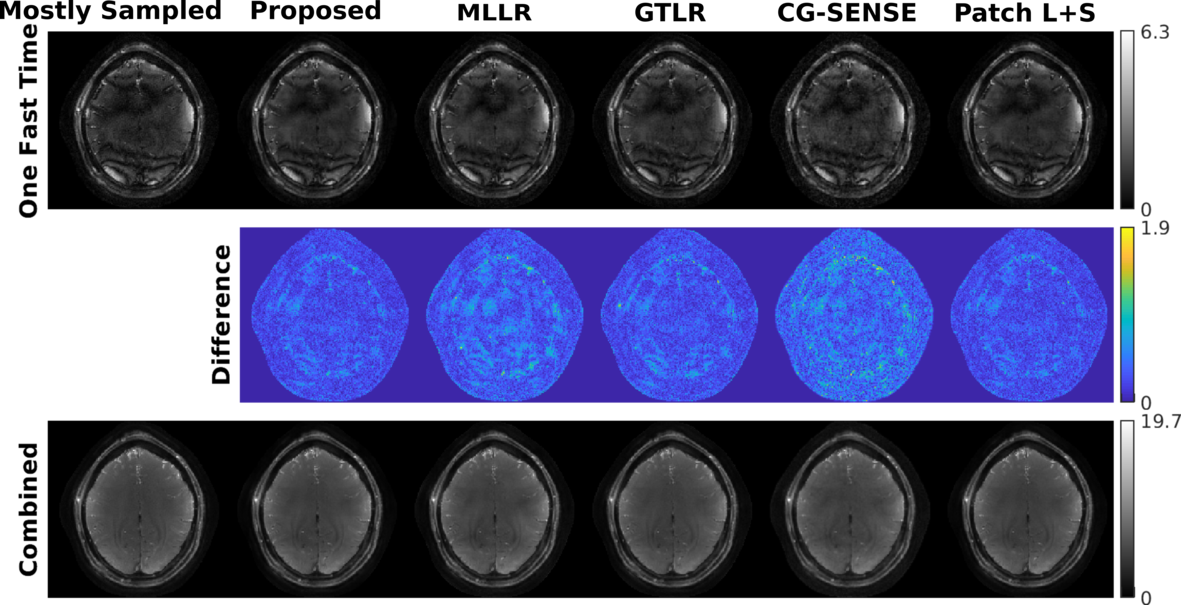}
\caption{Fast time images from the retrospectively undersampled reconstructions are compared to the mostly sampled results. The proposed approach outperforms other methods with less noisy fast time images, less structure in the difference maps before combination, and high-resolution 2-norm combined images.}
\label{2D1}
\end{figure}

\begin{figure}
\centering
\includegraphics[width=0.95\textwidth]{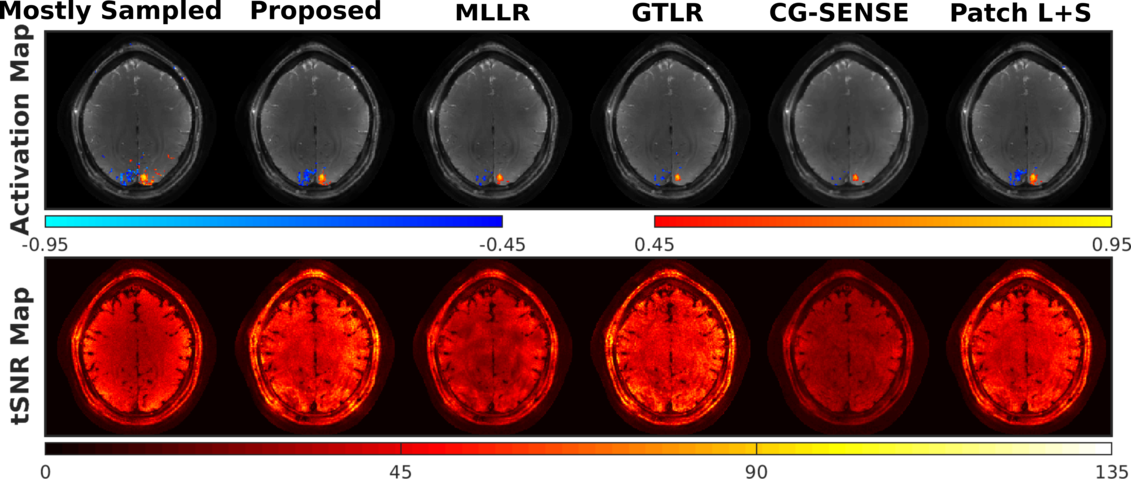}
\caption{Activation maps and temporal SNR maps from retrospectively undersampled reconstructions. A contiguity (cluster-size) threshold of 2 was applied for the activated regions. The proposed model provides more functional activation than other approaches with high temporal SNR, and shows similar results as the patch-tensor low-rank plus sparse model.}
\label{2D2}
\end{figure}

\begin{table}
\setlength{\tabcolsep}{4pt}
\renewcommand{\arraystretch}{1.2}
\caption{Quantitative comparisons of OSSI retrospectively undersampled reconstructions}
\label{tt1}
\centering
\begin{tabular}{ccccccc}
\hline\hline
& \begin{tabular}[c]{@{}c@{}}Mostly \\[-0.3em] Sampled\end{tabular} & Proposed & MLLR & GTLR & \begin{tabular}[c]{@{}c@{}} CG-\\[-0.3em] SENSE\end{tabular} & \begin{tabular}[c]{@{}c@{}}Patch\\[-0.3em] L$+$S\end{tabular} \\ \hline
\begin{tabular}[c]{@{}c@{}} 
NRMSD\\[-0.3em] Before Comb\end{tabular}        
& - & 0.17 & 0.22 & 0.18 & 0.25 & 0.17        \\ \hline
\begin{tabular}[c]{@{}c@{}}  
NRMSD\\[-0.3em]After Comb\end{tabular}        
& - & 0.05 & 0.06 & 0.07 & 0.07 & 0.06        \\ \hline
\begin{tabular}[c]{@{}c@{}}\# Activated\\[-0.3em] Voxels\end{tabular} 
& 229 & 168 & 73 & 68 & 46 & 153          \\ \hline
\begin{tabular}[c]{@{}c@{}}Average\\[-0.3em] tSNR\end{tabular}    & 37.1 & 43.6 & 32.4 & 44.1 & 25.6 & 41.1 \\ \hline\hline
\end{tabular}
\end{table}

\subsection{Retrospective and Prospective 2D Reconstructions}

OSSI 2D retrospectively and prospectively undersampled data were reconstructed using the proposed method and the comparison methods. OSSI 2D mostly sampled data were reconstructed using regularized CG-SENSE. For the proposed retrospectively undersampled reconstructions, the number of time points before combination = 330 for every overlapping time block, and the patch-tensor size = $64\, (8*8) \times 10 \times 33$. Similarly for prospectively undersampled data, the number of time points = 420 for each overlapping time block, and the patch-tensor size = $64\, (8*8) \times 10 \times 42$. We used $S$ = 2 outer iterations, $T$ = 11 inner iterations for ADMM, and 4 iterations for the CG update of $\mathbf{Z}$. The number of iterations for regularized CG-SENSE reconstruction was 19. All the OSSI reconstructions were initialized with data-shared images.

Figure~\ref{2D1} shows reconstructions from mostly sampled data, the proposed patch-tensor LR, MLLR, GTLR, regularized CG-SENSE, and patch-tensor L$+$S models. The fast time image reconstructed using the proposed approach is less noisy compared to the mostly sampled reference and other reconstructions. The oscillatory patterns and the high-resolution details of the fMRI image (after 2-norm combination of the fast time images) are also well preserved. The difference maps after combination is presented in Fig.~\ref{tsf11}.

Figure~\ref{2D2} gives functional results including activation maps and tSNR maps. The proposed model enables high-resolution fMRI with larger activated regions than other undersampled reconstructions, and maintains the SNR advantage of OSSI with tSNR values that are comparable to the mostly sampled reconstruction. patch-tensor LR regularization and the patch-tensor L$+$S model present similar results, suggesting that L$+$S decomposition and Fourier sparsity along the two time dimensions were not critical given the patch-tensor modeling of the data. 

The quality of the retrospectively undersampled reconstructions was further assessed with ROC analysis. 
ROC curves for the activation maps of different reconstruction approaches were compared with mostly sampled activation at the lower third of the brain as ground truth. 
Figure \ref{tsf12} shows that the proposed approach leads to the largest area under the ROC curve.

\begin{figure}
	\centering
	\includegraphics[width=0.9\textwidth]{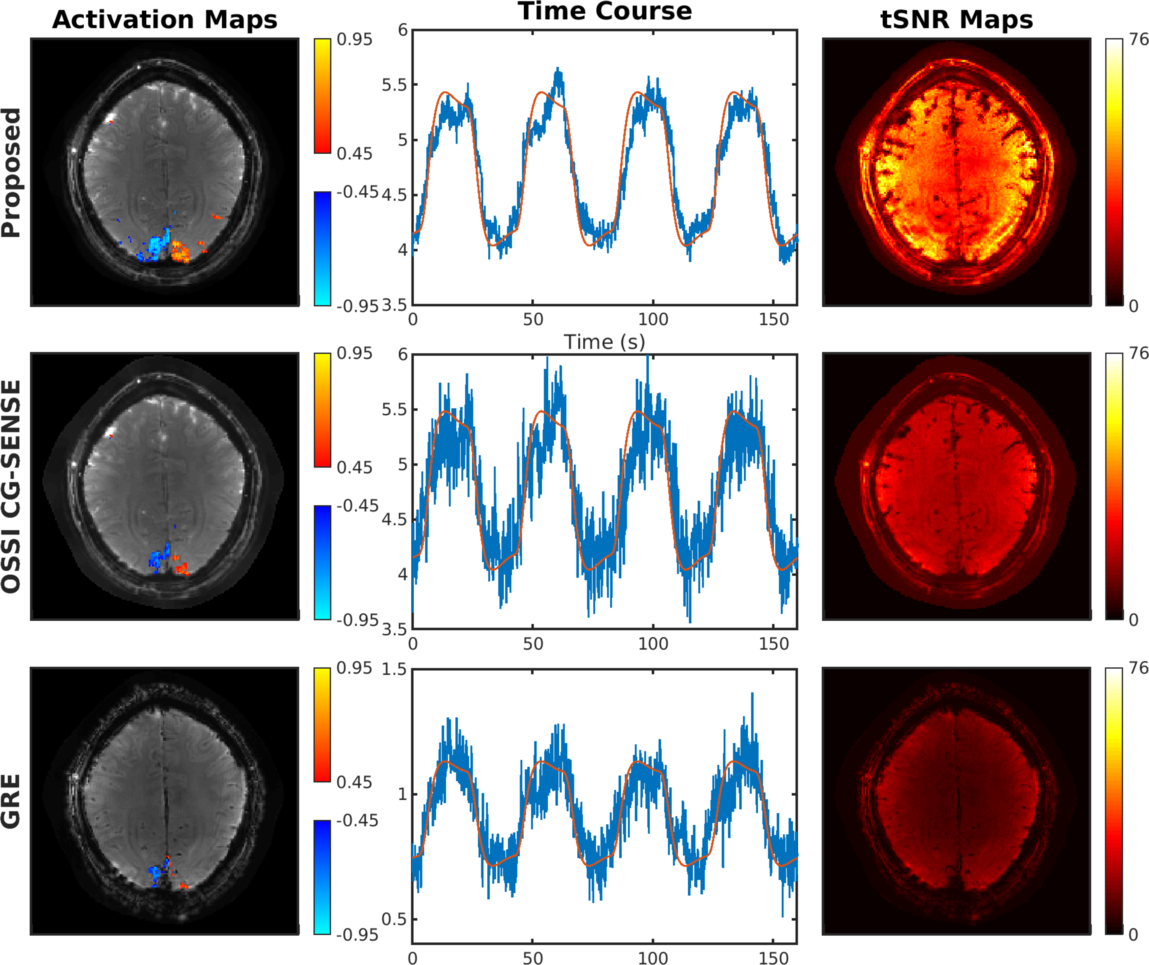}
	\caption{OSSI tensor model prospectively undersampled reconstruction demonstrating high-resolution and high SNR fMRI with high-resolution background and larger activated regions for the activation map, less noisy time course (red curve showing the reference waveform), and higher SNR for the temporal SNR map.}
	\label{2D3}
	\vspace{-0.05in}
\end{figure}

Figure~\ref{2D3} presents prospectively undersampled reconstructions. Compared to OSSI regularized CG-SENSE reconstruction and standard GRE fMRI, the proposed approach yields more functional activity, less noisy time courses, and higher tSNR with the largely improved spatial and temporal resolutions. 
Other qualitative and quantitative comparisons for 2D prospectively undersampled reconstructions are in Fig.~\ref{tsf15} and Table~\ref{tst1}.

Table~\ref{tt1} summarizes quantities from different reconstructions including NRMSD for the whole dataset before and after fast time combination, number of activated voxels, and average tSNR within the brain. The proposed patch-tensor modeling outperforms other reconstruction methods with more functional activation and a high average tSNR. 

Reconstruction comparisons of a different subject are presented in Figs.~\ref{tsf18}, \ref{tsf19}, \ref{tsf20}, and Table~\ref{tst4} for retrospectively undersampled data, and Fig.~\ref{tsf21} and Table~\ref{tst4} for prospectively undersampled data.

\begin{figure}
\centering
    \includegraphics[width=0.9\textwidth]{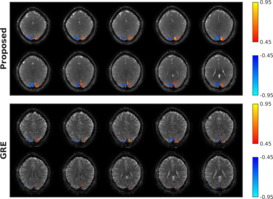}
    \caption{
3D OSSI (prospectively undersampled) and GRE activation maps of the central 10 slices. A contiguity (cluster-size) threshold of 2 was applied for the activated regions. With matched spatial and temporal resolutions, 3D OSSI acquired and reconstructed using the proposed method presents 2 times more activated voxels compared to multi-slice Ernst angle GRE imaging at TE = 30~ms.
    }\label{3D1}
\end{figure}

\begin{figure}
\centering
    \includegraphics[width=0.92\textwidth]{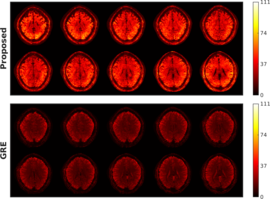}
    \caption{
3D OSSI (prospectively undersampled) and GRE temporal SNR maps of the central 10 slices. At the same spatial-temporal resolution, 3D OSSI acquired and reconstructed using the proposed method presents at least 2 times higher temporal SNR than standard multi-slice GRE imaging.
    }\label{3D2}
\end{figure}

\subsection{3D OSSI to GRE Comparison}

The 3D OSSI prospectively undersampled data were reconstructed using the proposed model with number of time points before combination = 120 for each non-overlapping time block. The patch-tensor size = $108\, (6*6*3) \times 10 \times 12$, and $\{\lambda_i\} = [1\, 1\, 2]$. Number of ADMM outer iterations $S$ = 2, inner iterations $T$ = 11, and number of CG iterations = 7 for every $\mathbf{Z}$ update. 
We used data-shared images to initialize each $\Xi$ and $\mathbf{Z}$.
The multi-slice GRE data were reconstructed with regularized CG-SENSE with 19 CG iterations for each slice. 

Figure~\ref{3D1} shows the activation maps of 3D OSSI and multi-slice GRE. The proposed tensor model almost fully recovers the high-resolution structures of the OSSI images with a factor of 10 acquisition acceleration, and presents larger activated regions than multi-slice Ernst angle GRE.

Figure~\ref{3D2} shows the 3D tSNR maps, where OSSI provides higher average tSNR than GRE. The OSSI acquisition combined with the proposed undersampling design and tensor model reconstruction enable high-resolution and high SNR fMRI.

Quantitatively as presented in Table~\ref{t2}, the proposed 3D OSSI tensor reconstruction improves the amount of functional activity and average tSNR within the brain by a factor of 2 more than standard GRE imaging at matched spatial and temporal resolutions. 

\begin{table}
\renewcommand{\arraystretch}{1.565}
\caption{Functional performances of proposed OSSI prospectively undersampled reconstruction and standard GRE imaging}
\label{t2}
\centering
\begin{tabular}{ccccc}
\hline\hline
\multirow{2}{*}{} & \multicolumn{2}{c}{\# Activated Voxels} & \multicolumn{2}{c}{Average tSNR} \\ \cline{2-5} 
& 2D & 3D & 2D & 3D 
\\ \hline 
OSSI & 322 & 2150 & 32.8 & 34.7           
\\ \hline
GRE & 83 & 947 & 9.8 & 15.9             
\\ \hline
Ratio & 3.9 & 2.3 & 3.3 & 2.2            
\\ \hline\hline
\end{tabular}
\end{table}

\section{Discussion}
\label{sec:dis}

To our knowledge, the patch-tensor LR model is new for fMRI data.
Reshaping and partitioning the data to patch-tensors
facilitates exploiting high-dimensional structures,
and considering all the unfoldings of the tensors better uses spatial-temporal low-rankness. 
Therefore, the model is flexible and adaptive to other high-dimensional image reconstruction problems
that satisfy the patch-tensor LR constraints.
Local models may be more valid than assuming low-rankness of the whole dataset. 

Other reconstruction methods such as MLLR account for the locality of low-rank representations
while treating the time dimension as a whole.
GTLR separates the fast and slow time dimensions but enforces the low-rankness globally on the images.
The proposed patch-tensor LR model combines the advantages of both methods
by exploiting local low-rankness with two time dimensions,
and improved the reconstruction and functional performances.

Another feature of the work is an incoherent sparse sampling scheme
formed by properly rotating VD spirals along fast time and slow time.
The angular dimension of the k-space can be mostly covered with different frames,
and the trajectory is well accommodated with the spatial-temporal regularizers used here.
Moreover, we noticed that for 3D undersampling,
increasing number of interleaves in the central $k_z$ planes
greatly improves the amount of functional activation recovered and reduces false positives.
The sampling pattern is practical, and the prospective undersampling is easy to implement.

We selected and vectorized patches of spatial size
$8 \times 8$ (2D) or $6 \times 6 \times 3$ (3D)
based on the empirical reconstruction performance.
The choice of spatial patch size is still an open question.
At one extreme, the spatially global GTLR preserves little activation
for 2D retrospectively undersampled reconstruction as presented in Fig.~\ref{2D2} and Table~\ref{tt1},
but performs similar to the proposed method for 2D prospective undersampling
as in Fig.~\ref{tsf15} and Table~\ref{tst1}. In both cases, GTLR used temporal patches.

We investigated multi-scale low-rank decomposition \cite{Ong2016BeyondDecomposition}
with multi-scale patch-tensors of the OSSI images
to explore the idea that different parts of the data
may have different density and different low-rankness;
however, it provided limited performance improvement
and made the reconstruction more time-consuming. 
We also tested a 4D patch-tensor LR model with two spatial dimensions and two temporal dimensions.
The cost function is the same as \eqref{costf} without vectorizing spatial dimensions in \Pm.
That model gave similar results as the 3D patch-tensor LR approach,
making it well suited for potential applications such as GRE fMRI.
The comparison results of the new models including functional maps, ROC curves,
and quantitative evaluations are in Figs.~\ref{tsf16}, \ref{tsf17}, and Table~\ref{tst2}.

We imposed low-rankness on all the unfoldings of all the patch-tensors.
However, some unfoldings of some patches are not very low-rank,
especially for the second unfolding that is greatly affected by the nonlinear fast time oscillations.
Therefore, nonlinear mapping approaches
such as kernel methods or neural networks,
that map the fast time data to a low-dimensional linear subspace \cite{Poddar2019ManifoldImaging},
may further improve the model capacity,
which might also help optimize combination of the OSSI fast time images
instead of combing with 2-norm to yield band-free post-combined images.
Because OSSI images are not very sparse in the Fourier domain,
as shown in Figs.~\ref{tsf3} and \ref{tsf4},
the patch-tensor L$+$S reconstruction results in a very small sparse component seen in Fig.~\ref{tsf14}. Therefore,
future work on adaptive sparsity 
\cite{Ravishankar2017Low-RankImaging} might be beneficial.

Because low-rank approaches might cause spatial-temporal smoothing
that makes tSNR comparisons less compelling,
we assessed and matched the amount of regularization for fast time image reconstructions
based on their impulse responses. 
To evaluate spatial resolutions of the fMRI dynamics for different reconstructions after combination,
we compared spatial autocorrelations of the different correlation maps
(at the center of the brain without activation).
Figure \ref{tsf13} demonstrates that the proposed approach has similar autocorrelation profiles
as the mostly sampled reconstruction and preserves fMRI spatial resolution. 
We also compared ROC curves of different approaches with varying activation thresholds;
these curves are invariant to the degrees of freedom for performance evaluation.
The effective degrees of freedom calculation for the nonlinear reconstructions
will be explored in the future as in \cite{Xue2020}.

The proposed sparse sampling uses fast VD spirals with designed rotations along the two time dimensions
to increase sampling incoherence for the spatial-temporal models.
However, the sampling incoherence from VD spirals is limited by the shape of the spiral,
and the non-Cartesian nature requires NUFFT that needs more computation than FFT for Cartesian sampling.
More importantly, designing the sampling pattern according to reconstruction models
can improve the performance \cite{Gozcu2018LearningBasedMRI,Bahadir2019AdaptiveLearning},
so we will further explore joint optimization of the sampling pattern and the reconstruction model.

\section{Conclusion}
\label{sec:con}

We proposed a novel fMRI reconstruction method based on patch-tensor low-rank for the oscillating nature of OSSI images.
We also introduced an incoherent variable-density sampling pattern that is easy to implement,
and retrospectively and prospectively undersampled the multi-coil data with less than 10\%
of the fully sampled k-space.
By exploiting the inherent high-dimensional structure
and local spatial-temporal low-rankness of OSSI images,
the proposed model was able to recover high-resolution images and preserve functional signals
compared to matrix local low-rank and tensor low-rank methods.
It further enabled 3D high SNR fMRI with 2 times more functional activity and 2 times higher tSNR
compared to standard GRE imaging.
\newpage

\section{Supporting Information}

This supplemental material presents: 
(1) OSSI signal properties including example OSSI images and time courses before and after Fourier transform, and tensor low-rankness for patch-tensors at different regions of the brain; 
(2) the incoherent trajectory rotation schemes for both retrospective and prospective undersampling;
(3) reconstruction details including effects of overlapping time blocks and regularization parameter selection based on impulse perturbation;
(4) reconstruction comparisons for 2D retrospective and prospective undersampling;
(5) other reconstruction methods including 4D patch-tensor low-rank and multi-scale tensor low-rank;
(6) reconstruction results of a different subject.

\subsection{OSSI Signal Properties}

\begin{figure}[htb!]
\centering
\includegraphics[width=0.96\textwidth]{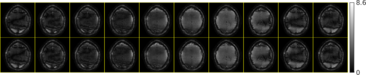}
\caption{Example OSSI fast time magnitude images for 2 cycles of the periodic oscillations.}\label{tsf1}
\end{figure}

\begin{figure}
\centering
\includegraphics[width=0.96\textwidth]{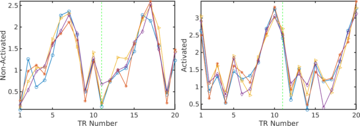}
\caption{OSSI fast-time time courses (magnitudes) of 4 different voxels
within a brain region that is not activated (left) or activated (right).
The signal oscillation pattern repeats every $n_c$ = 10 TRs,
as indicated by the vertical green dashed line.}
\label{tsf2}
\end{figure}

\begin{figure}
\centering
\includegraphics[width=0.96\textwidth]{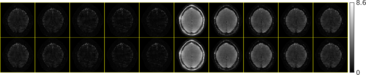}
\caption{Results after taking 1D Fourier transform along fast time of the OSSI images shown in \fref{tsf1}.
Magnitude is shown and
temporal frequency 0 is in ``middle'' (6th image from left).
OSSI fast time images are not very sparse in the Fourier domain.}\label{tsf3}
\end{figure}

\begin{figure}
\centering
\includegraphics[width=0.96\textwidth]{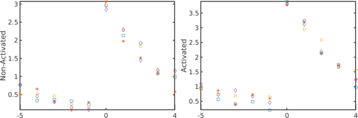}
\caption{Results after taking 1D Fourier transform along fast time (every $n_c$ = 10 TRs)
of the OSSI time courses in \fref{tsf2}.
Magnitude of one cycle is shown and temporal frequency 0 is in ``middle''.
OSSI fast time signals are not very sparse in the Fourier domain.
}\label{tsf4}
\end{figure}

\begin{figure}
\centering
\hspace{-1em}
\includegraphics[width=0.96\textwidth]{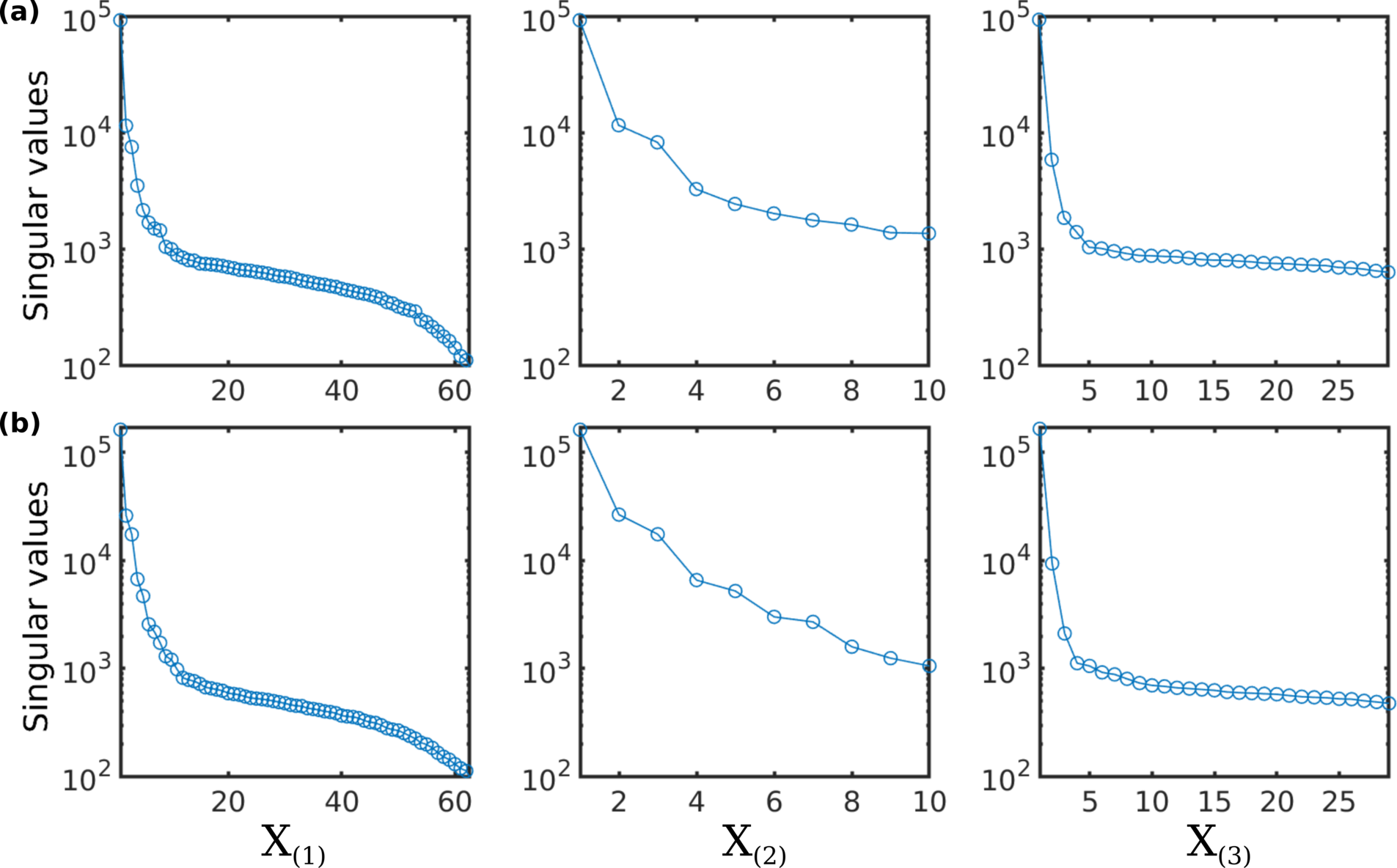}
\caption{
Log-scale singular value plots for all 3 unfoldings of a 3D patch-tensor (a) at the center of the brain with no activation (b) at the activation region.
For both activated and non-activated patch-tensors, the unfoldings show a similar pattern that $X_{(3)}$ has lower rank than $X_{(1)}$ and $X_{(2)}$.
}\label{tsf5}
\end{figure}

This section presents in-vivo OSSI images and time courses, and demonstrates local low-rankness of OSSI fMRI time-series.
\fref{tsf1} shows 2 cycles of OSSI fast time images with periodic oscillation patterns.
\fref{tsf2} provides example time courses from non-activated and activated ROIs of the OSSI images.
\fref{tsf3} gives 1D Fourier transformed (along fast time) results
for the complex time series corresponding to the images in \fref{tsf1},
and \fref{tsf4} presents the Fourier transformed time courses of \fref{tsf2}.
OSSI images are not very sparse before or after Fourier transform due to the nonlinear oscillations.
\fref{tsf5} gives log-scale singular value plots of non-activated and activated 3D patch-tensors
from an OSSI fMRI time block. 

\subsection{Incoherent Sampling Pattern}

\begin{figure}
\centering
\hspace{-1em}
\includegraphics[width=0.96\textwidth]{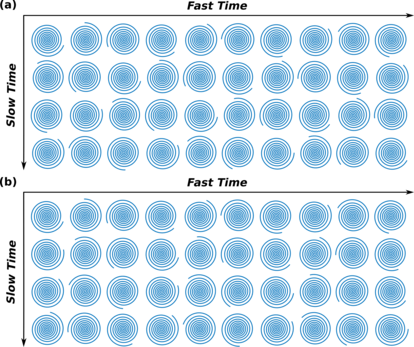}
\caption{Demonstration of the incoherent rotations for 2D prospective undersampling. The proposed scheme of $ga\cdot k + ga\cdot\left\lfloor k/n_c \right\rfloor$ in (a) increases the sampling incoherence along slow time compared to a baseline rotation scheme of $ga\cdot k$ in (b).}\label{tsf6}
\end{figure}

\begin{figure}
\centering
\hspace{-1em}
\includegraphics[width=0.96\textwidth]{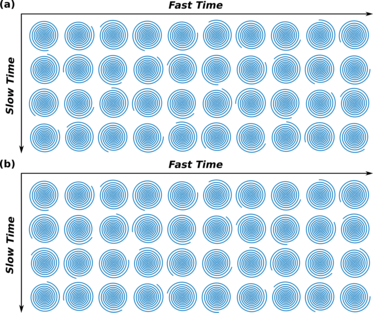}
\caption{Demonstration of the incoherent rotations for 2D retrospective undersampling. The proposed scheme of $ga\cdot k + 2\cdot ga\cdot\left\lfloor k/n_c/n_i \right\rfloor$ in (a) increases the sampling incoherence along slow time compared to a baseline rotation scheme of $ga\cdot k$ in (b).
}\label{tsf7}
\end{figure}

This section illustrates how the proposed spiral rotations
help increase temporal incoherence for OSSI acquisition. 
For prospective undersampling, the baseline rotation of $ga \cdot k$ for frame $k$ leads to an angle difference of $10ga \bmod 360^\circ = 32^\circ$ between consecutive slow time points. With the additional angle of $ga\cdot\left\lfloor k/n_c \right\rfloor$, the angle difference becomes $11ga \bmod 360^\circ = 144^\circ$, which increases sampling incoherence along slow time as compared in \fref{tsf6}.
Similarly for retrospective undersampling, the angle difference between undersampled slow time points changes from $90ga \bmod 360^\circ = -68^\circ$ to $92ga \bmod 360^\circ = 155^\circ$ with improved incoherence as in \fref{tsf7}.

\subsection{Reconstruction Adjustment}

This section presents practical adjustments to the reconstruction methods
including local impulse responses for regularization parameter selection
and structuring overlapping time blocks for the OSSI fMRI time course. 

\subsubsection{Regularization Parameter Selection}

\begin{figure}
\centering
\includegraphics[width=0.96\textwidth]{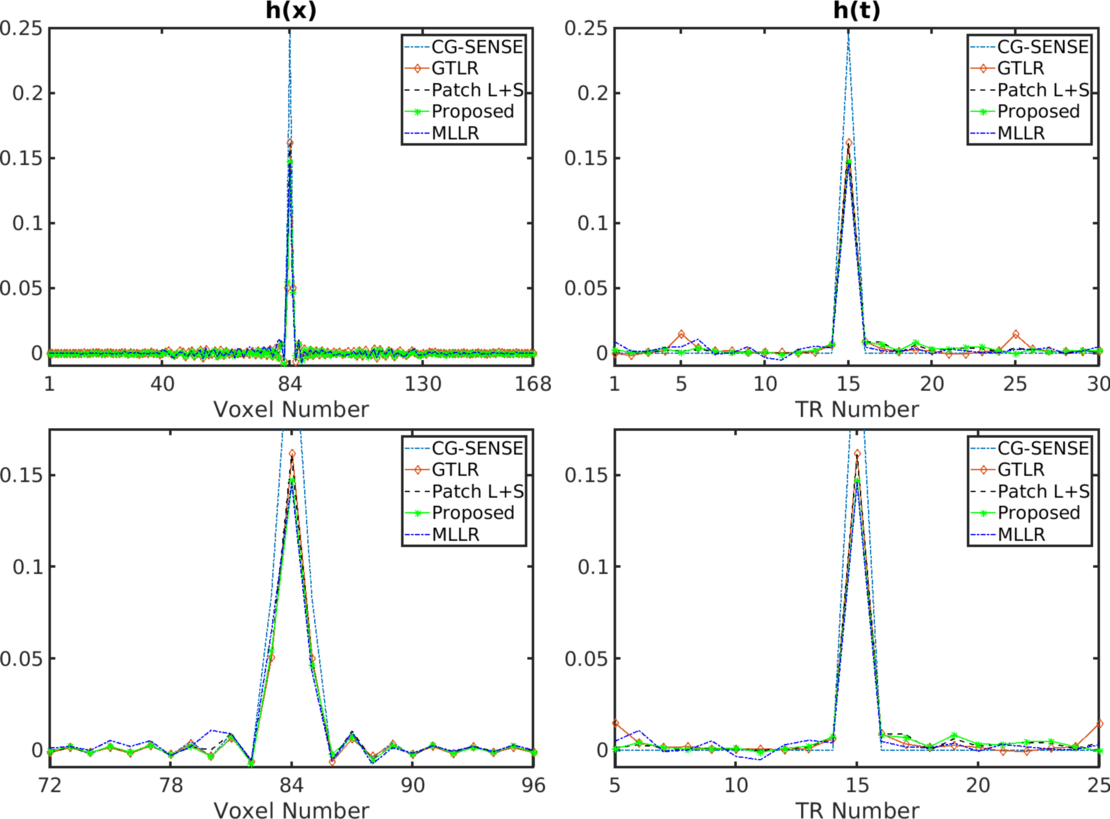}
\caption{
Impulse responses of different reconstructions along spatial dimension (left) and temporal dimension (right). Enlarging the central part of the impulse responses (bottom left and right) shows that impulse responses for different reconstruction models are of similar magnitudes and preserve spatial and temporal resolution with relatively small tails. Because the perturbation of $\delta(j,t)$ added to the image domain is real, and the imaginary parts of the impulse responses are small enough to be neglected, the real parts of the impulse responses are shown.
}\label{tsf8}
\end{figure}

The local impulse response profiles in \fref{tsf8}
demonstrate that we have tuned the different reconstruction methods
so that they are regularizing the data by similar amounts
without excessive spatial or temporal smoothing.

\subsubsection{Overlapping Time Blocks}

\begin{figure}
\centering
\includegraphics[width=0.96\textwidth]{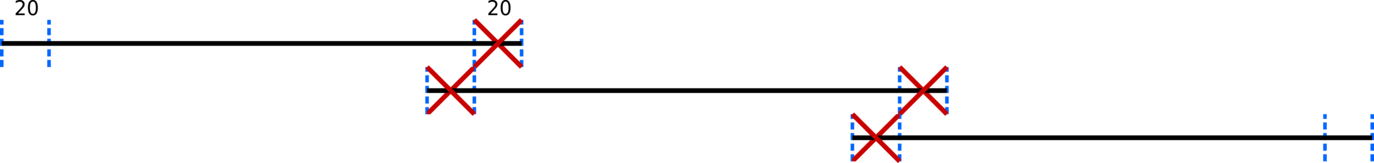}
\caption{
The OSSI fMRI time course is broken into overlapping time blocks of about 300 time points (denoted by black horizontal lines) for reconstruction. The overlapping portion of 20 time points at both ends of the time blocks (denoted by red crosses) are discarded after reconstruction except for the beginning and ending portions of the whole time series. 
}\label{tsf9}
\end{figure}

\begin{figure}
\centering
\includegraphics[width=0.96\textwidth]{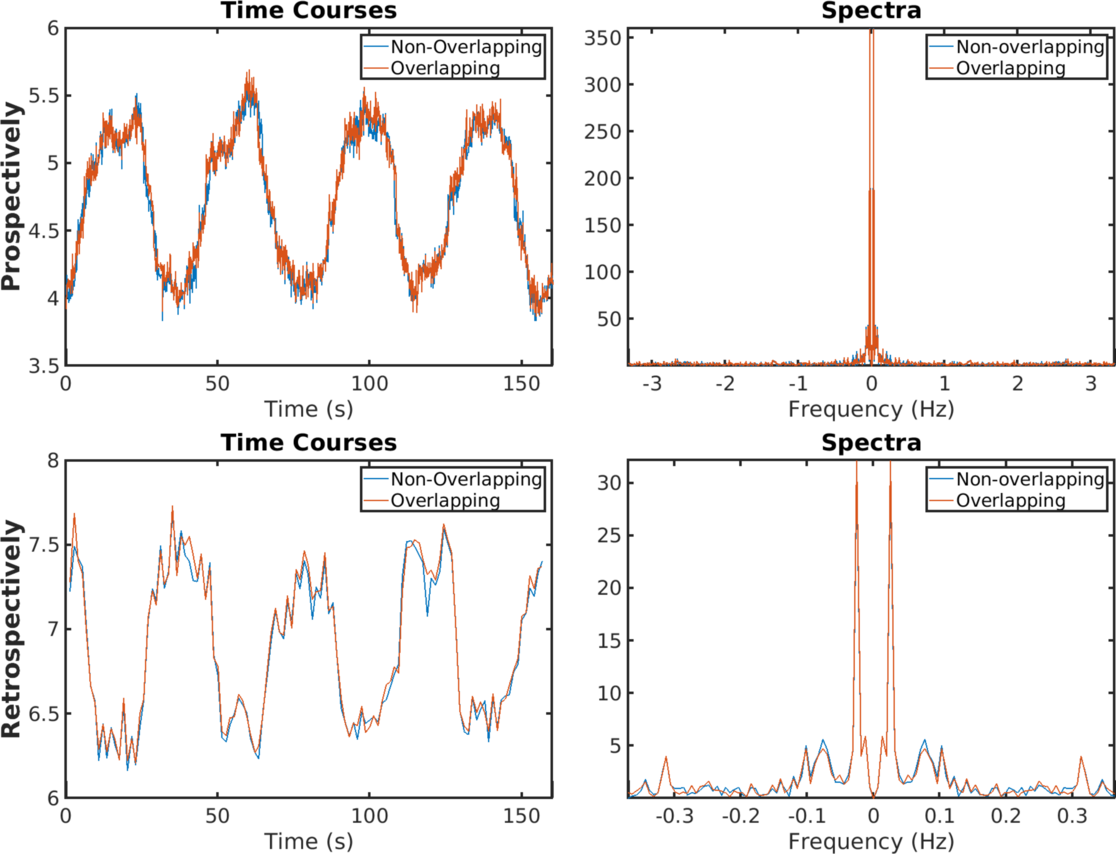}
\caption{
For both prospectively and retrospectively undersampled data, reconstructing overlapping time blocks or non-overlapping time of the whole OSSI fMRI time course leads to very similar time courses and spectra.
}\label{tsf10}
\end{figure}

\fref{tsf9} illustrates ranges of overlapping time blocks and the formation of the entire reconstructed time course after discarding the overlapping portions. \fref{tsf10} compares activated time courses and spectra from reconstructions using non-overlapping time blocks or overlapping time blocks.
With carefully adjusted regularization parameters,
reconstructing overlapping blocks or non-overlapping blocks led to similar results.

\subsection{Comparison and Results}

This section presents additional reconstruction results
for 2D retrospectively and prospectively undersampled data. 

\subsubsection{2D Retrospectively Undersampling}

\begin{figure}
\centering
\includegraphics[width=0.96\textwidth]{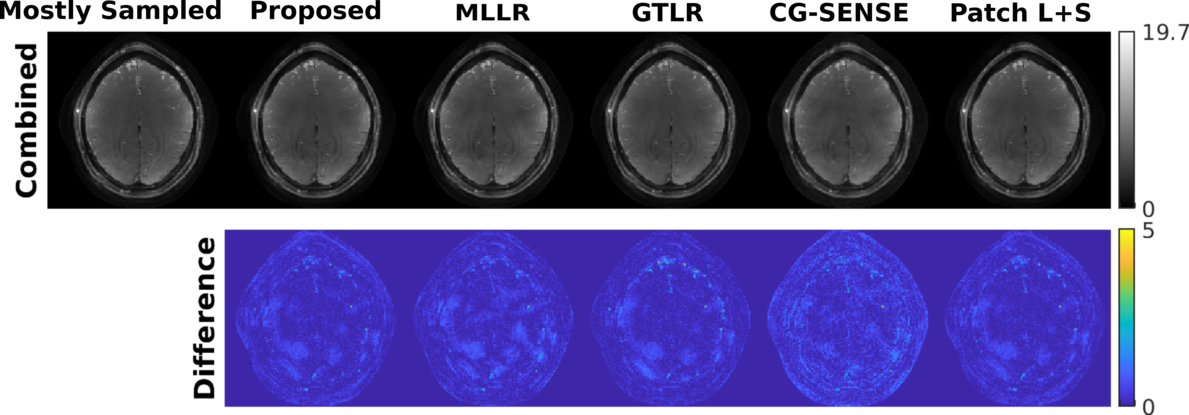}
\caption{
Reconstructed images and difference maps (compared to the mostly sampled reconstruction) of different models after 2-norm combination. The proposed approach presents less residual in the difference map.
}\label{tsf11}
\end{figure}

\begin{figure}
\centering
\includegraphics[width=0.96\textwidth]{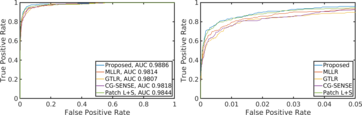}
\caption{
ROC curves of different reconstruction approaches with mostly sampled activation at the lower third of the brain as ground truth. The proposed method outperforms other approaches with the largest area under the ROC curve (left). The ROC curve of the proposed approach is also the closest to the top left corner, especially for the reasonable range with false positive rate less than 0.05 (right). 
}\label{tsf12}
\end{figure}

\begin{figure}
\centering
\includegraphics[width=0.96\textwidth]{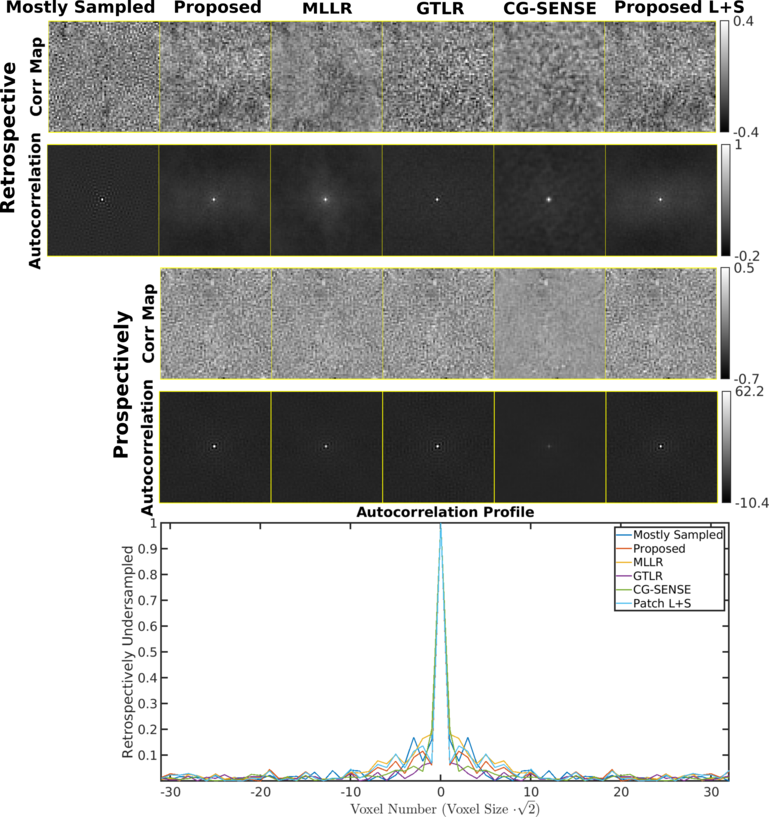}
\caption{
Correlation maps and normalized autocorrelations of the correlation map for the different reconstructions at the center of the brain without activation. 
The proposed model results in similar autocorrelation profiles along diagonal as the mostly sampled reconstruction.
}\label{tsf13}
\end{figure}

\begin{figure}
\centering
\includegraphics[width=0.98\textwidth]{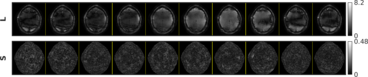}
\caption{
The low-rank and sparse components (first 10 fast time points) of the patch-tensor low-rank plus sparse reconstruction with 2D retrospectively undersampled data. The sparse component is very small and contain limited structural information.
}\label{tsf14}
\end{figure}

\fref{tsf11} shows difference maps of 2-norm combined reconstructions compared to the mostly sampled case. 
ROC curves for the activation maps of different reconstruction approaches in \fref{tsf12}
shows that the proposed approach leads to the largest area under the ROC curve (AUC).
Mostly sampled activation at the lower third of the brain was used as ground truth,
and the activation threshold ranges from -0.1 to 0.99 with a 0.001 spacing. 
\fref{tsf13} presents autocorrelations of the correlation maps for different reconstructions. It verifies that the proposed approach preserves spatial resolution for fMRI.
\fref{tsf14} shows
the low-rank and sparse components (10 fast time points) of the patch-tensor low-rank plus sparse reconstruction.
The sparse component is small and contains little information.

\subsubsection{2D Prospectively Undersampling}

\begin{figure}
\centering
\includegraphics[width=0.99\textwidth]{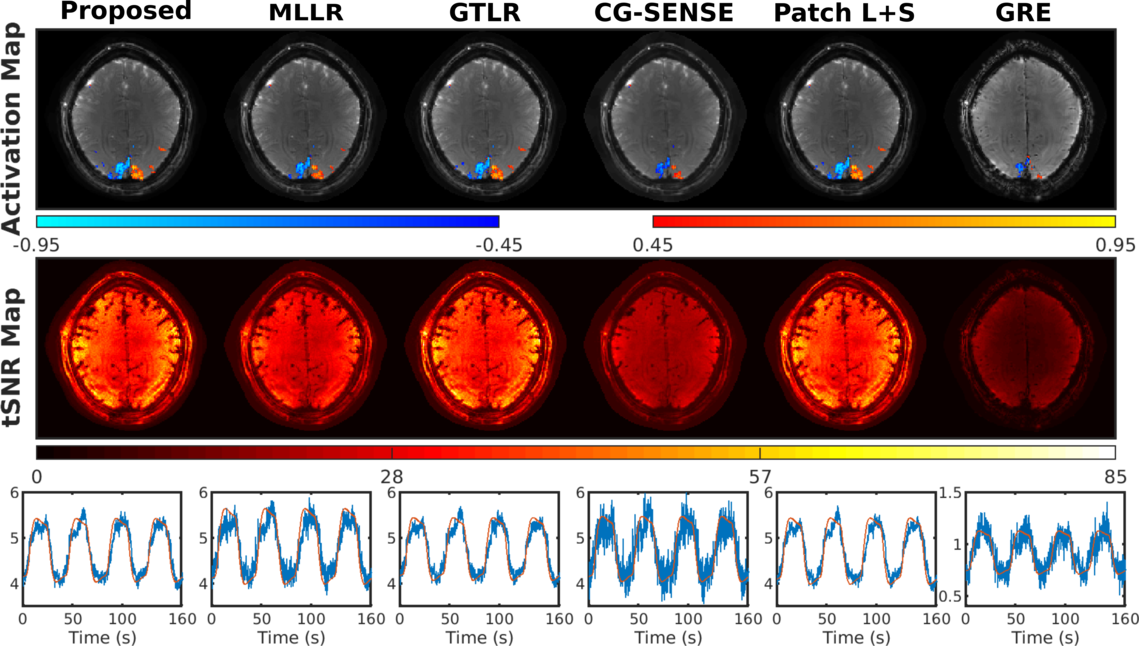}
\caption{
Activation maps, temporal SNR maps, and time courses in the activated regions from prospectively undersampled reconstructions and GRE fMRI. A contiguity threshold of 2 was applied for the activation maps. 
The patch-tensor low-rank, global tensor low-rank, and patch-tensor low-rank plus sparse reconstructions outperform other approaches with more functional activation and cleaner time courses.
}\label{tsf15}
\end{figure}

\begin{table}
\setlength{\tabcolsep}{4pt}
\renewcommand{\arraystretch}{1.2}
\caption{Quantitative comparisons of OSSI 2D prospectively undersampled reconstructions}
\label{tst1}
\centering
\begin{tabular}{ccccccc}
\hline\hline
& Proposed & MLLR & GTLR & \begin{tabular}[c]{@{}c@{}}CG-\\[-0.3em] SENSE\end{tabular} & \begin{tabular}[c]{@{}c@{}}Patch\\[-0.3em] L$+$S\end{tabular} & GRE 
\\ \hline
\begin{tabular}[c]{@{}c@{}}\# Activated\\[-0.3em] Voxels\end{tabular} 
& 322 &  233 &  311  & 149  & 324  &  83
\\ \hline
\begin{tabular}[c]{@{}c@{}}Average\\[-0.3em] tSNR\end{tabular}
&  32.8    &     25.6    &     32.1    &     18.2     &    32.4  &        9.8
\\ \hline\hline
\end{tabular}
\end{table}

\fref{tsf15} and \tref{tst1} give qualitative and quantitative results for 2D prospectively undersampled data reconstructed using patch-tensor LR, MLLR, GTLR, CG-SENSE, and patch-tensor L$+$S approaches with comparison to GRE fMRI. The patch-tensor LR, GTLR, and patch-tensor L$+$S models result in similar performances. The 2D prospectively undersampled data have better temporal resolution (by a factor of 9) than the 2D retrospectively undersampled data, which helps improve the quality of the data-shared initialization and thus the reconstructions. 

\subsection{4D Patch-Tensor and Multi-Scale Patch-Tensor Low-Rank Models}

\begin{figure}
\centering
\includegraphics[width=0.96\textwidth]{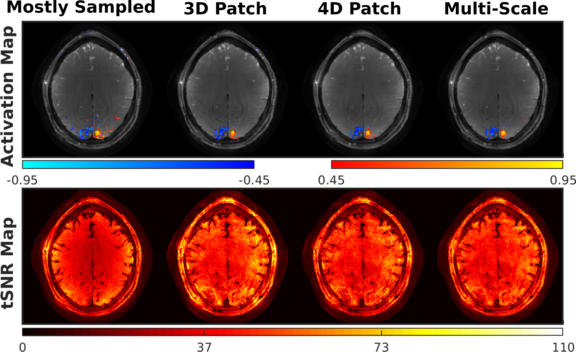}
\caption{
Activation maps and temporal SNR maps from retrospectively undersampled data and reconstruction models including the proposed 3D patch-tensor low-rank, 4D patch-tensor low-rank, and multi-scale tensor low-rank. A contiguity threshold of 2 was applied for the activated regions. All three approaches perform well with similar amounts of activation and temporal SNR.
}\label{tsf16}
\end{figure}

\begin{table}
\setlength{\tabcolsep}{4pt}
\renewcommand{\arraystretch}{1.2}
\caption{Quantitative comparisons of other OSSI 2D retrospectively undersampled reconstructions}
\label{tst2}
\centering
\begin{tabular}{ccccc}
\hline\hline
& \begin{tabular}[c]{@{}c@{}}Mostly \\[-0.3em] Sampled\end{tabular} & 3D Patch & 4D Patch & Multi-Scale \\ \hline
\begin{tabular}[c]{@{}c@{}} 
NRMSD\\[-0.3em]Before Comb\end{tabular}        
& - & 0.17 & 0.19 & 0.17 
\\ \hline
\begin{tabular}[c]{@{}c@{}}  
NRMSD\\[-0.3em]After Comb\end{tabular}        
& - & 0.05 & 0.06 & 0.05 
\\ \hline
\begin{tabular}[c]{@{}c@{}}\# Activated\\[-0.3em] Voxels\end{tabular} 
& 229 & 168 & 145 & 146 
\\ \hline
\begin{tabular}[c]{@{}c@{}}Average\\[-0.3em] tSNR\end{tabular}
& 37.1 & 43.6 & 41.4 & 41.2
\\ \hline\hline
\end{tabular}
\end{table}

\begin{figure}
\centering
\includegraphics[width=0.96\textwidth]{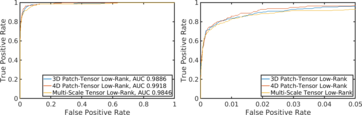}
\caption{
ROC curves of different reconstruction models including the proposed 3D patch-tensor low-rank, 4D patch-tensor low-rank, and multi-scale tensor low-rank. The activation of the mostly sampled data at the lower third of the brain is used as ground truth. 
All three models perform well with large areas under the ROC curve (left), and the ROC curve of the 4D patch-tensor low-rank model is slightly closer to the top left corner than other approaches, especially for the reasonable range with false positive rate less than 0.05 (right). 
}\label{tsf17}
\end{figure}

This section focuses on comparisons to other models including 4D patch-tensor low-rank and multi-scale patch-tensor low-rank. Instead of vectorizing the spatial dimensions as for the proposed 3D patch-tensor low-rank, 4D (or 5D for 3D OSSI fMRI with 2 time dimensions) patch-tensor low-rank model keeps all the spatial dimensions of the tensor for imposing low-rank constrains. The cost function is the same as equation (2) without vectorization of spatial dimensions in $\Pm$. 
The cost function for the multi-scale low-rank model we tested can be expressed as
\begin{equation}
\argmin{\mathbf{X}} \sum_{n = 1}^{3} \sum_{m = 1}^{M_n} \sum_{i = 1}^{\N}
\lambda_i\left\lVert\Pm(\mathbf{X}_n)_{(i)}\right\rVert_*
+ \frac{1}{2}\left\lVert
\mathcal{A}\left(\sum\nolimits_{n = 1}^{3}\mathbf{X}_n\right)-\mathbf{y}
\right\rVert_2^2,
\end{equation}
where $\mathbf{X}_n$ is composed of scale-$n$ patch-tensors.
Specifically, we imposed tensor low-rank on patches of different spatial dimension $4 \times 4$, $8 \time 8$,
and $14 \time 14$.
Here, $\mathcal{P}(\mathbf{\cdot})$ partitions and reshapes the input into $M_n$ low-rank patch-tensors
for different scale $n$.
The regularization parameters for the new models were also selected
based on their impulse responses with similar magnitudes to the 3D patch-tensor LR model. 

All three models are of similar reconstruction and functional performance.
\fref{tsf16} provides activation maps and tSNR maps of 3D patch-tensor LR, 4D patch-tensor LR,
and multi-scale patch-tensor LR with comparison to the mostly sampled reconstruction.
Quantitative evaluations including NRMSD and functional activation are in \tref{tst2}.
\fref{tsf17} shows the ROC curves for the models. 

\subsection{Other Subjects}

\begin{figure}
\centering
\includegraphics[width=0.96\textwidth]{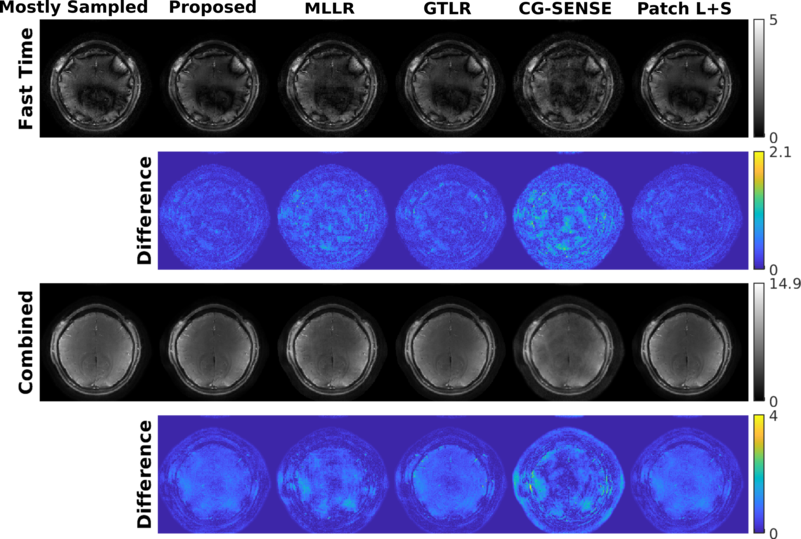}
\caption{
The retrospectively undersampled reconstructions of a different subject are compared to the mostly sampled results. The proposed approach outperforms other methods with less noisy fast time images and less structure in the difference maps before and after combination.
}\label{tsf18}
\end{figure}

\begin{figure}
\centering
\includegraphics[width=0.96\textwidth]{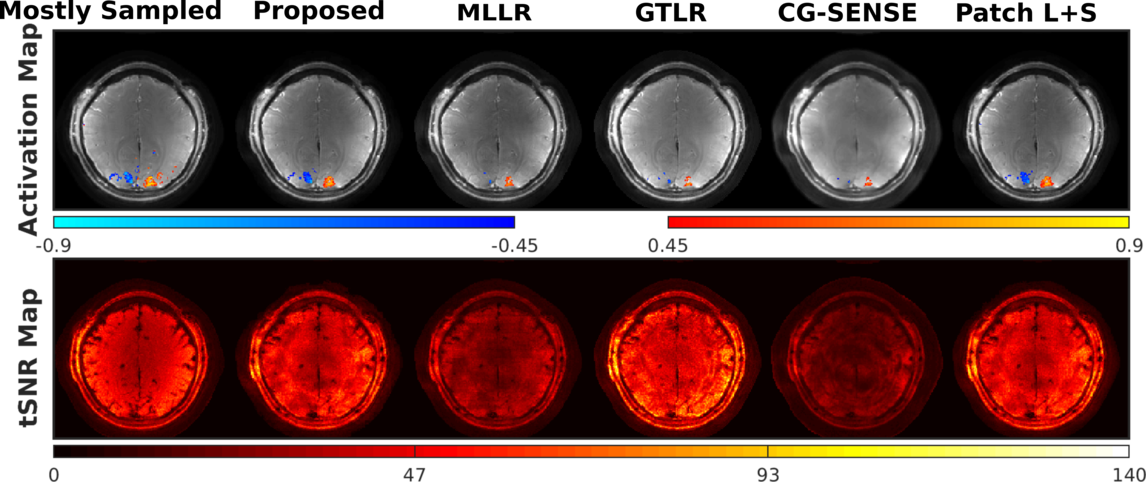}
\caption{
Activation maps and temporal SNR maps from retrospectively undersampled reconstructions of a different subject. A contiguity (cluster-size) threshold of 2 was applied for the activated regions. The proposed model provides more functional activation than other approaches and shows similar results as the patch-tensor low-rank plus sparse model.
}\label{tsf19}
\end{figure}

\begin{figure}
\centering
\includegraphics[width=0.96\textwidth]{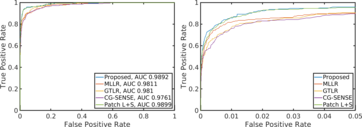}
\caption{
ROC curves for a different subject with mostly sampled activation at the lower third of the brain as ground truth. The proposed method outperforms other approaches with the largest area under the ROC curve (left). The ROC curve of the proposed approach is also the closest to the top left corner, especially for the reasonable range with false positive rate less than 0.05 (right). 
}\label{tsf20}
\end{figure}

\begin{table}
\setlength{\tabcolsep}{4pt}
\renewcommand{\arraystretch}{1.2}
\caption{Retrospectively undersampled reconstructions of a different subject}
\label{tst3}
\centering
\begin{tabular}{ccccccc}
\hline\hline
& \begin{tabular}[c]{@{}c@{}}Mostly \\[-0.3em] Sampled \end{tabular} & Proposed & MLLR & GTLR & \begin{tabular}[c]{@{}c@{}}CG-\\[-0.3em] SENSE\end{tabular} & \begin{tabular}[c]{@{}c@{}}Patch\\[-0.3em] L$+$S\end{tabular} \\ \hline
\begin{tabular}[c]{@{}c@{}} 
NRMSD\\[-0.3em]Before Comb\end{tabular}        
& - &    0.19   &      0.28    &      0.2    &     0.36    &      0.2   
\\ \hline
\begin{tabular}[c]{@{}c@{}}  
NRMSD\\[-0.3em]After Comb\end{tabular}        
& - & 0.12      &   0.13    &     0.13    &     0.14    &     0.13      
\\ \hline
\begin{tabular}[c]{@{}c@{}}\# Activated\\[-0.3em] Voxels\end{tabular} 
& 225  & 166  &  52  &  48  &  34  & 164      
\\ \hline
\begin{tabular}[c]{@{}c@{}}Average\\[-0.3em] tSNR\end{tabular}    
& 40.2      &     41      &   25.2    &     46.1     &      19     &    42.1 \\ \hline\hline
\end{tabular}
\end{table}

\begin{figure}
\centering
\includegraphics[width=0.96\textwidth]{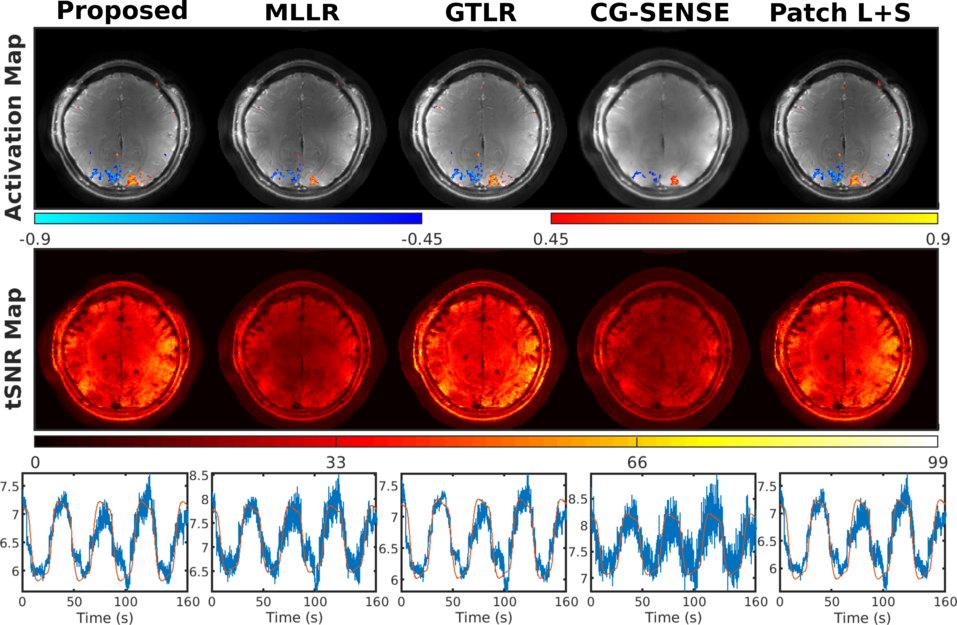}
\caption{
Activation maps, temporal SNR maps, and activated time courses from prospectively undersampled reconstructions of a different subject. A contiguity (cluster-size) threshold of 2 was applied for the activation maps. The patch-tensor low-rank, global tensor low-rank, and patch-tensor low-rank plus sparse reconstructions outperform other approaches with more functional activation and cleaner time courses.
}\label{tsf21}
\end{figure}

\begin{table}
\setlength{\tabcolsep}{4pt}
\renewcommand{\arraystretch}{1.2}
\caption{Prospectively undersampled reconstructions of a different subject}
\label{tst4}
\centering
\begin{tabular}{cccccc}
\hline\hline
& Proposed & MLLR & GTLR & \begin{tabular}[c]{@{}c@{}}CG-\\[-0.3em] SENSE\end{tabular} & \begin{tabular}[c]{@{}c@{}}Patch\\[-0.3em] L$+$S\end{tabular}  
\\ \hline
\begin{tabular}[c]{@{}c@{}}\# Activated\\[-0.3em] Voxels\end{tabular} 
& 225 &  120 &  223  &  89  & 227
\\ \hline
\begin{tabular}[c]{@{}c@{}}Average\\[-0.3em] tSNR\end{tabular}
& 33.5     &    21.1    &     34.9    &     20.6      &     34
\\ \hline\hline
\end{tabular}
\end{table}

This section presents 2D reconstruction results of a different subject. Both retrospectively and prospectively undersampled data were acquired with spiral-out trajectories. Retrospectively undersampled reconstruction results before and after 2-norm combination, and difference maps compared to the mostly sampled data are in \fref{tsf18}. \fref{tsf19} presents functional activation maps and tSNR maps demonstrating that the proposed model outperforms other approaches with more activation. \tref{tst3} summarises quantitative values of different reconstructions. \fref{tsf20} provides ROC curves of the activation maps. 2D prospectively undersampled reconstruction results including activation maps, tSNR maps, and example time courses are given in \fref{tsf21}. \tref{tst4} gives the corresponding quantitative evaluations.

\chapter{Manifold Model for High-Resolution fMRI Joint Reconstruction and Dynamic Quantification}
\label{chap:3manifold}
In \autoref{chap:2tensor}, we exploited high-dimensional spatial-temporal similarities of OSSI fMRI time series using the patch-tensor low-rank model. The structured patch-tensor has one vectorized spatial dimension from a local patch and two time dimensions (fast and slow times). We noticed that due to the nonlinear oscillations in OSSI images, the unfolded matrix along fast time is not very low-rank compared to other unfoldings as presented in \cref{rank}. 
This chapter focuses on OSSI fast time images, and aims to accurately model the nonlinearity of fast time signals with a manifold model. We compare the proposed manifold model to a subspace model that imposes matrix low-rankness on fast time images with global spatial and fast time dimensions.

OSSI
is a recent fMRI acquisition method that exploits a large and oscillating signal,
and can provide high SNR fMRI.
However, the oscillatory nature of the signal leads to an increased number of acquisitions.
To improve temporal resolution and accurately model the nonlinearity of OSSI signals,
we build the MR physics for OSSI signal generation
as a regularizer for the undersampled reconstruction
rather than using subspace models that are not well suited for the data.
Our proposed physics-based manifold model
turns the disadvantages of OSSI acquisition into advantages
and enables joint reconstruction and quantification. 
OSSI manifold model (OSSIMM) outperforms subspace models
and reconstructs high-resolution fMRI images
with a factor of 12 acceleration and without spatial or temporal resolution smoothing.
Furthermore, OSSIMM can dynamically quantify important physics parameters,
including $R_2^*$ maps,
with a temporal resolution of 150 ms. \footnote{This chapter is based on \cite{2021manifold,dictionary,dictionary2}.
}

\section{Introduction}

Functional magnetic resonance imaging (fMRI) is an important tool for brain research and diagnosis.
In its most common form, it detects functional activation by acquiring a time-series of MR images with blood-oxygen-level-dependent (BOLD) contrast \cite{Ogawa1990}. However, the BOLD effect has a relatively low signal-to-noise ratio (SNR) \cite{noll2001}, and the SNR further decreases with improved spatial resolution. Because the functional units (cortical columns) of the brain are on the order of 1 mm, high resolution with high SNR is critical for some fMRI experiments.
This chapter focuses on Oscillating Steady-State Imaging (OSSI),
a recent fMRI acquisition approach
that provides higher SNR signals than standard gradient-echo (GRE) imaging \cite{ossi2019}. 

The SNR advantage of OSSI comes at a price of spatial and temporal resolutions. 
OSSI acquisition requires a quadratic RF phase cycling with cycle length $n_c$
(e.g., $n_c = 10$).
The corresponding OSSI signal oscillates with a periodicity of $n_c \cdot$ TR,
and the frequency-dependent oscillations result in oscillatory patterns in OSSI images.
Therefore, every image in a regular fMRI time course is acquired $n_c$ times with different phase increments in OSSI,
and combining the $n_c$ images eliminates oscillations for fMRI analysis.
Acquiring $n_c$ times more images compromises temporal resolution,
and the short TR necessary for OSSI acquisition can limit single-shot spatial resolution. 

To improve the spatial-temporal resolution,
we previously used a patch-tensor low-rank model for the sparsely undersampled reconstruction \cite{tensor}.
While low-rank regularization fits data to linear subspaces,
OSSI images are not very low-rank because of the nonlinear oscillations \cite{dictionary}.
Instead of imposing low-rankness and/or sparsity that may or may not suit the data,
this chapter proposes a nonlinear dimension reduction approach for OSSI reconstruction
that uses a MR physics-based manifold as a regularizer,
inspired by parameter map reconstruction methods for MR fingerprinting \cite{Zhao2016,Asslander2018}.

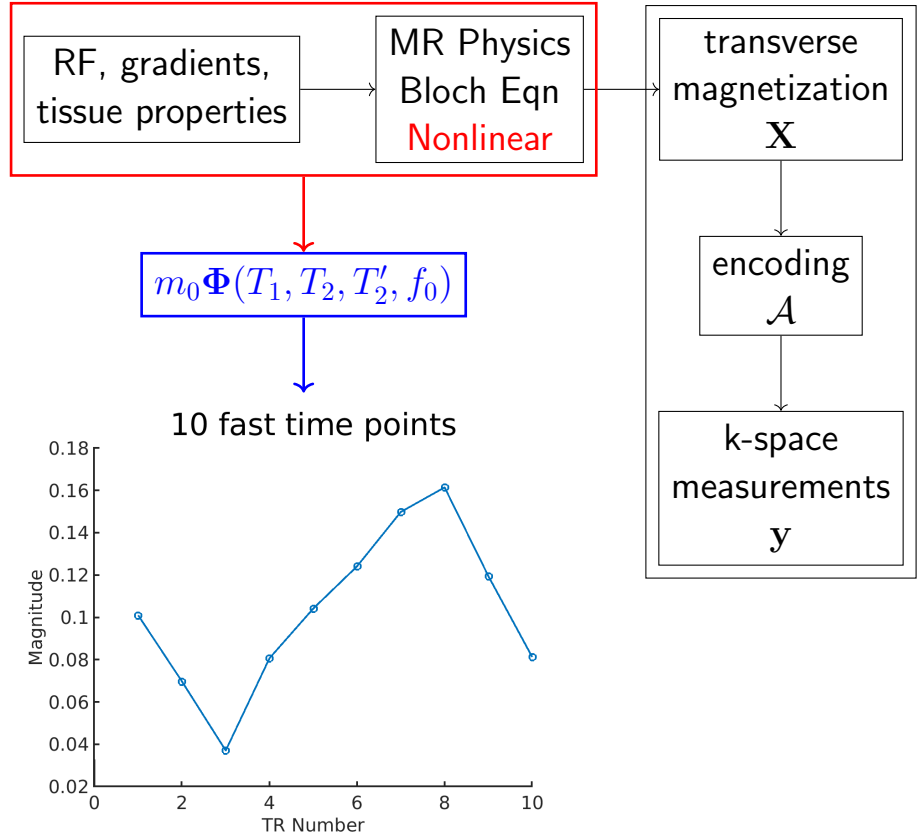
\begin{figure}
\centering
\begin{tikzpicture}[node distance=1cm]
\tikzset{every node}=[font=\large\sffamily]
\node [draw] (A) {\Longstack[c]{RF, gradients,\\tissue properties}};
\node [draw, right=of A] (B) {\Longstack[c]{MR Physics\\ Bloch Eqn\\ \color{red}{Nonlinear}}};
\node [draw, right=of B] (C) {\Longstack[c]{transverse \\ magnetization\\ $\mathbf{X}$}};
\node [draw, below=of C] (D) {\Longstack[c]{encoding\\ $\mathcal{A}$}};
\node [draw, below=of D] (E) {\Longstack[c]{k-space\\ measurements\\ $\mathbf{y}$}};
\draw [->] (A) edge (B) (B) edge (C) (C) edge (D) (D) edge (E);
\node[draw,fit=(C) (D) (E)] {};
\node[draw,red,line width=1pt,fit=(A) (B)] (M) {};
\node[draw,blue,line width=1pt,below=of M] (N) {$m_0 \mathbf{\Phi} (T_1, T_2, T_2', f_0)$};
\path[->,red,line width=1pt] (M) edge (N);
\node [draw, below=of N, white] (F) {\includegraphics[height=0.25\textheight]{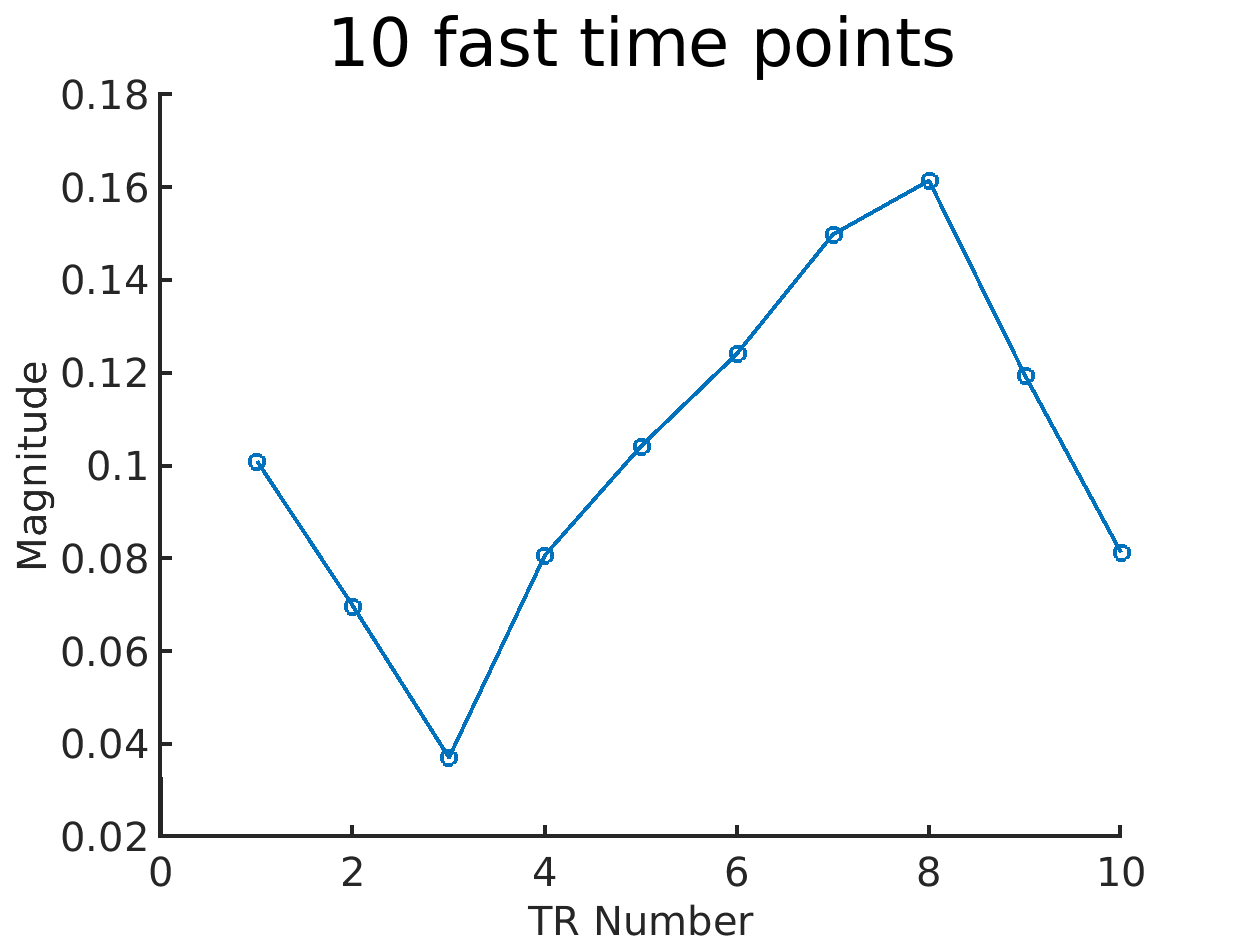}};
\path[->,blue,line width=1pt] (N) edge (F);
\end{tikzpicture}
\caption{
The proposed manifold model uses the MR physics for signal generation
as a regularizer for the undersampled reconstruction.
}\label{mf0}
\end{figure}

As outlined in Fig.~\ref{mf0},
the manifold model focuses on MR physics for OSSI signal generation.
It represents $n_c$ OSSI signal values per voxel
by just 3 physical parameters, via Bloch equations.
The nonlinear nature of the Bloch equations
enables nonlinear representations of the data and nonlinear dimension reduction.
We further introduce a near-manifold regularizer
that 
encourages the reconstructed signal values
to lie near the manifold.
Compared to quantitative imaging works that enforce the reconstructed images
to be exactly equal to the physics-based representations
\cite{Zhao2016,Asslander2018,Dong2019,Tamir2020},
the proposed near-manifold regularizer encourages the images to be near the manifold
while also allowing for potential model mismatch.

Standard $T_2^*$-weighted magnitude images only assess relative signal changes due to BOLD effects
and are not quantitative in terms of the blood oxygenation level, $T_2^*$ or $T_2'$ \cite{Speck1998,Wennerberg2001,Olafsson2008}.
Quantifying $T_2^*$ is important because of its sensitivity to iron concentration for disease monitoring \cite{Wang2019}.
By constructing a $T_2'$ manifold based on BOLD-induced intravoxel dephasing,
our work demonstrates the utility of the OSSI manifold model
for dynamic quantification of $T_2^*$/$R_2^*$.

This chapter shows that the proposed $T_2'$ manifold and near-manifold regularizer
can jointly optimize OSSI images and quantitative maps.
The manifold model enables high-resolution OSSI fMRI with 12-fold acquisition acceleration,
outperforms low-rank regularization with more functional activation,
and provides quantitative and dynamic assessment of tissue $R_2^*$ maps
and off-resonance $f_0$,
with a temporal resolution of 150~ms.

\section{OSSI Manifold Model (OSSIMM)}
\begin{figure}
\centering
    \includegraphics[width=0.9\textwidth]{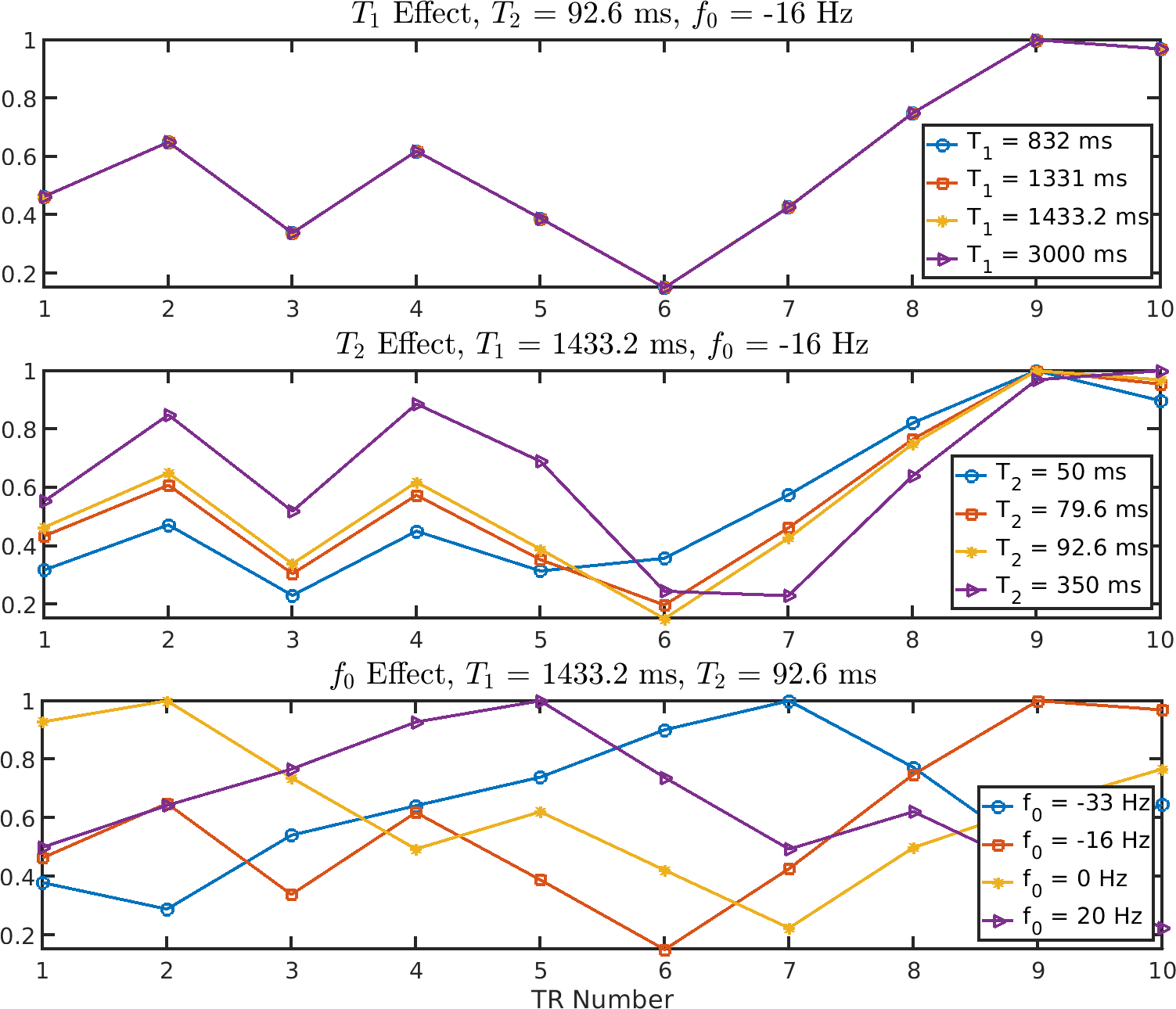}
   \caption{
   Normalized OSSI fast time signal magnitude for one isocromat with nonlinear oscillations determined by physics parameters $T_2$ and $f_0$. The change of $T_1$ only scales OSSI signal values.
   }\label{mf1}
\end{figure}

OSSI signal oscillates with a periodicity of $n_c$TR, and the OSSI fMRI time course contains $n_c$ images for every image in a regular fMRI time series. We refer to the fast acquisition dimension of size $n_c$ as ``fast time’’ and the regular fMRI time dimension as ``slow time’’ as presented in supplemental Fig.~\ref{msf0}.
OSSI fast time signals can have different shapes
and change nonlinearly
with respect to MR physics parameters,
as illustrated in Fig. \ref{mf1}.
To accurately model the nonlinear oscillations,
we propose a MR-physics based manifold model for the undersampled reconstruction.

\subsection{Physics-Based Manifold}

In OSSI, the steady-state transverse magnetization of one isocromat at observation time $t$ is
\[
m_0 \, \Bp (t; T_1, T_2, f_0), 
\]
where $m_0 \in \Cp$ is the equilibrium magnetization,
$\Bp(\cdot) \in \Cp^{n_c}$ represents MR physics calculated by Bloch equations,
$T_1$ and $T_2$ are tissue relaxation times,
and $f_0$ denotes central off-resonance frequency from $B_0$ field inhomogeneity.

$T_2'$-weighted OSSI signal in a voxel with an intra-voxel spreading of off-resonance frequencies $f$ can be modeled as:
\desa{
\begin{aligned}
m_0  & \BP (t; T_1, T_2, T_2', f_0) = \int m_0 \Bp (t; T_1, T_2, f_0 + f) \, p(f; T_2') \df
.
\end{aligned}
}
The $T_2'$ exponential decay corresponds to a Cauchy distribution
for $f$ with a probability density function (PDF)
$p(f) = \gamma / \pi (\gamma^2 + f^2)$, and scale parameter $\gamma = 1 / (2 \pi T_2')$.

The isocromat signal at time $t > 0$ presents increased $T_2$ decay and increased off-resonance dephasing due to field inhomogeneity and BOLD-related field changes,
\desa{
\begin{aligned}
m_0 & \Bp (t; T_1, T_2, f_0) = m_0 \Bp (t = 0; T_1, T_2, f_0) 
\Expn{t / T_2}
\Expni{2\pi f_0 t}
,
\end{aligned}
}
where $t = 0$ denotes the time right after the excitation.


As OSSI TR is relatively short (e.g., TR = 15 ms),
we neglect the intravoxel dephasing during the readout
and approximate the signal at $0 \leq t \leq \textrm{TR}$
with the signal at the echo time TE.
The $T_2'$-weighted signal becomes
\desa{
\begin{aligned}
m_0 \BP (T_1, T_2, T_2', f_0)
& \approx \int m_0 \Bp (\textrm{TE}; T_1, T_2, f_0 + f) 
\Expn{\textrm{TE}/ T_2}
\Expni{2\pi (f_0 + f)\textrm{TE}}
p(f; T_2') \df
\label{T2s}
.
\end{aligned}
}

Accordingly, $T_2'$-weighted OSSI fast time signals lie on the physics-based manifold:
\desa{ 
\{m_0 \BP (T_1, T_2, T_2', f_0) \in \Cp^{n_c}:
m_0 \in \Cp, \, T_1, T_2, T_2', f_0 \in \Re
\},
}
The manifold maps a limited number of physics parameters to the $n_c$-dimensional oscillating signals via MR physics. 

\subsection{Near-Manifold Regularization}

The physics-based manifold models the generation of MR signals,
enables nonlinear dimension reduction,
and can be an accurate prior for the undersampled reconstruction. 
Because the physics parameters are location dependent,
and because OSSI signal values change drastically with varying parameters as shown in Fig.~\ref{mf1},
we model the fast time signals in a voxel-by-voxel manner.
Furthermore, to account for potential mismatches due to model simplifications
and nonidealities in experiments (e.g., flip angle inhomogeneity),
we propose a near-manifold regularizer that encourages the signal values in each voxel
to be close to the manifold estimates but not necessarily exactly the same. 

The proposed $T_2'$ manifold-based image reconstruction problem
uses the following optimization formulation:
\desa{
\begin{array}{ll}
\hat{\X} =
\argmin{\X}
\frac{1}{2}\lVert\mathcal{A}(\X)-\y \rVert_2^2
+ \beta \sum_{n=1}^N \mathcal{R} \left(\X[n,:]\right)
, \quad
\\[1.5em]
\mathcal{R}(\bv) = \underset{m_0,T_2', f_0}{\textrm{min}}\ \lVert \bv-m_0 \BP (T_2', f_0; T_1, T_2)\rVert_2^2,
\end{array}
\label{costmf}
}
where
$\X \in \Cp^{N \times n_c}$
denotes $n_c$ fast time images to be reconstructed. The vectorized spatial dimension $N$ is $N_{xy}$ for 2D OSSI fMRI. 
$\mathcal{A}(\mathbf{\cdot})$ is a linear operator consisting of coil sensitivities and the non-uniform Fourier transform including undersampling,
$\y$ represents sparsely sampled k-space measurements.
$\beta$ is the regularization parameter.
$\bv \in \Cp^{n_c}$ is a vector of fast time signal values for each voxel in $\X$, 
$m_0\BP(T_2', f_0; T_1, T_2) \in \Cp^{n_c}$ denotes the manifold estimates.
The regularizer minimizes the Euclidean distance between $\bv$ and $m_0\BP(T_2', f_0; T_1, T_2)$.
$T_1$ and $T_2$ are not directly estimated by the model.
$T_1$ has a signal scaling effect
that can be absorbed in $m_0$,
as illustrated in Fig.~\ref{mf1}.
Section~\ref{sec:simu} describes
the choices of baseline $T_2$ values for $T_2^*$ estimation.

The voxel-wise parametric regularizer $\mathcal{R}(\bv)$
not only performs regularization for the ill-posed reconstruction problem,
but also involves parameter estimation and can provide quantitative maps for $T_2'$ and $f_0$. 

\subsection{Optimization Algorithm} 

To solve
\eqref{costmf},
we alternate between a regularization update and a data fidelity update for the reconstruction. 
The minimization of the voxel-wise parametric regularizer is a nonlinear least-square problem
that we solve using the variable projection (VARPRO) method \cite{Golub1973,Golub2003}. 
Let $\ba = [T_2', f_0]$ denote the two nonlinear tissue parameters;
the calculation of $\ba$ using VARPRO simplifies to
\desa{
\hat{\ba} 
= \argmax{\ba}
\frac{\abs{{\BP(\ba)}'\bv}^2}{\norm{\BP(\ba)}_2^2},
\label{up1}
}
where
$\bv = \X[:,n] \in \Cp^{n_c}$.
Instead of solving \eqref{up1} for the explicit and sophisticated $\BP(\ba)$, 
we construct a dictionary consisting of discrete $\BP(\ba)$ realizations
with varying $\ba$ parameters using Bloch simulations,
and then perform grid search
to find $\hat{\ba}$
for which $\BP(\hat{\ba})$ best matches $\bv$. 

Updating $m_0$ is a least-squares problem with closed-from solution:
\desa{
\hat{m}_0 = \frac{{\BP(\hat{\ba})}'\bv}{\normr{\BP(\hat{\ba})}_2^2}
\label{up2}
.}
We parallelize the regularization update across different voxels.

The update step for $\X$
involves a quadratic least-squares problem
that we solve using the conjugate gradient method
as implemented in the Michigan Image Reconstruction Toolbox \cite{fesslermirt}.
This data fidelity update is easily parallelized over different fast time images or different fast time images sets
to speed up the fMRI time series reconstruction. 

\subsection{Comparison Method}

We compare the manifold approach to a low-rank reconstruction approach
that models the fast time signals using linear subspaces.
The cost function for this low-rank comparison method is
\desa{
\begin{array}{ll}
\hat{\X} =
\argmin{\X}
\frac{1}{2}\lVert\mathcal{A}(\X)-\y \rVert_2^2
+ \alpha \lVert \X \rVert_*
\end{array}
\label{costlr}
}
where
$\X \in \Cp^{N \times n_c}$ represents every $n_c$ fast time images,
and $\alpha$ is the regularization parameter.
We solve the optimization problem \eqref{costlr}
using the proximal optimized gradient method (POGM)
with adaptive restart
\cite{kim2018adaptive,Taylor2017,lin:19:edp}.

\section{Simulation Investigations}
\label{sec:simu}
We generated OSSI signals via Bloch simulation
using pulse-sequence parameters that matched the actual data acquisition.
We used TR = 15 ms, TE = 2.7 ms (spiral-out trajectory), RF excitation pulse length = 1.6 ms,
quadratic RF phase cycling with $\Phi(n) = {\pi n^2}/{n_c}$ for $n$th TR, $n_c = 10$,
and flip angle = $10^\circ$ \cite{ossi2019}. 

\begin{figure}
     \centering
     \subcaptionbox{\label{mf2a}}{\includegraphics[width=.49\textwidth]{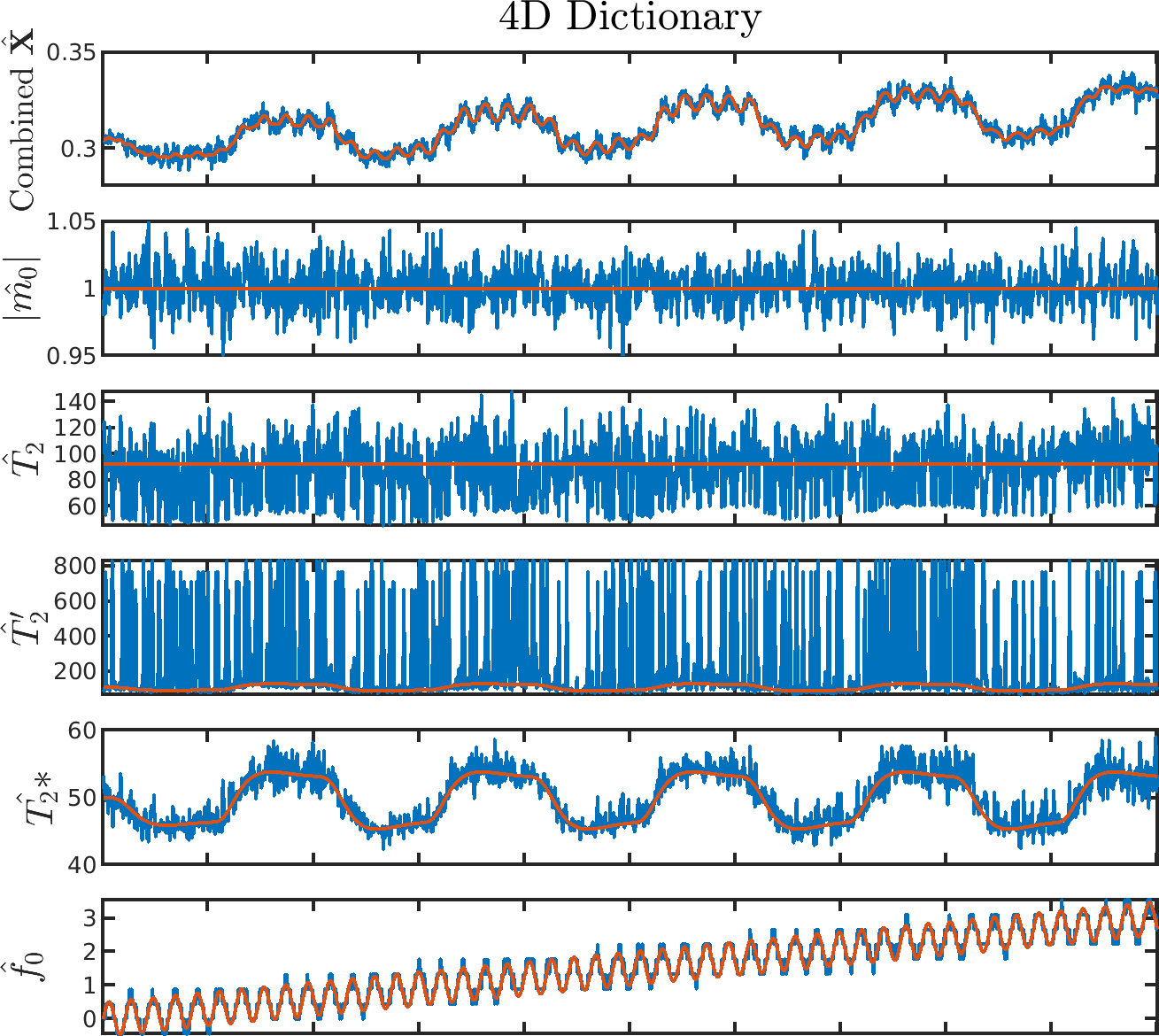}}
     \subcaptionbox{\label{mf2b}}{\includegraphics[width=.49\textwidth]{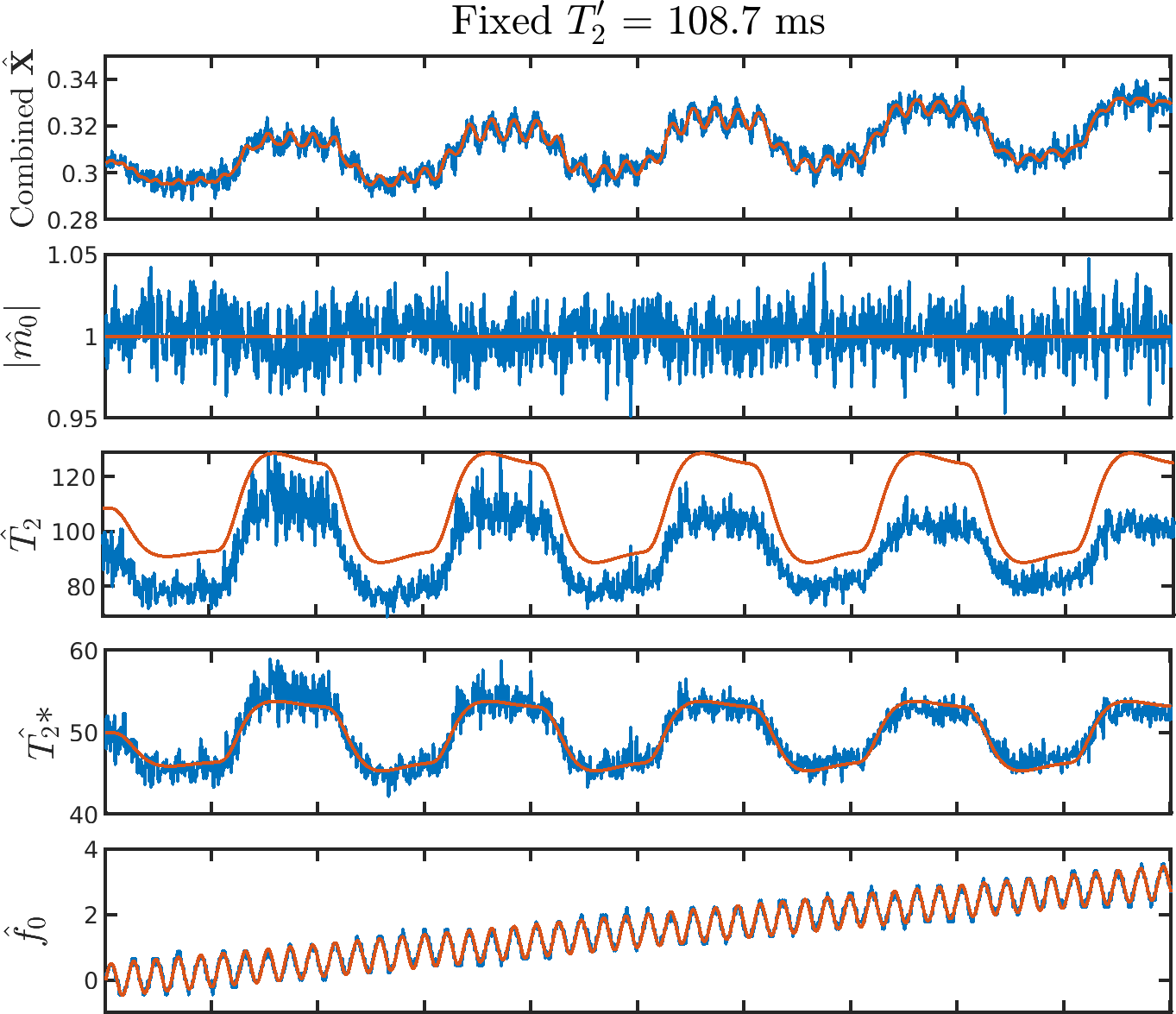}}
     \subcaptionbox{\label{mf2c}}{\includegraphics[width=.49\textwidth]{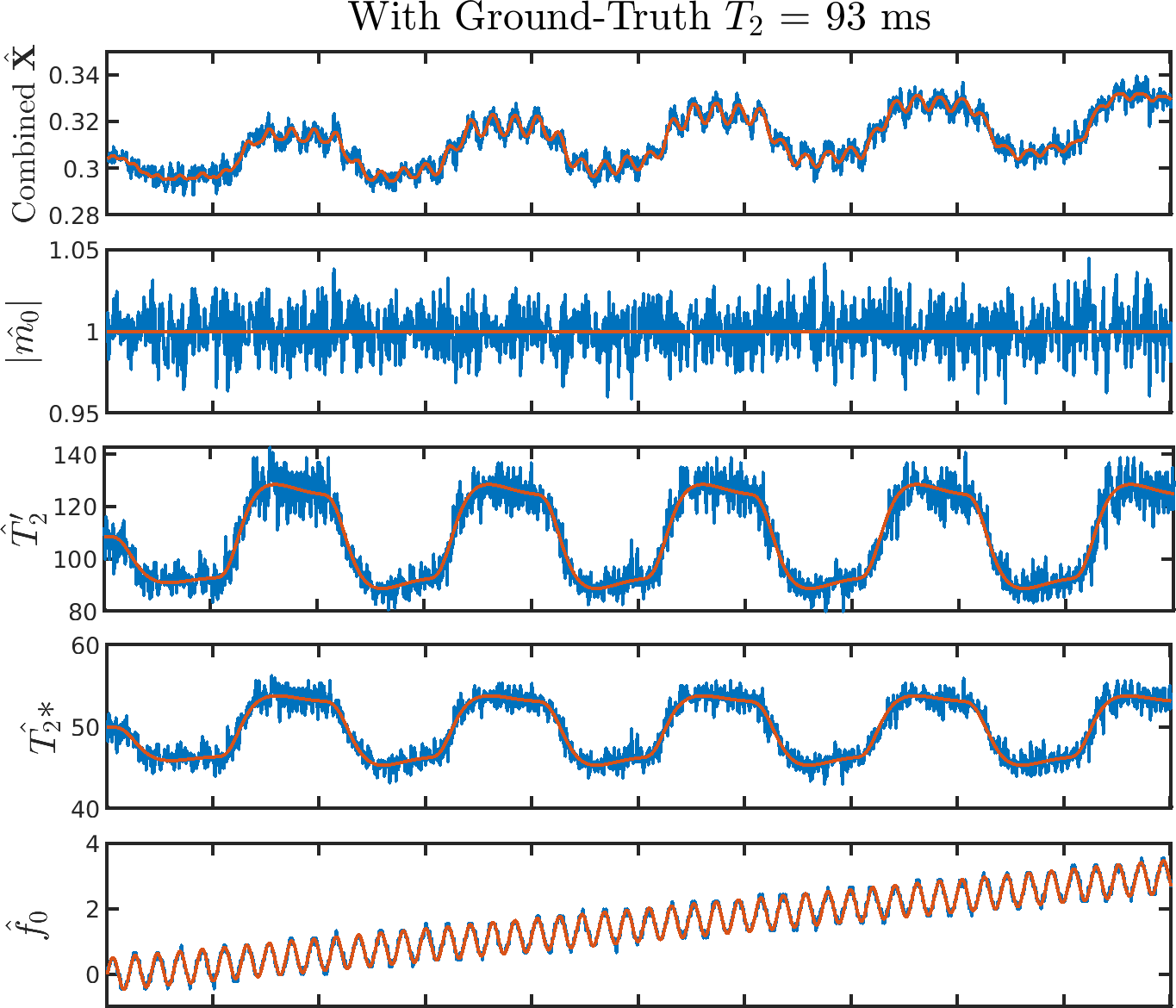}}
     \subcaptionbox{\label{mf2d}}{\includegraphics[width=.49\textwidth]{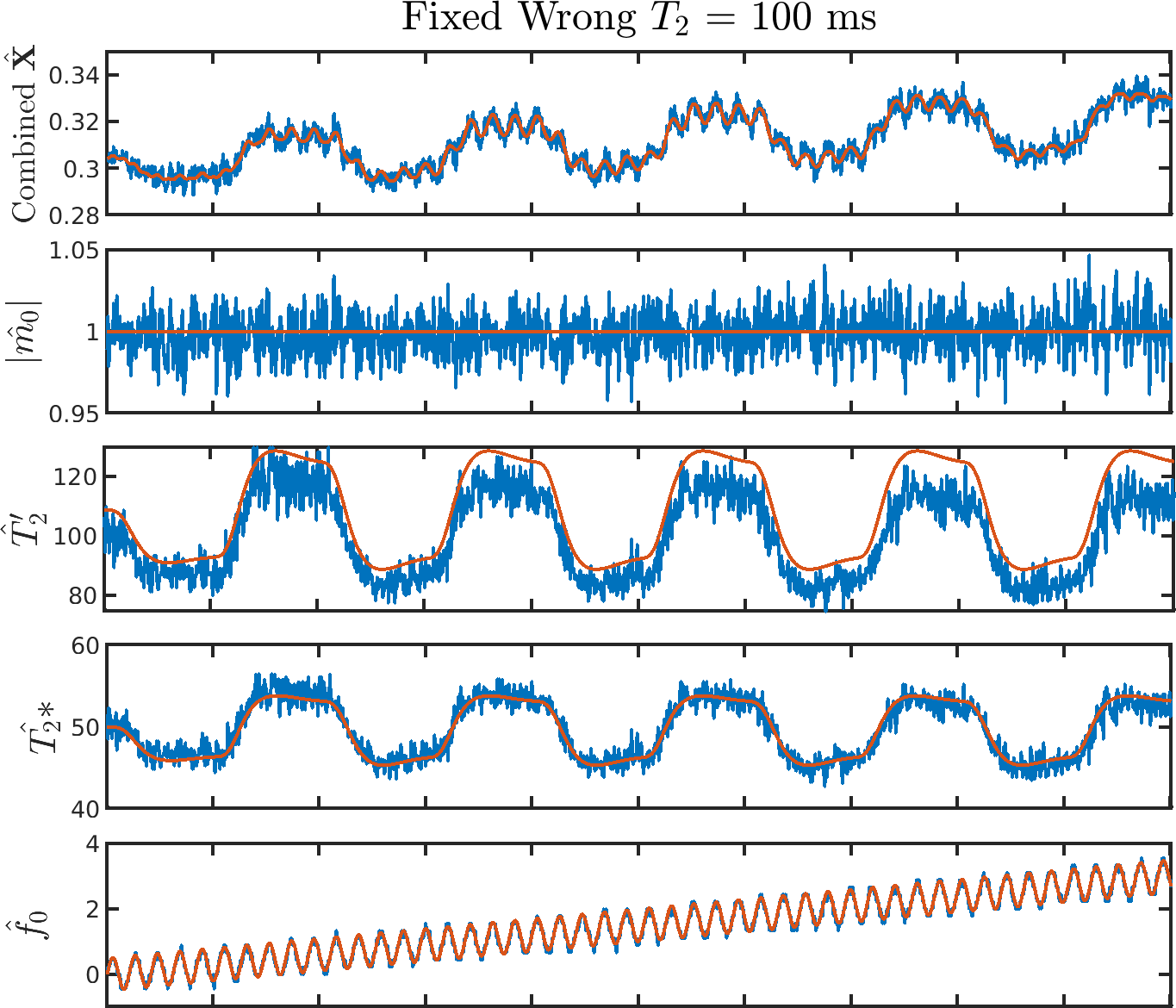}}
     \caption{
     Quantification results for a simulated OSSI fMRI voxel using the manifold model with 4 different choices of the manifold. Because $T_2$ and $T_2'$ effects to OSSI signals are correlated (Fig.~\ref{mf2a}), and a $T_2$ manifold is not good enough for capturing BOLD-induced $T_2'$ changes (Fig.~\ref{mf2b}), we use a $T_2'$ manifold for quantification. We can estimate $T_2^*$ and $T_2'$ with known $T_2$ values (Fig.~\ref{mf2c}), or use a biased guess of $T_2$ for quantifying $T_2^*$ (Fig.~\ref{mf2d}).
     }
     \label{mf2}
\end{figure}

\subsection{OSSI Signals}

The OSSI signal $\in \mathbb{C}^{n_c}$ for one isocromat is determined by physics parameters $T_1$, $T_2$, and $f_0$. 
Fig. \ref{mf1} presents example OSSI isocromat signals (normalized by the maximum magnitude)
with varying physics parameters selected based on gray matter relaxation parameters:
$T_1$ = 1400 ms, $T_2$ = 92.6 ms \cite{Parameters}. 
As an approximation of \eqref{T2s},
we simulated $T_2'$-weighted OSSI signal in a voxel with Riemann sum of numerous OSSI isocromat signals
at different off-resonance frequencies.
Specifically, we calculated a weighted sum of OSSI signals
from 4000 isocromats at off-resonance frequency $f_0+f$,
where $f$ uniformly ranged from -200 Hz to 200 Hz,
and the weighting function was the PDF of the Cauchy distribution.

We further simulated a fMRI time course for one voxel with time-varying $T_2'$ values. 
The $T_2'$ waveform is the convolution of the canonical hemodynamic response function (HRF) \cite{spm}
and the fMRI task waveform. 
Because fMRI percent signal change
$\Delta \% \approx \Delta R_2' \cdot \textrm{TE}_\textrm{eff}$ \cite{Jin2006}
and OSSI $\textrm{TE}_\textrm{eff}$ = 17.5 ms \cite{ossi2019},
we set $\Delta T_2'$ = 15.4 ms to produce a typical percent signal change of 2\%.
The fMRI time course is also affected by scanner drift and respiration induced $f_0$ changes.
We simulated $f_0$ with a linearly increasing scanner drift of about 1 Hz per minute
and a sinusoidal waveform (magnitude of 0.5 Hz and period of 4.2 s) to model the respiratory changes.
We also added complex Gaussian random noise for a typical temporal SNR (tSNR) value of 38 dB.

\subsection{Dictionary Selection}

We represented OSSI manifold using a signal dictionary,
and each dictionary atom is a point on the manifold.
Because $T_2$, $T_2'$, and $f_0$ affect OSSI signals in different ways
while $T_1$ has a scaling effect,
we constructed a 4D dictionary by varying $T_2$, $T_2'$, $f_0$,
for $T_1$ = 1400 ms.
The $T_2$ grids were in the 40 to 150 ms range with a 1 ms spacing.
The $T_2'$ grids were calculated by uniformly changing $R_2^*$ from 12 to 38 Hz \cite{Peran2007}
with a step size of 0.1 Hz and a fixed $T_2$ of 92.6 ms.
We set central off-resonance frequency $f_0$ to [-33.3,33.3] Hz
with a 0.22 Hz spacing as OSSI signals are periodic
with off-resonance frequency period = 1/TR = 66.7 Hz \cite{ossi2019}.

We reconstructed the functional signal and physics parameters from the simulated noisy fMRI time courses
using the near-manifold regularizer in \eqref{costmf} and the 4D dictionary.
The reconstructions were performed by
(a) simultaneously estimating $T_2$ and $T_2'$ using the 4D dictionary,
(b) assuming $T_2'$ is fixed and estimating $T_2$
using the 3D subset of the 4D dictionary based on the assumed $T_2'$ value,
(c) estimating $T_2'$ with the actual $T_2$ value and the corresponding 3D dictionary,
(d) assuming $T_2$ is fixed and estimating $T_2'$ with a biased $T_2$ value and the corresponding 3D dictionary.

As shown in Fig.~\ref{mf2},
because of the strong coupling between $T_2$ and $T_2'$ values,
it is infeasible to simultaneously estimate $T_2$ and $T_2'$ (see Fig.~\ref{mf2a}).
Using a biased $T_2'$ value for $T_2$ estimation (Fig.~\ref{mf2b})
or a biased $T_2$ value for dynamic $T_2'$ estimation (Fig.~\ref{mf2d})
results in noticeable bias,
whereas Fig.~\ref{mf2c} presents accurate $\hat{T}_2'$ when the ground truth $T_2$ is provided. 
However, all the different estimation approaches lead to relatively good $T_2^*$ estimates.
Because $m_0$ and $T_2^*$ estimates are more accurate in Figs.~\ref{mf2c} and~\ref{mf2d},
we propose to use assumed $T_2$ values
or to measure accurate baseline $T_2$ maps
to use for dynamic $T_2^*$ quantification.
The latter approach also provides $T_2'$ estimates.
Notably,
the quality of the combined functional signals is insensitive to the choice of manifold for reconstruction.

\section{Experiments}
We collected resolution phantom data and human fMRI data to evaluate the potential of the manifold model for joint reconstruction and quantification.
All the data were acquired with a 3T GE MR750 scanner (GE Healthcare, Waukesha, WI) and a 32-channel head coil (Nova Medical, Wilmington, MA). 

\subsection{Data Acquisition}

OSSI acquisition parameters were the same as in Simulation Investigations
with 10~s discarded data points to ensure the steady state.
We selected a 2D oblique slice passing through the visual cortex
with FOV = $220 \times 220 \times 2.5 \, \text{mm}^3$,
matrix size = $168 \times 168 \times 1$,
and spatial resolution = $1.3\times 1.3 \times 2.5 \, \text{mm}^3$. 
For OSSI, both ``mostly sampled’’ data (for retrospective undersampling) and prospectively undersampled data were acquired.
The sampling trajectories were undersampled VD spirals with golden-angle based rotations between time frames
as in \cite{tensor}. 
The ``mostly sampled’’ data used number of interleaves $n_i = 9$ VD spirals with approximately a 1.5 undersampling factor,
and temporal resolution = 1.35 s = $\text{TR} \cdot n_c \cdot n_i$.
The retrospective undersampling used the first interleave out of 9 for each time frame of the ``mostly sampled’’ data.
The prospective undersampling used $n_i = 1$ with temporal resolution = 150 ms = $\text{TR} \cdot n_c$
Both retrospective and prospective undersampling provided $12\times$ acceleration.

For quantification evaluation, we acquired multi-echo GRE images to get standard estimations of $f_0$ and $R_2^*$ values. 
GRE images were collected with a spin-warp sequence with TR = 100 ms, Ernst flip angle = 16$^\circ$, and different TEs = 5.9, 13, 26, and 40 ms.
$R_2^*$ maps were estimated based on the exponential decay of $T_2^*$. 
The field map $f_0$ was estimated using fully sampled GRE images at TE = 30 and 32 ms \cite{noll1995spiral}. 
For the phantom data, we additionally acquired spin-echo images with a spin-warp sequence at TR = 400 ms and different TEs = 20, 40, 60, and 80 ms to get $\hat{T}_2$ maps.  

For coil sensitivity map calculation, we collected spin-warp images and generated ESPIRiT sensitivity \cite{Uecker2014,Bart} after compressing the 32-channel coil images to 16 virtual coils using PCA \cite{Huang2008AChannels}. The coil images were 2-norm combined for brain region extraction using the Brain Extraction Tool \cite{Smith2002}.

For human data, the functional task was a left vs. right reversing-checkerboard visual stimulus with 10 s rest followed by 5 cycles of left or right stimulus (20 s L/20 s R $\times$ 5 cycles). The 10 s resting-state data ensured the oscillating steady state and were discarded. The number of time frames (both fast time $n_c$ and slow time) was 1490 for ``mostly sampled’’ data and was 13340 for prospectively undersampled data. 

\subsection{Performance Evaluation}

Every non-overlapping set of $n_c = 10$ fast time images were reconstructed and 2-norm combined for fMRI analysis.
To avoid modeling error from the HRF of the initial rest period, the data for the first 40 s task block were discarded.
The data were detrended using the first 4 discrete cosine transform basis functions to reduce effects of scanner drift.

We evaluated the functional performance of OSSIMM and comparison approaches using activation maps and tSNR maps. 
The backgrounds of activation maps were the mean of time-series of images.
The activated regions of activation maps were determined by correlation coefficients above a 0.45 threshold.
The correlation coefficients were generated by correlating the reference waveform (task and HRF related)
with the fMRI time course for each voxel.
For each voxel, dividing the mean of the time course by the standard deviation of the time course residual
(mean and task removed) provided the tSNR map.
We further calculated numbers of activated voxels at the bottom third of the brain
(where the visual cortex is located) and the average tSNR values within the brain (after skull stripping).

For quantification, parameter estimations at regions with little or no signal are masked out.
Specifically, we generated a mask with the first-echo GRE image (TE = 5.9 ms and after skull stripping)
for signals larger than 10\% of the signal magnitude and GRE $\hat{R}_2^*\, <$ 50 Hz.
Regions with GRE $\hat{R}_2^*\, >$ 50 Hz are concentrated at the edge of the brain as shown in Fig.~\ref{msf1}. 
The quantitative accuracy of OSSI $\hat{R}_2^*$ was evaluated by RMSE with multi-echo GRE $\hat{R}_2^*$ as the standard.
Because OSSI $\hat{f}_0$ estimates are in the range of [-33.3, 33.3] Hz,
we mapped the GRE $\hat{f}_0$ to the same range for comparison.

\begin{figure}
\centering
     \includegraphics[angle=-90,origin=c,width=0.89\textwidth]{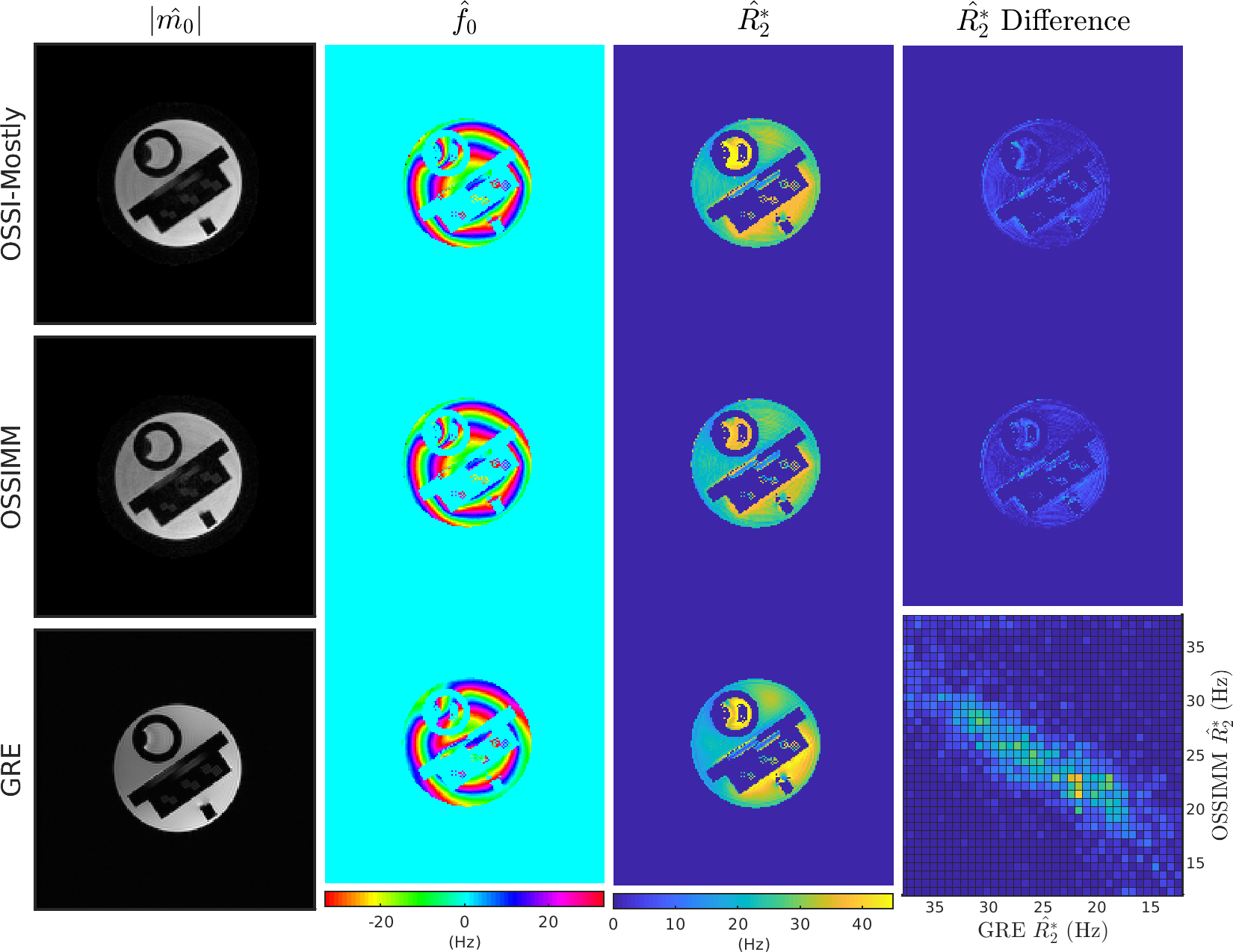}
   \caption{
   Phantom quantification of $m_0$, $f_0$, and $R_2^*$ from mostly sampled OSSI data, retrospectively undersampled OSSI data (reconstructed and quantified using OSSIMM), and multi-echo GRE. The $\hat{m}_0$ estimates are on arbitrary scales. The GRE $\hat{R}_2^*$ map is used as the standard for difference map calculation. The $\hat{R}_2^*$ maps and $\hat{R}_2^*$ difference maps use the same color scale. The 2D histogram (bottom right) compares OSSIMM and GRE $\hat{R}_2^*$ within the 12-38 Hz range. OSSI $\hat{R}_2^*$ and GRE $\hat{R}_2^*$ demonstrates similar contrasts.
   }\label{mf3}
\end{figure}

\section{Reconstruction, Quantification, and Results}
The proposed OSSIMM method jointly reconstructed high-resolution images and quantitative maps using the near-manifold regularization. For both phantom and human experiments, we used the $T_2'$ manifold with a fixed $T_2$ = 100 ms unless otherwise specified. After reconstructing fast time images with mostly sampled data (OSSI-Mostly), or other models such as low-rank (OSSI-LR) and regularized cgSENSE (OSSI-cgSENSE), we further estimated their corresponding parameter maps using the same manifold as in OSSIMM. 

\subsection{Implementation Details}

We selected the regularization parameters based on the Lipschitz constant $\sigma(A)$ calculated with power iteration. We set the regularization parameter $\beta$ in \eqref{costmf} to be a fraction of $\sigma(A)$ that the condition number of the cost function was about 10 to 20 and the performance of the functional maps are maximized. $\alpha$ in \eqref{costlr} was selected to enforce that the rank $\approx$ 4 for the fast time image sets. 

In OSSIMM, we used 4 iterations of alternating minimization, and 2 iterations of conjugate gradient for the data fidelity update. We used 15 iterations of POGM for the LR reconstruction and 19 iterations of conjugate gradient for cgSENSE reconstruction and the mostly sampled data. We generated data-shared images as the initialization for the undersampled reconstructions by utilizing the sampling incoherence between fast and slow time \cite{tensor2020} and combining k-space data of every 10 slow time points.

\subsection{Results}

\begin{table}
\renewcommand{\arraystretch}{1.6}
\caption{Phantom quantification comparison of OSSI $\hat{R}_2^*$ to GRE with or without a known $\hat{T}_2$ map}
\label{mt1}
\centering
\begin{tabular}{ccccc}
\hline\hline
\multirow{2}{*}{} & \multicolumn{2}{c}{Fixed $T_2$ = 100 ms} & \multicolumn{2}{c}{Known $\hat{T}_2$ map} \\ \cline{2-5} 
& \begin{tabular}[c]{@{}c@{}}$\hat{R}_2^*$ RMSE \\[-0.6em] (Hz)\end{tabular}
& \begin{tabular}[c]{@{}c@{}}Additional \\[-0.6em] Mask\end{tabular} 
& \begin{tabular}[c]{@{}c@{}}$\hat{R}_2^*$ RMSE \\[-0.6em] (Hz)\end{tabular}
& \begin{tabular}[c]{@{}c@{}}Additional \\[-0.6em] Mask\end{tabular}
\\ \hline 
OSSI-Mostly & 4.9 & 4.3 & 5.0 & 4.6           
\\ \hline
OSSIMM & 5.5 & 4.6 & 5.3 & 4.5             
\\ \hline\hline
\end{tabular}
\end{table}

For the phantom study,
Fig.~\ref{mf3} and Fig.~\ref{msf2} present OSSI quantification results
with a fixed $T_2$ of 100 ms and a known $\hat{T}_2$ map, respectively.
OSSIMM quantifies parameters from retrospectively undersampled data,
and results in similar maps as mostly sampled reconstruction and multi-echo GRE.
The 2D histogram demonstrates a close to a linear relationship between OSSI and GRE $\hat{R}_2^*$ values.
As summarized in Table~\ref{mt1},
OSSIMM with a known $\hat{T}_2$ map produces similar results as OSSIMM with a fixed $T_2$ value.
Demonstrated by RMSE values with additional masking in Table \ref{mt1}, OSSI $\hat{R}_2^*$ RMSE improves by 0.5-1 Hz when a GRE $12 < \hat{R}_2^* < 38$ mask (within OSSIMM $R_2^*$ dictionary range) is applied.

\begin{figure*}
\centering
    \includegraphics[width=\textwidth]{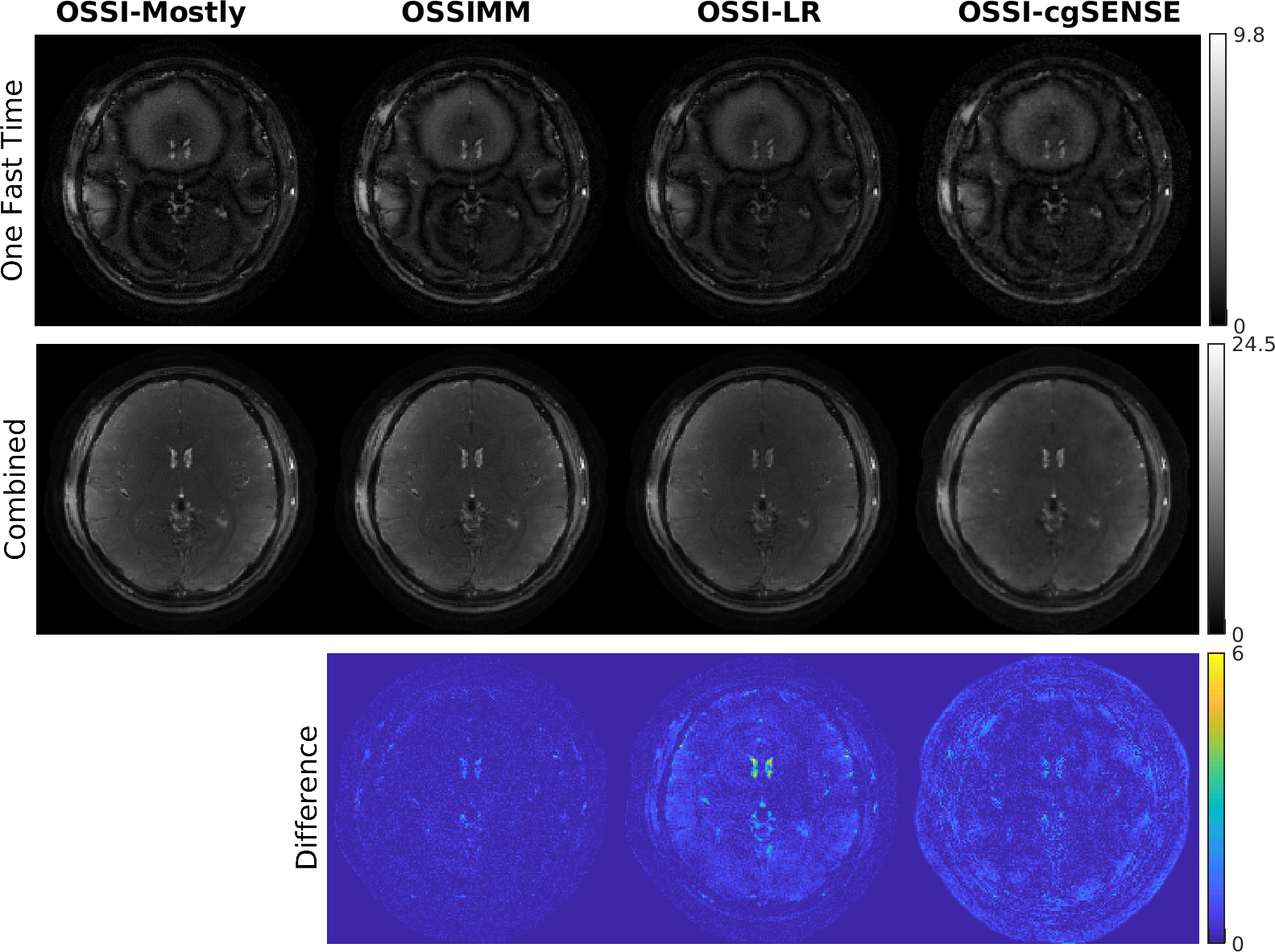}
    \caption{
   Manifold, low-rank, and cgSENSE reconstructions for retrospectively undersampled OSSI data are compared to the mostly sampled reconstruction. The example fast time images present spatial variation in OSSI. OSSIMM outperforms other approaches with cleaner high-resolution details and less structure in the difference map.
    }\label{mf4}
\end{figure*}

Figure \ref{mf4} compares retrospectively undersampled reconstructions to the mostly sampled reference.
OSSIMM reconstruction well preserves high-resolution structures in oscillatory fast time images and combined images,
and leads to less residual in the difference map than LR and cgSENSE approaches.

\begin{figure*}
\centering
     \includegraphics[width=\textwidth]{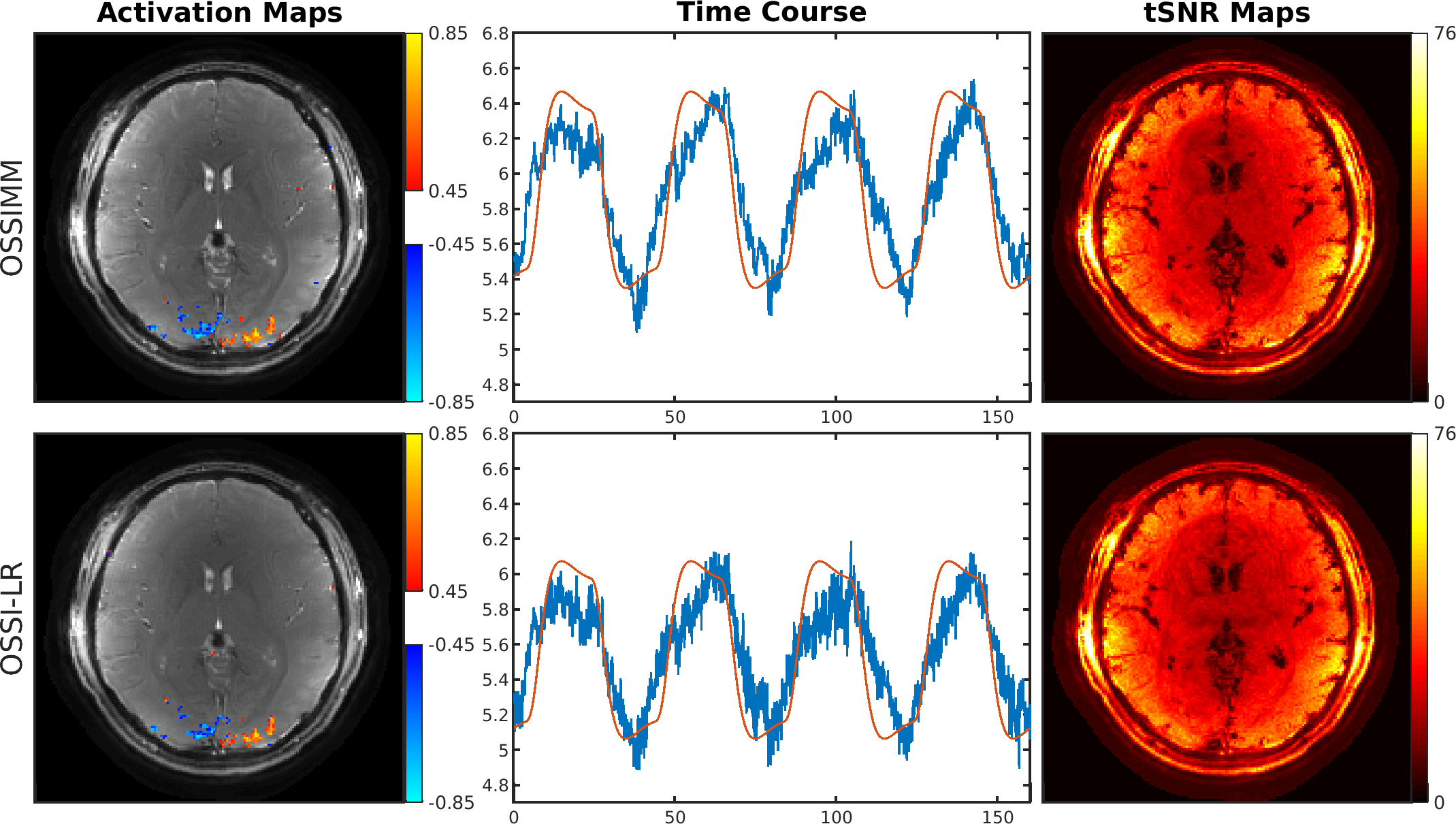}\\
     \includegraphics[width=\textwidth]{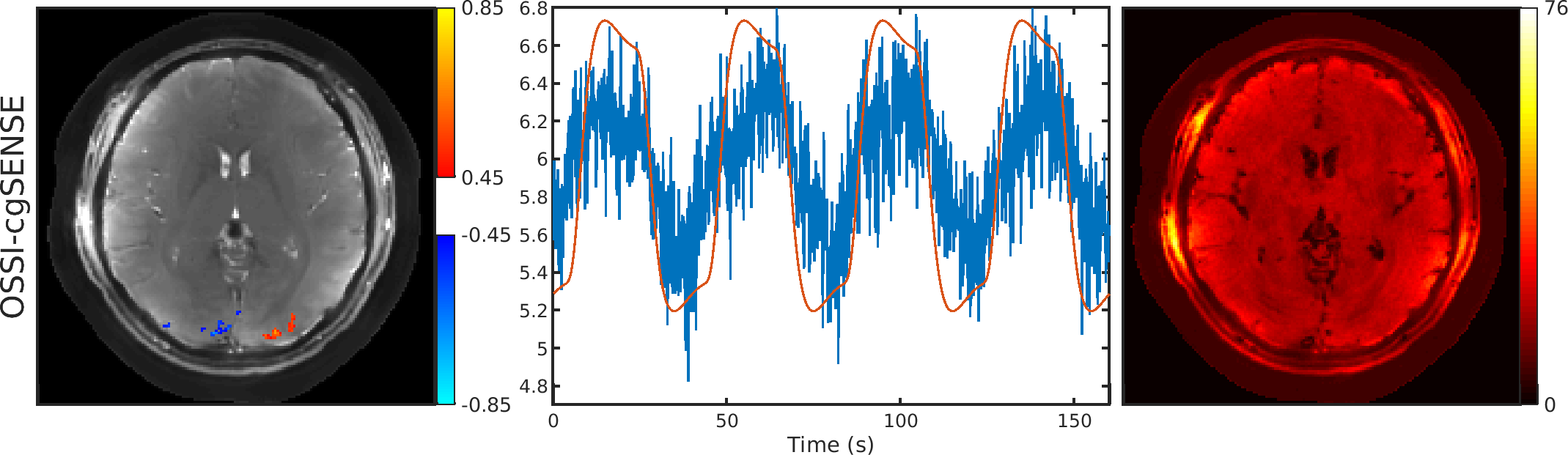}
     \caption{
     Functional results for prospectively undersampled data with spatial resolution of 1.3 mm
     and temporal resolution of 150 ms.
     The proposed OSSIMM reconstruction provides an activation map
     with high-resolution background image and larger activated regions,
     and time course (reference waveform in red) and temporal SNR map with higher SNR than other methods.
     }
     \label{mf5}
\end{figure*}

Figure \ref{mf5} presents prospectively undersampled reconstructions (temporal resolution = 150 ms)
using OSSIMM, LR, and cgSENSE.
OSSIMM demonstrates activation map with more activated voxels, time course with higher SNR,
and sharper tSNR map than other methods. 
The functional maps from the mostly sampled reconstruction (temporal resolution = 1.35 s)
are included in supplemental Fig.~\ref{msf3} for reference. 

\begin{figure*}
    \centering
    \begin{minipage}{0.495\textwidth}
        \includegraphics[width=\linewidth]{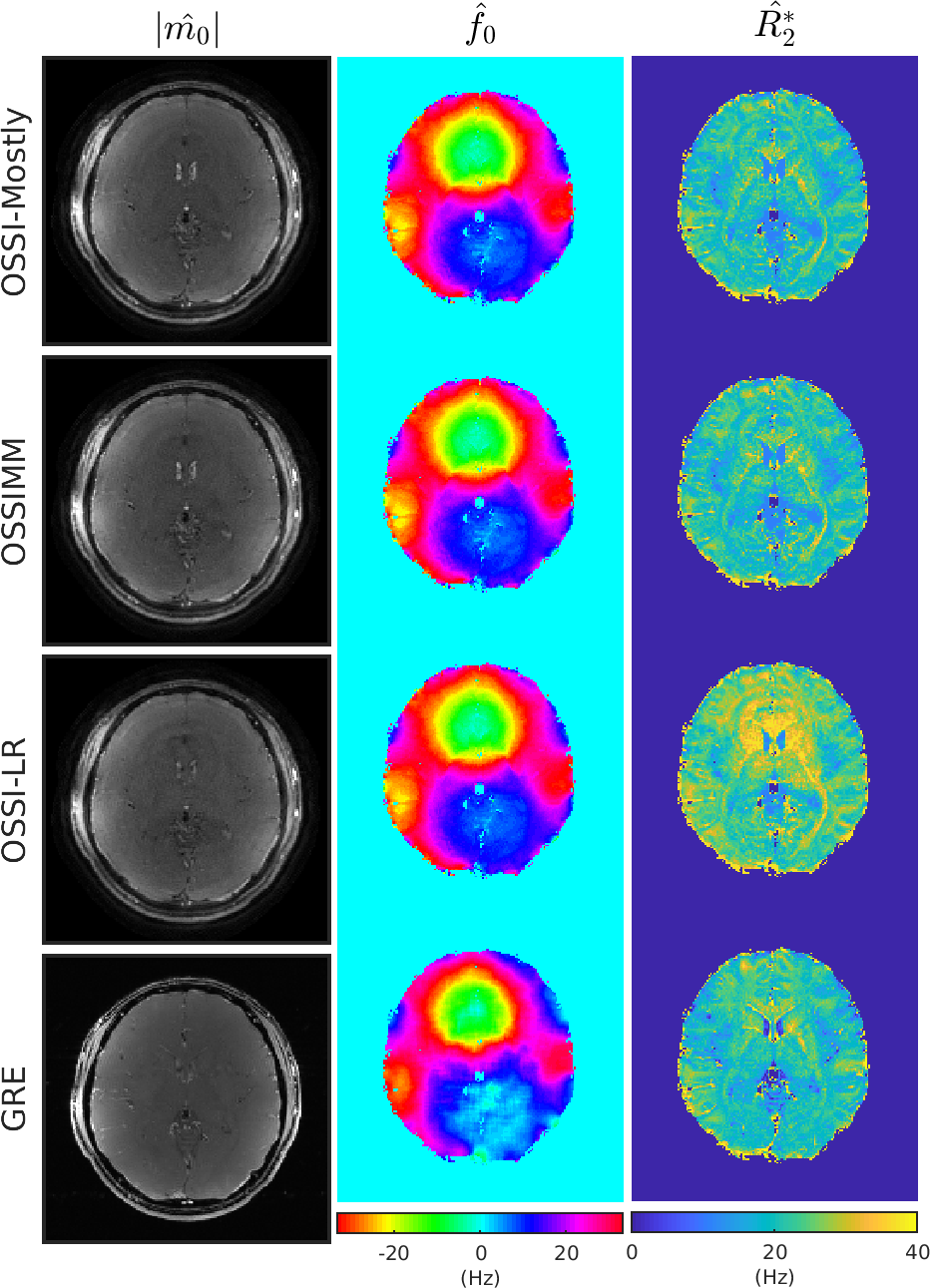}
        \includegraphics[width=\linewidth]{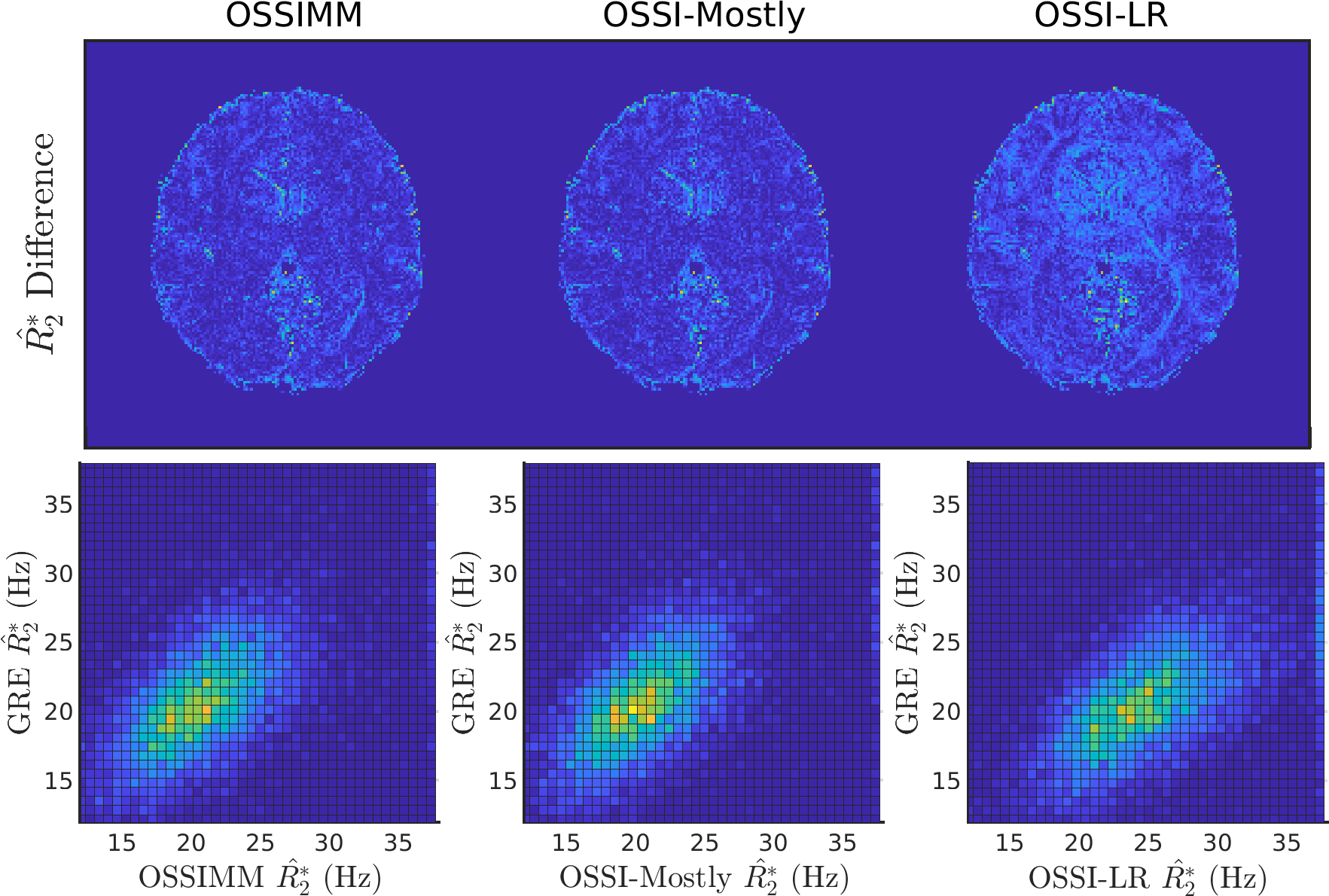}
        \caption{
   Retrospectively undersampled quantifications and comparison to multi-echo GRE estimates. OSSIMM presents similar results as the mostly sampled data. $\hat{R}_2^*$ difference maps (using GRE $\hat{R}_2^*$ as standard and of same color scale as $\hat{R}_2^*$ maps) and 2D histogram of $\hat{R}_2^*$ values show that OSSIMM provides comparable quantitative maps to GRE.
        }
        \label{mf6}
    \end{minipage}
    \hfill
    \begin{minipage}{0.495\textwidth}
        \includegraphics[width=\linewidth]{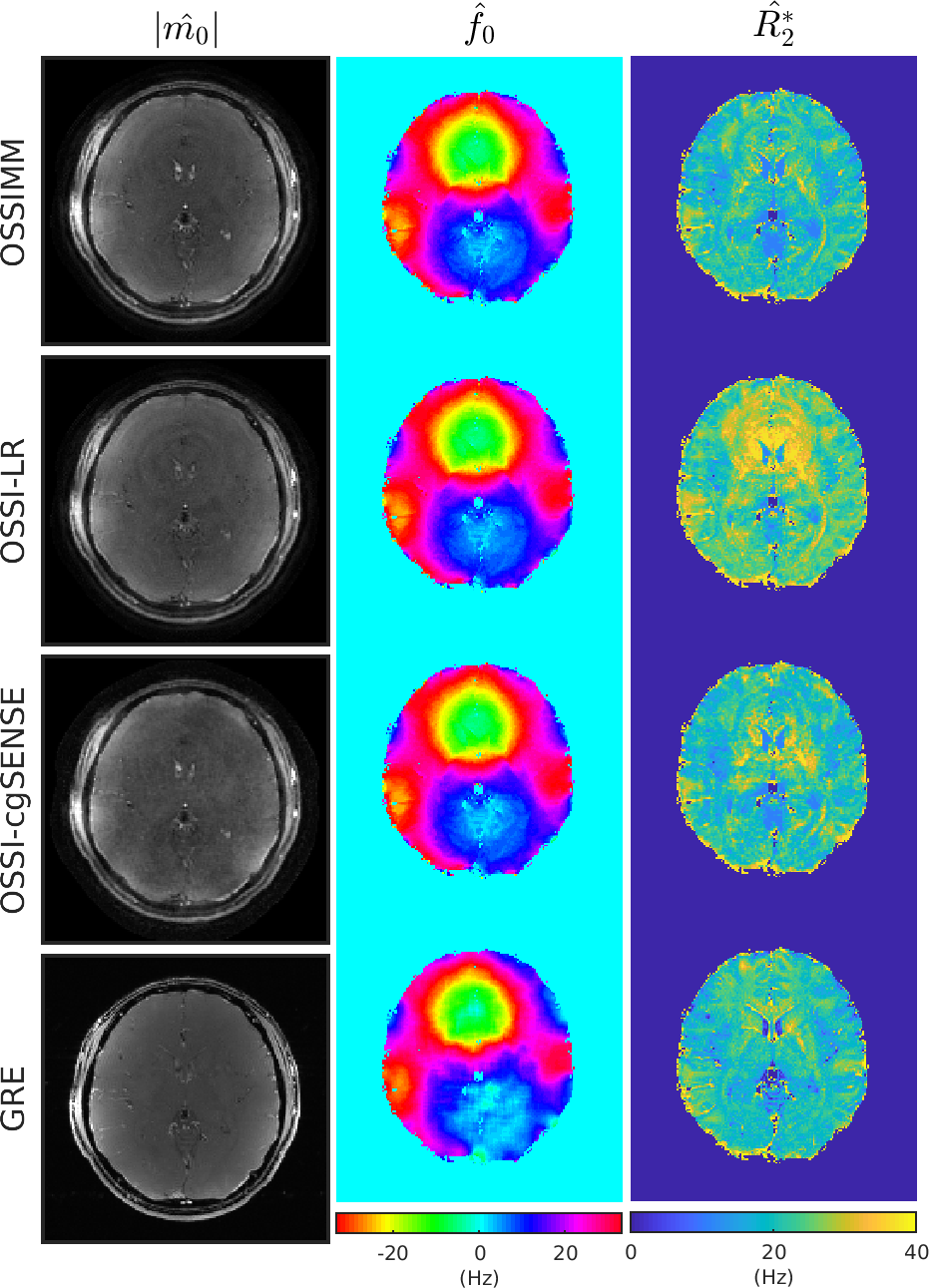}
        \includegraphics[width=\linewidth]{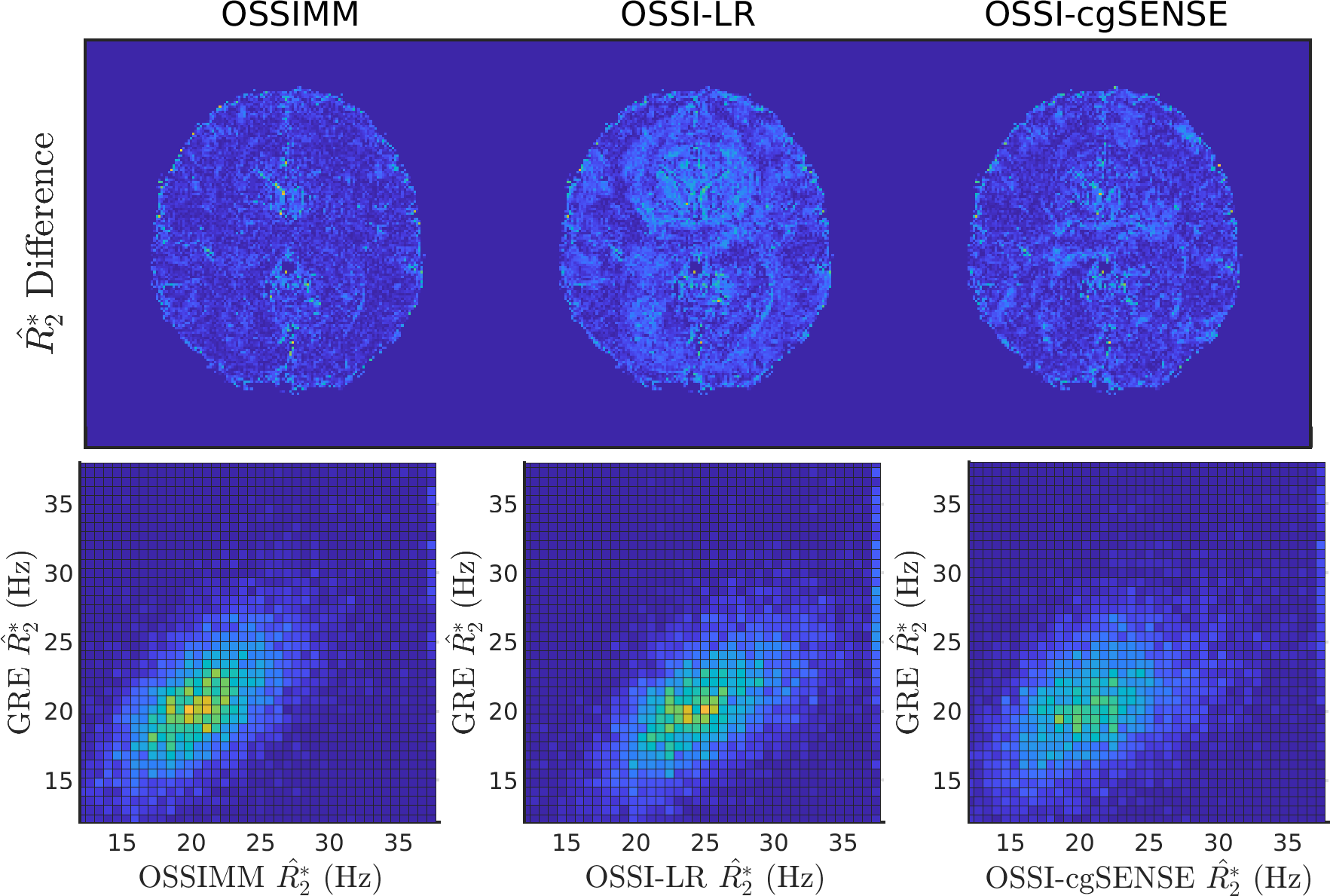}
        \caption{
   Prospectively undersampled quantifications compared to multi-echo GRE. OSSIMM results in reasonable parameter maps with 1.3 mm spatial resolution and a 150 ms acquisition time. OSSIMM also outperforms low-rank and cgSENSE reconstructions with less residual in the $\hat{R}_2^*$ difference map (same color scale as $\hat{R}_2^*$ maps).
        }
        \label{mf7}
    \end{minipage}
\end{figure*}

Figure \ref{mf6} gives retrospectively undersampled and mostly sampled OSSI quantification results
with comparison to multi-echo GRE.
OSSIMM with $12\times$ undersampling leads to $\hat{m}_0$, $\hat{f}_0$, and $\hat{R}_2^*$ estimates
that are almost identical to the mostly sampled case and have finer structures than OSSI-LR.
OSSIMM also provides comparable $\hat{R}_2^*$ maps to GRE
and demonstrates a similar distribution of $\hat{R}_2^*$ values within the brain as GRE
according to the 2D histogram.
Because of field drift and respiratory changes between different scans,
the OSSI-Mostly and OSSIMM $\hat{f}_0$ maps are close to GRE $\hat{f}_0$ but not exactly the same.

Figure \ref{mf7} compares prospectively undersampled quantification results to multi-echo GRE. 
OSSIMM enables high-resolution quantification of $m_0$, $R_2^*$ and $f_0$ with a 150~ms acquisition,
and yields parameter estimates more similar to GRE than LR and cgSENSE reconstructions.

\begin{figure*}
\centering
     \includegraphics[width=0.96\textwidth]{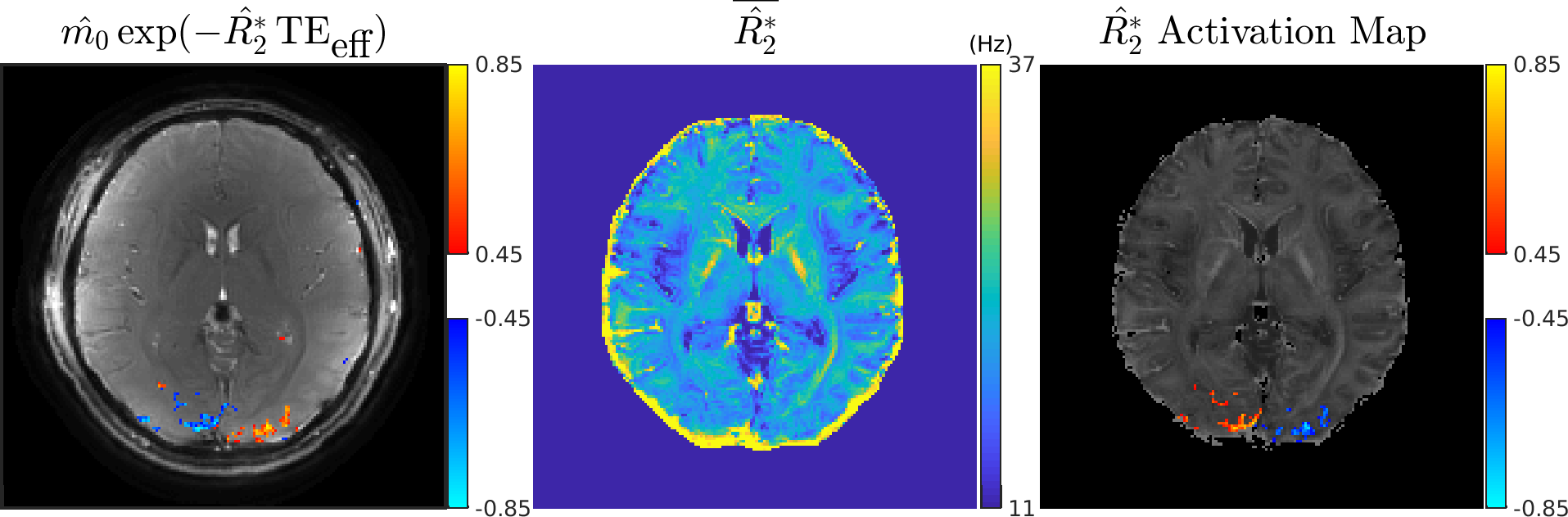}
   \caption{
   Activation maps from OSSIMM $\hat{m}_0$ and $\hat{R}_2^*$ with prospective undersampling demonstrating the dynamic quantification capacity of OSSIMM. Both time series of $\hat{m}_0\exp(-\hat{R}_2^*\,\textrm{TE}_\textrm{eff})$ (left) and $\hat{R}_2^*$ (right) almost fully recover the functional activation. The $\overline{\hat{R}_2^*}$ (middle) is the mean of $\hat{R}_2^*$ time series after skull stripping (without any other mask) and well preserves the $R_2^*$ contrast. 
   }\label{mf8}
\end{figure*}

The parameter maps in Figs. \ref{mf6} and \ref{mf7} are from a single set of $n_c = 10$ fast time images,
while OSSIMM jointly reconstructs undersampled measurements and quantifies physics parameters
for every 10 fast time images of the OSSI fMRI time course.
To demonstrate the dynamic quantification capacity of OSSIMM,
Fig.~\ref{mf8} shows activation maps
for $\hat{m}_0\exp(-\hat{R}_2^*\,\textrm{TE}_\textrm{eff})$
and $\hat{R}_2^*$,
where $\hat{m}_0$ and $\hat{R}_2^*$ are quantified using OSSIMM and prospectively undersampled data.
OSSI $\textrm{TE}_\textrm{eff}\,\approx$ 17.5 ms with a 2.6 ms actual TE \cite{ossi2019}. 

The activation maps based on $\hat{m}_0\exp(-\hat{R}_2^*\,\textrm{TE}_\textrm{eff})$ images
well preserves $R_2^*$ contrast of OSSI
and has the same activated regions as the activation map from 2-norm combined OSSI images (in Fig.~\ref{mf5}).
The activation map from $\hat{R}_2^*$ maps recovers the activation and reduces false positives
(negative activation in the positive activation region and vice versa).
The colors of the activation are the opposite of activation in Fig. \ref{mf5}
due to the negative correlation between $m_0\exp(-R_2^*\,\textrm{TE}_\textrm{eff})$ and $R_2^*$.
The mean $R_2^*$ map ($\overline{R_2^*}$) of the time series,
when compared to GRE, leads to a smaller RMSE value of 4.4 Hz.
The RMSE value = 3.7 Hz with a GRE $12 < \hat{R}_2^* < 38$ Hz mask.  

\begin{table}
\setlength{\tabcolsep}{4pt}
\renewcommand{\arraystretch}{1.6}
\caption{Human reconstruction and $R_2^*$ quantification evaluation for different sampling patterns and models}
\label{mt2}
\centering
\begin{tabular}{ccccc}
\hline\hline
& OSSIMM & OSSI-LR & OSSI-cgSENSE & OSSI-Mostly  
\\ \hline\hline
\multicolumn{5}{c}{Retrospectively Undersampled}
\\ \hline
$\hat{R}_2^*$ RMSE (Hz) 
& 5.1 & 6.6 & 5.4 & 5.1  
\\ \hline
Additional Mask 
& 4.5 & 6.1 & 4.9 & 4.5 
\\ \hline
\multicolumn{5}{c}{Prospectively Undersampled}
\\ \hline
$\hat{R}_2^*$ RMSE (Hz) 
& 4.9 & 6.7 & 5.5 & -     
\\ \hline
Additional Mask
& 4.3 & 6.4 & 5.0 & -     
\\ \hline
\# Activated Voxels
& 181 & 159 & 68 & -     
\\ \hline
Average tSNR
& 26.4 & 26.5 & 18.8 & -     
\\ \hline\hline
\end{tabular}
\end{table}

Table~\ref{mt2} summarizes quantitative evaluations of different sampling schemes and reconstruction models.
OSSI $\hat{R}_2^*$ RMSE values compared to GRE
for retrospectively (Fig. \ref{mf6})
and prospectively (Fig. \ref{mf7}) undersampling are presented.
As demonstrated by RMSE values with additional masking, 
OSSI RMSE decrease by about 0.5 Hz with the GRE $12 < \hat{R}_2^* < 38$ mask.
The last two rows of the table correspond to Fig.~\ref{mf5}
and are numbers of activated voxels and average tSNR within the brain for prospectively undersampled reconstructions.
The proposed OSSIMM jointly reconstructs high-resolution images
with more functional activation and parameter maps with smaller $\hat{R}_2^*$ RMSE than other approaches.

\section{Discussion}
We propose a novel manifold model OSSIMM that uses MR physics for the signal generation as the regularizer
for image reconstruction from undersampled k-space data.
The proposed model simultaneously provides high-resolution fMRI images
and quantitative maps of important MRI physics parameters. 

The proposed near-manifold regularizer has the advantage
of allowing for potential imperfections of the manifold model.
Instead of requiring the signal values to lie exactly on the manifold,
it provides a balance between fitting the fast-time images to the noisy k-space data and to the manifold. 
For reconstruction, OSSIMM outperforms low-rank and cgSENSE models
by providing more functional activation, without spatial or temporal smoothing. 

For quantification, OSSIMM dynamically tracks
$m_0$, $R_2^*$, and $f_0$ changes with a temporal resolution of 150 ms in our experiments.
The OSSIMM estimates\\  $\hat{m}_0\exp(-\hat{R}_2^*\,\textrm{TE}_\textrm{eff})$ or $\hat{R}_2^*$
contain most of the functional information of fMRI time series,
and may be well-suited for examining quantitative changes in longitudinal studies.
Moreover, OSSIMM quantification is faster than other quantification methods such as \cite{Wang2019}.
The manifold model and the near-manifold regularization
can be generalized to other sparsely undersampled datasets for joint reconstruction and quantification.  

There are multiple factors that contribute to slight mismatches between OSSI $\hat{R}_2^*$ and GRE $\hat{R}_2^*$.
We noticed that OSSI and GRE images were not exactly aligned
due to different gradient delays or the movement of the brain between different scans,
especially around the edge of the brain.
It is also possible that through-plane gradients change signals slightly differently between OSSI and GRE.
The OSSIMM implementation could be improved with a larger dictionary with a larger range of $R_2^*$ values
and finer spacing of the varying physics parameters.
The RF inhomogeneity in the brain may influence the accuracy of the dictionary fitting
due to inaccuracy of the flip angle.

We have neglected the readout length effect for simplicity and have not performed field map correction for human data.
The field map correction improves quantification for resolution phantom,
but would increase computation for human fMRI time series.
One interesting extension would be to dynamically quantify $f_0$ and correct for field inhomogeneity
using the time-series of OSSI $\hat{f}_0$ maps.
Because OSSI $\hat{f}_0$ maps are in the range of [-33.3, 33.3] Hz,
we could use an initial estimate of $\hat{f}_0$ from two-echo GRE,
and dynamically update the initial $\hat{f}_0$ based on OSSI $\hat{f}_0$ changes along time as in \cite{Olafsson2008}.

We believe that the reconstruction performance can be further improved
with spatial-temporal modeling of OSSI fMRI image series.
We will combine OSSIMM with the patch-tensor low-rank model \cite{tensor}
to exploit different aspects of prior information (linear and nonlinear),
enlarge the capacity of regularization,
and enable more aggressive undersampling.
We will also extend the OSSIMM dynamic quantification to 3D fMRI.
Because a known $\hat{T}_2$ map can be helpful for $\hat{R}_2^*$ estimation,
one might considering modifying the OSSI sequence as in \cite{Wang2019}
with slowly varying flip angles and other changes to simultaneously quantify $T_2$ and $T_2'$.

\section{Conclusion}

This chapter proposes OSSIMM,
a novel reconstruction and quantification model for nonlinear MR signals.
With a factor of 12 undersampling and without spatial or temporal smoothing,
OSSIMM outperforms other reconstruction models
with high-resolution structures and more functional activation.
OSSIMM also provides dynamic $R_2^*$ maps that are comparable to GRE $\hat{R}_2^*$ maps
with a 150 ms temporal resolution.

\newpage

\section{Supporting Information}

\begin{figure}[!htbp]
\centering
     \includegraphics[width=0.92\textwidth]{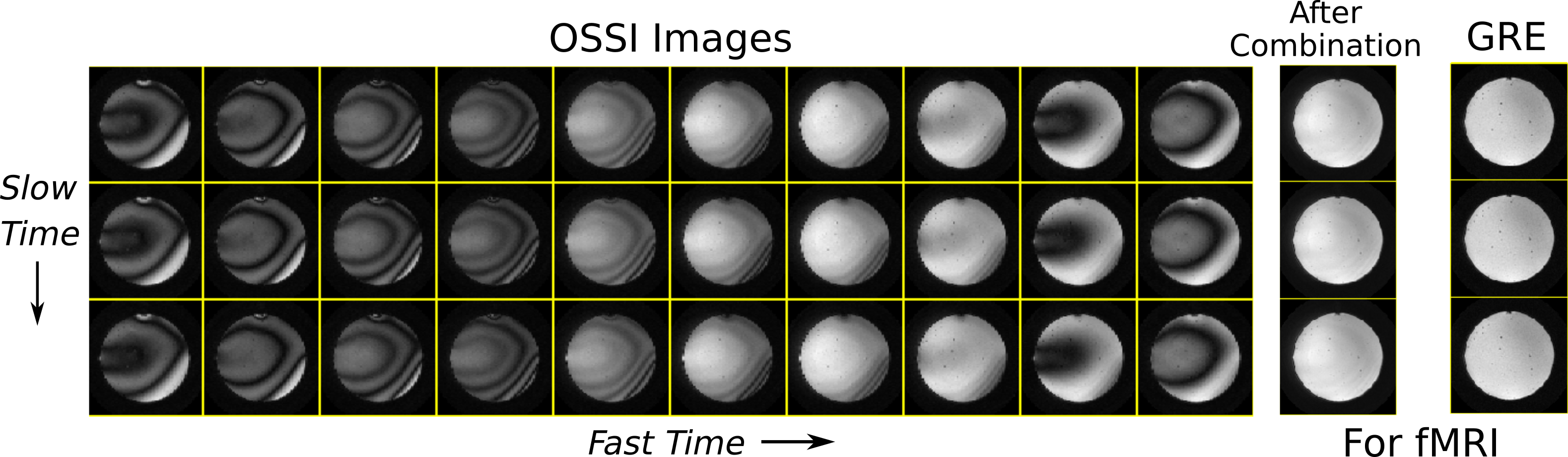}
     \caption{
OSSI images with periodic and nonlinear oscillation patterns are structured along ``fast time” and ``slow time”. Every $n_c$ fast time images can be 2-norm combined to generate fMRI images that have comparable $T_2^*$-sensitivity as standard GRE fMRI.
     }
     \label{msf0}
\end{figure}

\begin{figure}[!htbp]
\centering
     \includegraphics[width=0.45\textwidth]{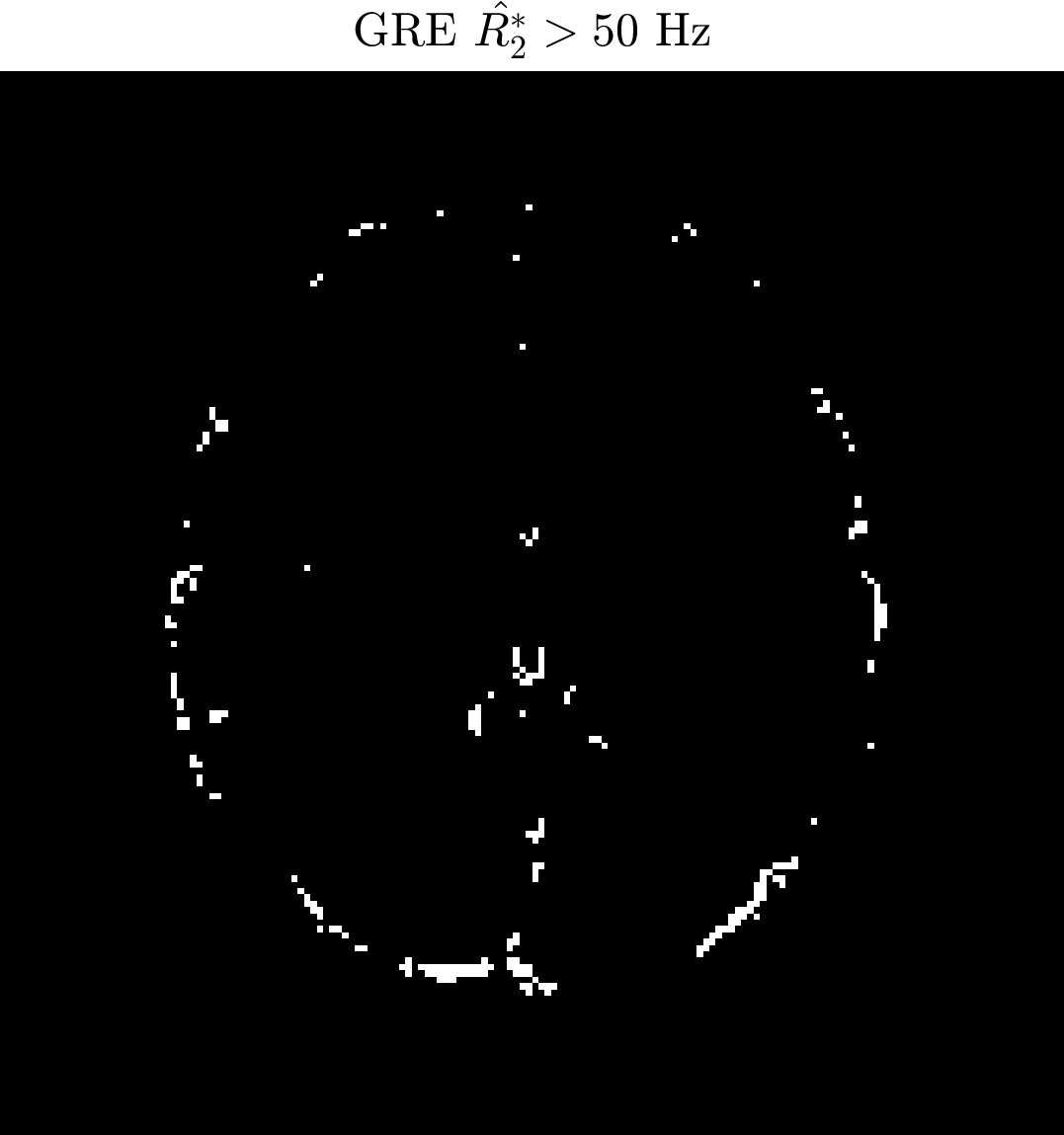}
     \caption{
Most voxel locations with GRE $\hat{R}_2^*\, >$ 50 Hz are around the edges of the brain.
     }
     \label{msf1}
\end{figure}

\begin{figure}[!htbp]
\centering
    \includegraphics[width=0.95\textwidth]{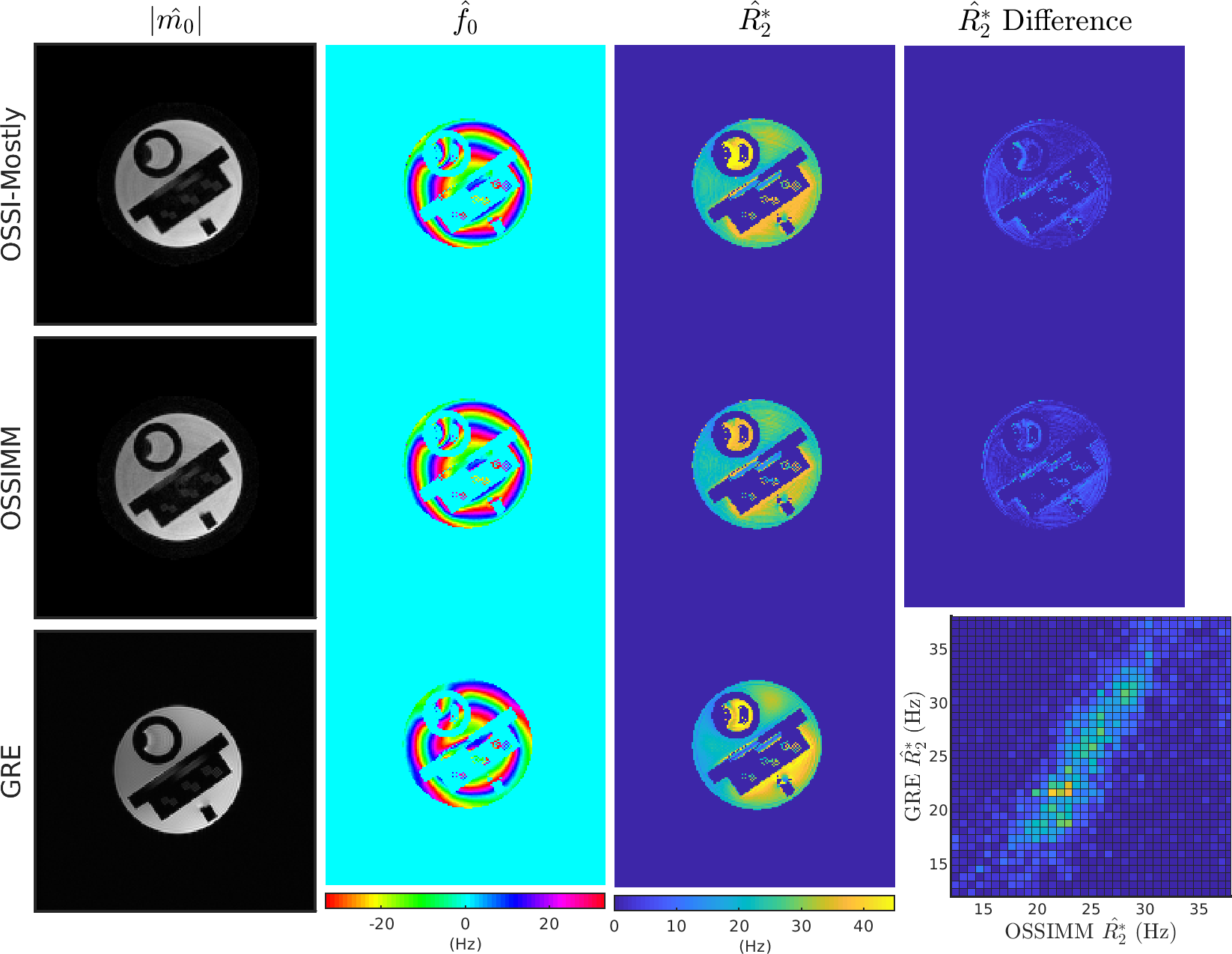}
   \caption{
   Phantom quantification of $m_0$, $f_0$, and $R_2^*$ from mostly sampled OSSI data,
   retrospectively undersampled OSSI data (reconstructed and quantified using OSSIMM with a known $\hat{T}_2$ map),
   and multi-echo GRE. The $\hat{m}_0$ estimates are on arbitrary scales.
   The GRE $\hat{R}_2^*$ map is used as the standard for difference map calculation.
   The $\hat{R}_2^*$ maps and $\hat{R}_2^*$ difference maps use the same color scale.
   The 2D histogram (bottom right) compares OSSIMM and GRE $\hat{R}_2^*$ within the 12-38 Hz range.
   OSSI $\hat{R}_2^*$ and GRE $\hat{R}_2^*$ have similar contrasts.
   }\label{msf2}
\end{figure}

\begin{figure}[!htbp]
\centering
     \includegraphics[width=0.95\textwidth]{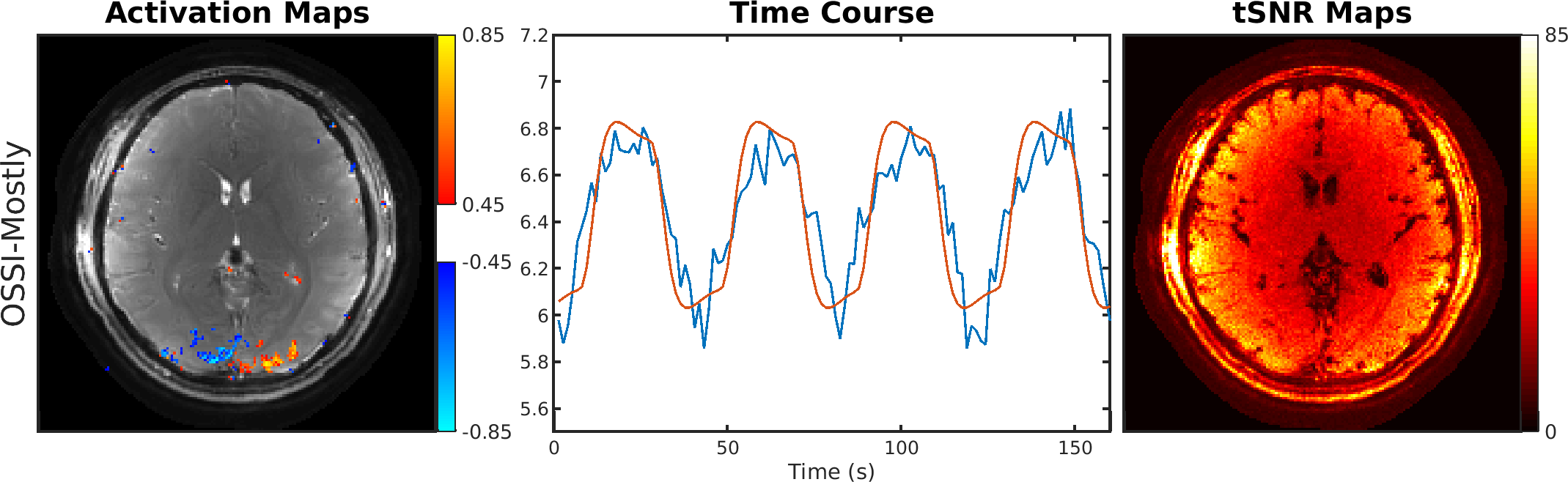}
     \caption{
     Functional results for mostly sampled data with spatial resolution of 1.3 mm and temporal resolution of 1.35 s. The number of activted voxels is 236, and the average temporal SNR within the brain is 31.3.
     }
     \label{msf3}
\end{figure}

Figure~\ref{msf0} illustrates OSSI ``fast time" and ``slow time".
Figure~\ref{msf1} demonstrates voxel locations with GRE $\hat{R}_2^*\, >$ 50 Hz.
Figure~\ref{msf2} presents phantom quantification results,
and OSSI quantitative maps that were calculated with a known $\hat{T}_2$ map.
Figure~\ref{msf3} presents fMRI results for mostly sampled human data.

\chapter{Voxel-wise Temporal Attention Network and Simulation-Driven Dynamic MRI Sequence Reconstruction}
\label{chap:4attnet}
Deep learning-based approaches have been successful for structural MRI undersampled reconstruction \cite{hammernik2018learning, aggarwal2018modl}. However, there are fewer works on learning-based dynamic MRI reconstruction \cite{schlemper2017deep, qin2018convolutional} with two main open questions: 1) what would be a good learning-based approach for temporal or spatial-temporal signal modeling, 2) for dynamic MRI sequence of images, how to get enough training data for the learning schemes that are data hungry? Inspired by these two questions, we propose a voxel-wise attention network based on the emerging attention mechanism \cite{bahdanau2014neural, vaswani2017attention} for temporal modeling, together with a matched transfer learning approach to handle the problem of limited amounts of training data. 

Our work has three novel contributions: 1) incorporate an attention mechanism for temporal learning and mapping, 2) propose a voxel-wise network architecture based on attention and Transformers for spatial-temporal undersampled reconstruction, 3) propose a two-stage learning scheme that pretrains the network with voxel-wise simulated data, and then fine-tunes with human temporal data for dynamic MRI. \footnote{This chapter is based on \cite{2022attnet}.
}

\section{Introduction}

Previous models \cite{tensor2020,2021manifold} for undersampled OSSI MRI sequence reconstruction focus on hand-crafted features of the data,
whereas learning features using neural networks has proven to be very useful and successful
for vision tasks \cite{krizhevsky2012imagenet,he2016deep}
and undersampled MRI reconstruction \cite{hammernik2018learning,aggarwal2018modl} in recent years.
Another advantage of neural network approaches is fast computation in the testing stage,
whereas iterative methods can be slow for reconstructing high-resolution images.

For dynamic MRI reconstruction,
previous works used a cascade of convolutional neural networks (CNN)
with designed data consistency layers \cite{schlemper2017deep}
and convolutional recurrent neural network (RNN)
for temporal dependence between images \cite{qin2018convolutional}. 
However, the convolutional operations in CNNs are local and fail to capture long-range dependencies \cite{liu2020transposer,mardani2020neural}, and the shared weights for the spatial/temporal dimensions could potentially lead to spatial/temporal smoothness effects.  
Moreover, 3D CNNs for video inputs \cite{carreira2017quo} or sequences of images often require large amounts of data for training due to the increased number of learnable parameters, and would be computationally expensive for long sequences. 
On the other hand, recurrent inference for images in a sequence is not ideal
for modeling the temporal redundancy between images because of the causal nature;
in particular,
when all the input images in a sequence are undersampled and aliased,
the recurrent inference will likely pass on the aliased features and noise. 

We found that none of the existing methods work well for functional MRI with small BOLD signals \cite{runet},
whereas the attention mechanism \cite{bahdanau2014neural,wang2018non} and Transformers \cite{vaswani2017attention,dosovitskiy2020image}
that naturally exploit long-range dependence in image sequences could be a great fit for MRI spatial-temporal modeling.

A Transformer is a new neural network structure first presented in \cite{vaswani2017attention}.
It consists of self-attention \cite{bahdanau2014neural},
multilayer perceptron, residual connections, and layer normalization \cite{ba2016layer}.
Transformers have been the building block for important natural language processing networks
such as BERT \cite{devlin2018bert} and GPT-3 \cite{brown2020language}.
For visual tasks, recent works have demonstrated great potential of Transformers
as they match or outperform state-of-the-art CNNs
for different visual tasks \cite{dosovitskiy2020image,chen2020pre,liu2021swin}.
The self-attention mechanism and Transformer architecture map a sequence of inputs to a sequence of outputs where each output in the sequence is a learned combination of all the inputs. We hypothesize that this attention design is beneficial for modeling spatial-temporal dependency in dynamic MRI image series and has the potential to outperform CNN and RNN methods.

Another major issue of the learning-based approach is the need for large amount of training data. In dynamic MRI, the object is changing as it is being imaged, so it is impossible to collect truly fully sample reference data. All dynamic MRI data is inherently undersampled. So high quality "ground truth" training data is never available. This problem is especially acute for novel acquisition methods that have not accumulated a large database of human subjects for training.

We aim to tackle two open questions of dynamic MRI reconstruction - the model and the data - in two steps. We propose a novel voxel-wise attention network for temporal modeling of the image sequences to be reconstructed. The voxel-wise design of the network enables voxel-wise training, and we further propose a transfer learning scheme that pretrains the network with a large amount of voxel-wise simulated data to alleviate the demand for human fMRI data during training.

\section{Methods}

\subsection{Attention Mechanism}

\begin{figure*}
\centering
    \includegraphics[width=0.9\textwidth]{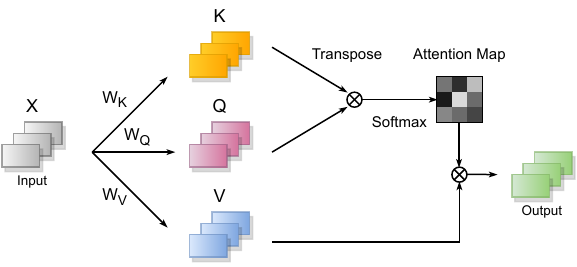}
    \caption{
Illustration of the attention mechanism. A sequence of input vectors is mapped to a sequence of output vectors. Each vector in the output sequence is a weighted combination of all the vectors in the input sequence, and the weights are determined by the learned attention map.
    }\label{af0}
\end{figure*}

The self-attention mechanism \cite{bahdanau2014neural, vaswani2017attention} maps a sequence of input vectors to a sequence of output vectors by computing a weighted combination of all the input vectors for each of the output vector. 
The weights are determined by similarity between pairs of feature representations. \cref{af0} illustrates the attention mechanism. 
For an input sequence 
$X\in\mathbb{R}^{t \times d}$ 
of $t$ vectors of dimension $d$ ($d$ can be vectorized spatial dimensions or spatial representations for a sequence of 2D MRI images, and the complex MRI data is formed into real and imaginary channels),
the self-attention mechanism first 
extracts feature representations  
by multiplying $X$ with three learnable parameter matrices:
$W_\Q \in \mathbb{R}^{d \times d_k}$, 
$W_\K \in \mathbb{R}^{d \times d_k}$, 
$W_\V \in \mathbb{R}^{d \times d_v}$,
where $d = d_k = d_v$ in our reconstruction task. The relative size of $d_k$ and $d$ depends on the implementation of the network. 
The three resultant matrices are called
query, key, and value,
and are calculated by
\begin{equation}
Q = X W_\Q, \quad K = X W_\K, \quad V = X W_\V.
\end{equation}
The output sequence of the attention mechanism
is then formed as
\begin{equation}
A(Q, K, V) = \mathrm{softmax}\left(\frac{Q K^T}{\sqrt{d}}\right)V
\in \mathbb{R}^{t \times d},
\end{equation}
where $\mathrm{softmax}(\cdot) : \mathbb{R}^{t} \rightarrow (0, 1)^{t}$
denotes the softmax function
$$
\sigma(\z)_i = \frac{e^{z_i}}{\sum_{j = 1}^{t}e^{z_j}},
$$
applied row-wise to the input matrix, $i = 1, ..., t$ and $\z = (z_1, ..., z_{t})$.

De-aliasing of a dynamic sequence with undersampling artifacts can be viewed as mapping a temporal sequence with aliasing to a sequence without aliasing. 
We use the attention mechanism as a key component for the dynamic sequence reconstruction.

\subsection{Proposed Voxel-Wise Attention Network}
We formulate the cost function of our reconstruction problem with two alternating minimization steps as
\begin{align}
\hat{\w}_i 
&= \label{ae1}
\argmin{\w}
\lVert \mathbf{\Phi} (\w; \mathbf{X}_{i-1}) - \mathbf{X}_{\mathrm{true}} \rVert_1
, \quad
\\[0.6em]
\hat{\mathbf{X}}_i 
&= \label{ae2}
\argmin{\mathbf{X}}
\frac{1}{2}\lVert\mathcal{A}(\mathbf{X})-\mathbf{y} \rVert_2^2
+ \beta
\lVert \mathbf{X}-\mathbf{\Phi} (\mathbf{X}_{i-1}; \w_{i})\rVert_2^2
,
\end{align}
where $\mathbf{X} \in \mathbb{C}^{t \times H \times W}$ denotes the dynamic sequence of images to be reconstructed, and the index $i$ corresponds to the $i$th iteration.
The attention network is parameterized as
$\mathbf{\Phi} (\cdot\, ; \w_i)$, and 
$\w_i$ denotes the network weights
for the $i$th ``outer'' iteration of the alternating minimization process.
$\mathbf{X}_0 \in \mathbb{C}^{t \times H \times W}$ is data-shared initialization, and $\mathbf{X}_{\mathrm{true}} \in \mathbb{C}^{t \times H \times W}$ is the ground truth labels for training.
$\mathcal{A}(\mathbf{\cdot})$ is a linear operator representing the MR physics, $\mathbf{y}$ denotes undersampled k-space measurements, and
$\beta$ is the regularization parameter.

\begin{figure*}
\centering
    \includegraphics[width=\textwidth]{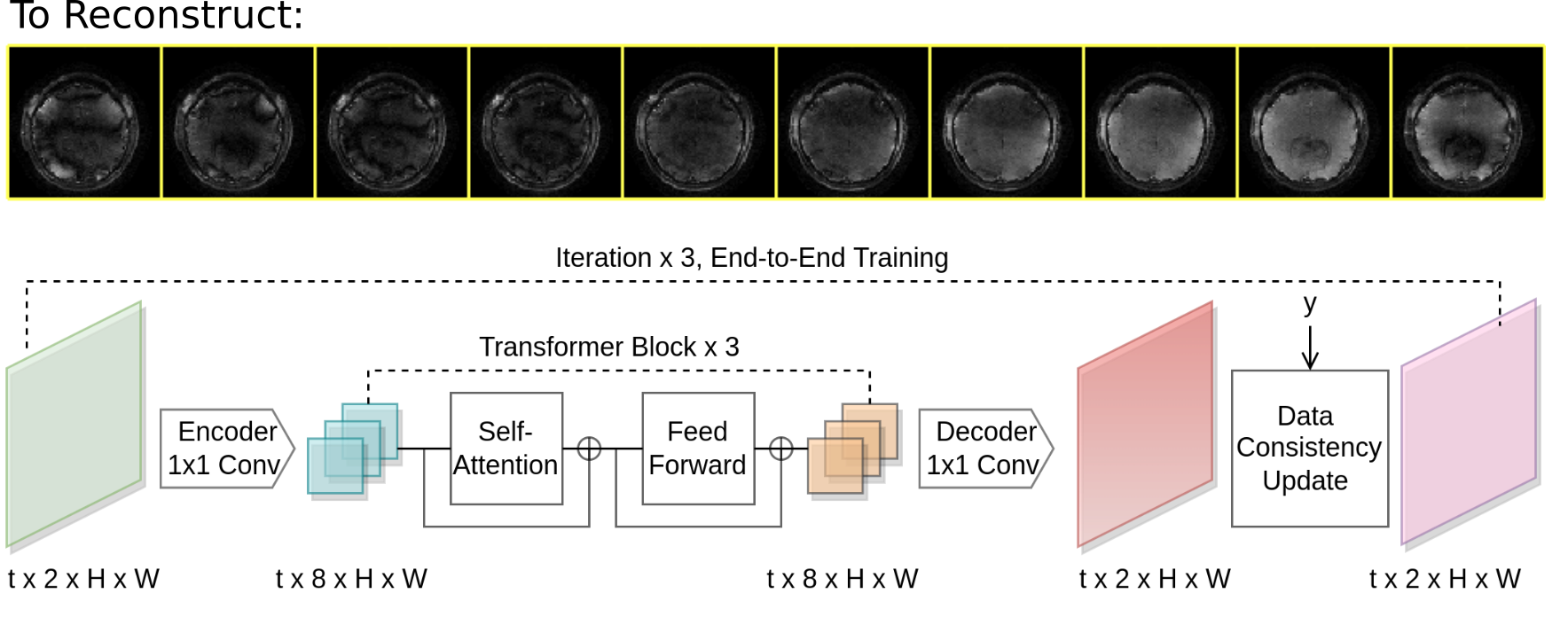}
    \caption{
Our proposed voxel-wise temporal attention network architecture and the dynamic OSSI MRI images (with temporal dimension = 10) to be reconstructed. The data fidelity contains 2 iterations of CG-SENSE for multi-coil NUFFT reconstruction. The main part of the network (encoder-Transformer-decoder) can take voxel-wise simulations or spatial images/patches from human data as inputs.
    }\label{af1}
\end{figure*}

We propose a voxel-wise attention network $\mathbf{\Phi} (\w)$ with Transformers as building blocks for the method of \eqref{ae1}.
The voxel-wise attention network is composed of three components: 1) an encoder that brings the input time-series to the feature domain, 2) consecutive Transformer blocks \cite{vaswani2017attention, dosovitskiy2020image} that consist of attention, feed-forward operations, and residual connections, 3) a decoder that brings the Transformed sequence to the image domain. We use convolutions with 1$\times$1 kernels in both the encoder and decoder to ensure the voxel-wise operations of the network. \cref{af1} presents the overall framework with an attention network and data consistency. 

The input complex MRI image sequence is formed with 2 channels of real and imaginary data, and the dimension is $t \times C \times H \times W$ with $C$ = 2. The encoder consists of 2 convolutional layers (followed by leakyReLU) and encodes the input to richer feature representations with an increased channel dimension of 8 and preserves the same spatial dimensions with 1$\times$1 convolutions. The temporal sequence of feature maps from the encoder is transformed to another temporal sequence of maps with three consecutive Transformer blocks. The learnable weights $W_\Q$, $W_\K$, and $W_\V$ are implemented with 1$\times$1 convolutions as in \cite{zhang2019self,zeng2020learning}, and the feed-forward operation in each Transformer block is performed with 2 convolutional layers with 3$\times$3 and instance normalization \cite{ulyanov2016instance}. The decoder brings the transformed feature maps back to the image domain of size $t \times 2 \times H \times W$ with 2 convolutional layers. 

The voxel-wise attention network is followed by a data fidelity layer for solving \eqref{ae2}. Equation~\eqref{ae2} is a quadratic least-squares problem that regularizes $\X$ to be close to the attention network output $\mathbf{\Phi} (\mathbf{X}_{i-1}; \w_{i})$ by minimizing the Euclidean distance. We solve \eqref{ae2} using the conjugate gradient (CG) method and form a data fidelity layer that takes attention network transformed images $\mathbf{\Phi} (\mathbf{X}_{i-1}; \w_{i})$ as part of the inputs, and performs 2 iterations of the CG update. The linear operator $\mathcal{A}$ with multi-coil and NUFFT operation is implemented using \cite{muckley:20:tah}. The output of the data fidelity layer becomes new inputs for the voxel-wise attention network, and we repeat this step 3 times as 3 outer iterations for the alternating minimization of $\mathbf{X}$ and $\w$. We choose the numbers of CG and outer iterations empirically. 

\subsection{Two-Stage Training and Data Simulation}

We performed two-stage training to handle the problem with limited human data for learning. We pretrain the attention network with voxel-wise simulated temporal sequences (which could be easier to simulate than spatial-temporal sequences). After pretraining, we fine-tune the attention network together with data consistency using human data as training data and train the whole framework in an end-to-end fashion.

For simulated data, we generated the ground truth sequence \cite{2021manifold} using Bloch simulation with varying physics parameters. The inputs for the network are complex Gaussian noise corrupted sequences with a standard deviation of 0.2 to very roughly model the aliasing artifacts. The pretraining of the attention network maps noisy input sequences to noiseless ground truth sequences, and trains the network to denoise.

\subsection{Implementation Details}

The human data were acquired with OSSI sequence \cite{ossi2019}. We formed a human data training set with 10 oscillatory temporal images for each 2D slice and 22 distinct slices in total. 
We augmented the training data 10 times by circularly shifting every image set of 10 oscillating images with 10 different shift positions along the dynamic time dimension. Ground truth images were reconstructed from mostly sampled data.
The k-space data were multi-coil and undersampled using variable-density spiral trajectories with an undersampling factor of 12 as in \cite{tensor2020}. We preprocessed the data by normalizing them using the maximum absolute value of data-shared images. We used data-shared initializations as inputs for the network.

The simulated data contains 8,662,000 voxel-wise time courses of dimension 10$\times$1. We pretrained the network with simulated data for 60 epochs and fine-tuned the network with human data patches for 60 epochs. The testing data has 1480 dynamic images for a 200 s OSSI fMRI scan. In the testing stage, we reconstructed sets of 10 dynamic images of the OSSI fMRI data using the proposed network, and $l_2$-combined each set of 10 reconstructed images to get a sequence of fMRI images for evaluation. The functional task was a left/right reversing-checkerboard visual stimulus for 5 cycles (20 s L/20 s R).

\section{Comparisons and Results} 

\begin{figure*}
\centering
    \includegraphics[width=\textwidth]{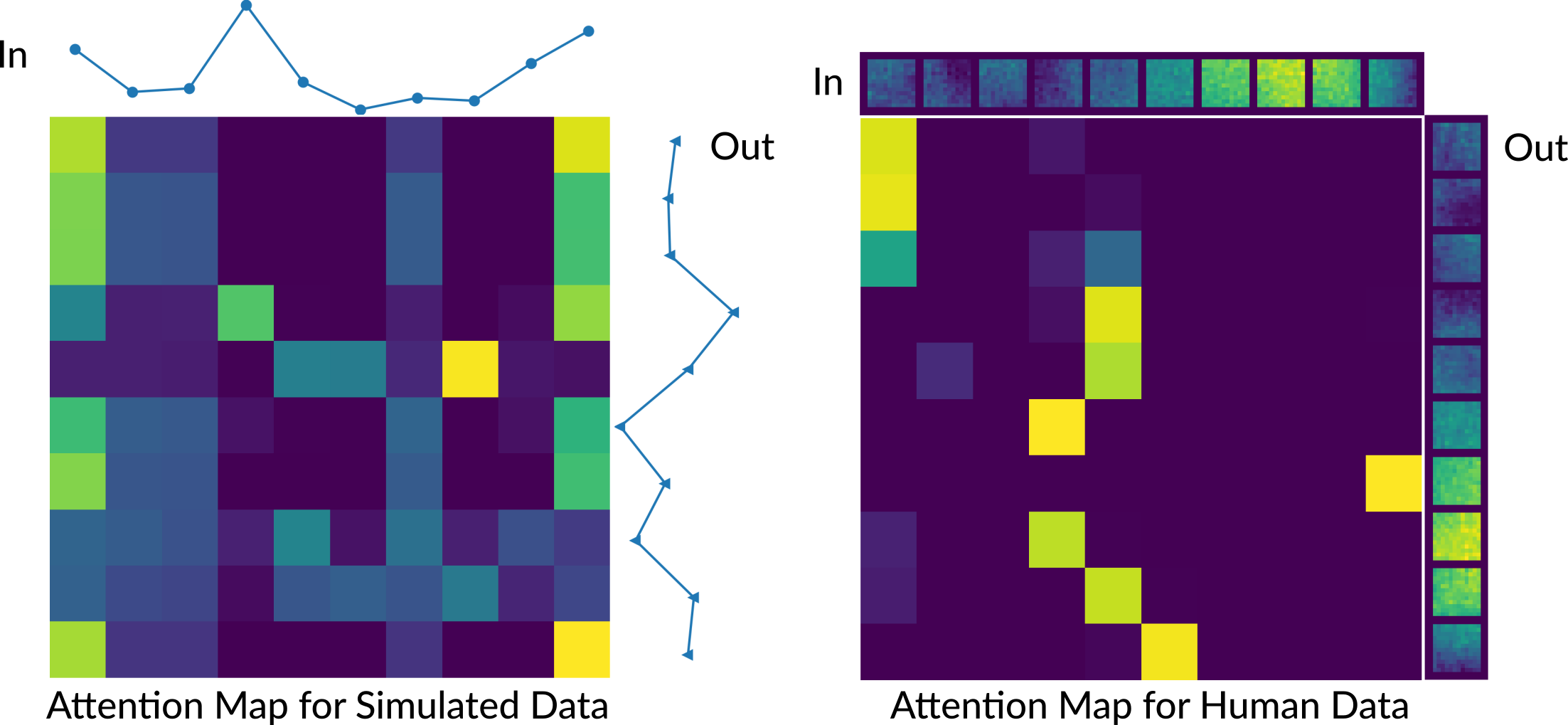}
    \caption{
Attention map visualization at the testing stage for voxel-wise simulation data (left) and human data patch mapping (right). In the attention mechanism, each output value in a 10$\times$1 sequence is generated with a weighted combination of all the values in the input sequence, and the learned weights are given by each row of the 10$\times$10 attention maps for each output value. The figure presents absolute values of the complex input/output for illustration while the proposed network inputs real and imaginary parts and uses deep representations from the encoder for attention calculation.
    }\label{af2}
\end{figure*}

We compared our proposed approach to a 3D U-Net \cite{cciccek20163d} that takes sets of 10 dynamic images as 3D volumes for processing. The network was trained with human fMRI data. Because 3D U-Net is a spatio-temporal network, we cannot easily pretrain the network with simulated data. 

\cref{af2} presents attention map visualization for simulated temporal sequence mapping and human temporal sequence mapping, respectively. Each sample of the output sequence is formed based on a weighted combination of all the samples in the input sequence, and the weights are given by the rows of the learned attention map. Specifically, for every row of the attention map, all the values in the row sum to 1, and each value in the row represents the weight for the corresponding input vector of the input sequence. For the $i$th row, a weighted combination of all the input vectors produces the $i$th output vector of the output sequence.  

\begin{figure*}
\centering
    \includegraphics[width=\textwidth]{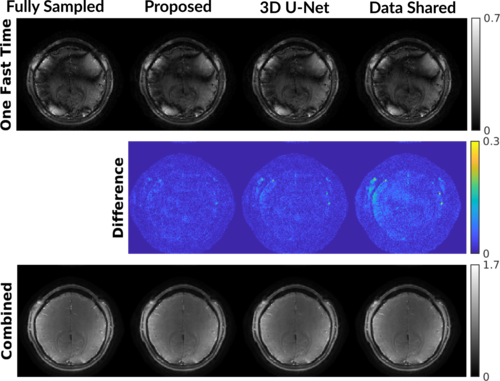}
    \caption{
The proposed voxel-wise model presents less residual in the difference maps than spatial-temporal reconstruction using 3D U-Net.
    }\label{af3}
\end{figure*}

\begin{figure*}
\centering
    \includegraphics[width=\textwidth]{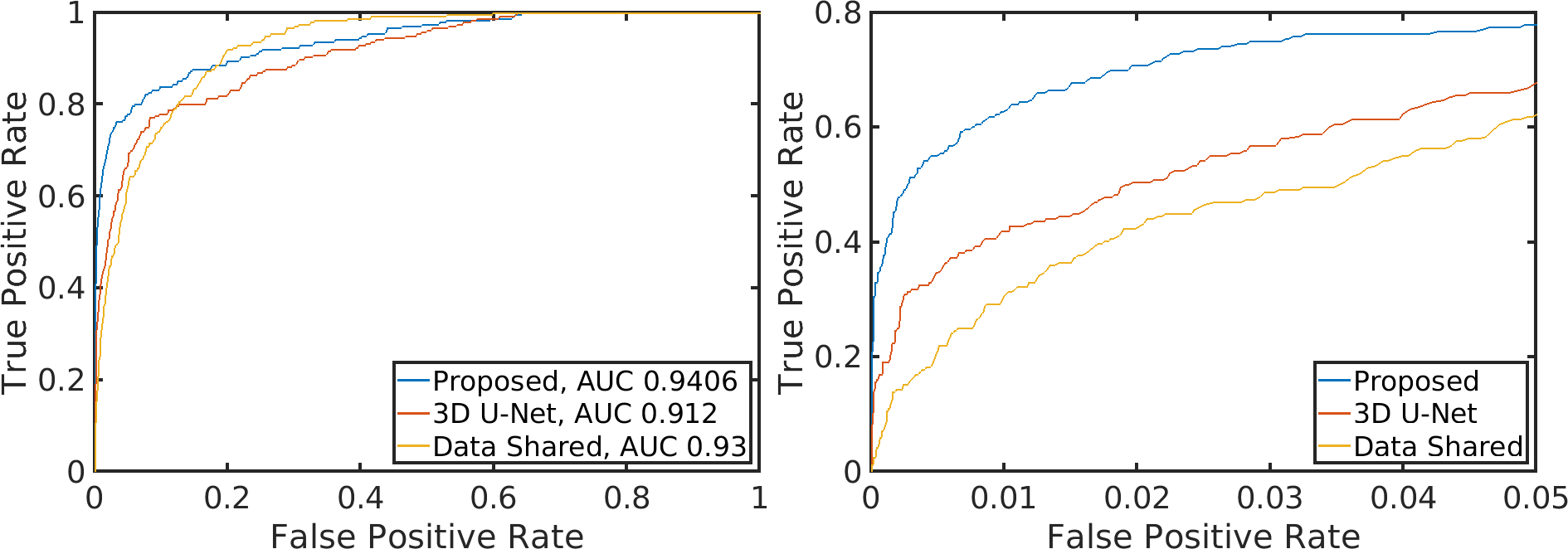}
    \caption{
The ROC curves for fMRI demonstrate that the proposed model outperforms other reconstructions.
    }\label{af5}
\end{figure*}

For reconstruction, \cref{af3} shows that the proposed method leads to less structure in the difference maps than other reconstruction methods such as 3D U-Net. Every 10 reconstructed images are combined with $l_2$-norm for fMRI. \cref{af4} provides functional maps for the reconstructions. The proposed model results in fewer false positives and cleaner time course compared to the fully sampled data. \cref{at1} and \cref{af5} summarizes quantitative evaluations of the reconstruction and functional performance. The proposed model outperforms other methods with lower NRMSE values, and also provides the largest area under the ROC curve.

\section{Conclusions}

We propose a novel voxel-wise attention network for dynamic MRI temporal modeling. The voxel-wise network design enables pretraining with voxel-wise simulated data that can be easier to obtain than spatial-temporal data, and resolves the training data limitation for dynamic imaging. Our proposed model reconstructs dynamic MRI images with a factor of 12 undersampling, and provides high-quality reconstruction and functional maps. The proposed learning-based reconstruction approach is at least 4$\times$ faster than the manifold model-based reconstruction method in \autoref{chap:3manifold}. The proposed voxel-wise, attention-based model can potentially be used for MR fingering reconstruction and other dynamic reconstruction applications.

\begin{figure*}
\centering
    \includegraphics[width=\textwidth]{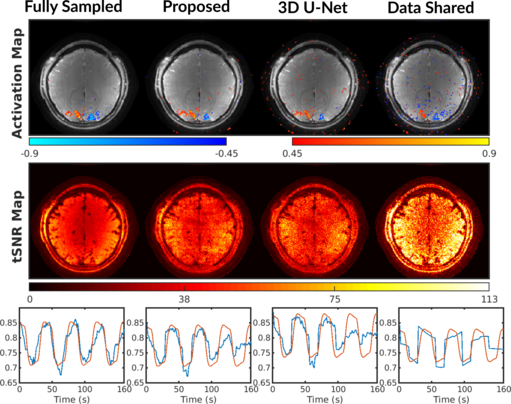}
    \caption{
The proposed approach results in fewer false positives in the activation map, less noisy temporal SNR map, and a time course more similar to the ground truth.
    }\label{af4}
\end{figure*}

We can observe some horizontal line artifacts in the tSNR map of the proposed approach in \cref{af4} because the network was trained with patches while being tested with the whole brain images. In future work, we plan to design a hierarchical network or a multi-stage training scheme to help the network process the whole brain images.
The network might take voxel-wise simulated time courses for the first stage, then fine-tune with sequences of image patches, finally refine with whole brain images. The proposed pipeline could potentially be improved with more sophisticated Transformer network designs for increased representation capacity and a GAN loss for sharpness of the images. 

\newpage

\begin{table}
\caption{Quantitative evaluation for dynamic undersampled reconstructions}
\label{at1}
\centering
\begin{tabular}{c|c|c|c}
\hline
Reconstruction & Proposed & 3D U-Net & Data Shared\\ \hline
NRMSE & \textbf{0.13} & 0.15 & 0.16 \\ \hline
AUC & \textbf{0.94} & 0.91 & 0.93 \\ \hline
\end{tabular}
\end{table}

\chapter{Future Work}
\label{chap:5future}
In previous chapters, we have proposed three different models:
patch-tensor low-rank model, physics-based manifold model, and attention network model for undersampled MRI sequence reconstruction. All of them preserve the high SNR advantage of OSSI, and outperform other reconstruction models with higher spatial-temporal resolution and more functional activation. In future work, we propose an extended approach that combines the advantages of the tensor model and the manifold model, and propose new ideas for learning-based reconstruction and directions for other types of models.

\section{Linear plus Nonlinear (L+N) Model for 3D OSSI fMRI Acceleration and Dynamic Quantification}

The patch-tensor low-rank approach
from \autoref{chap:2tensor}
and physics-based manifold models
from \autoref{chap:3manifold}
fit the OSSI sequence data to linear subspaces and a nonlinear manifold, respectively. 
The patch-tensor model exploits local spatial-temporal similarities with 3D patch tensors of vectorized spatial dimension, and fast and slow time dimensions. The physics-based manifold model focuses on nonlinear modeling of voxel-wise fast time signals and enables physics parameter quantification.
Because the two models exploit different properties of the OSSI images,
we propose a linear subspace plus nonlinear manifold (L+N) model that combines the advantages of the patch-tensor model with the manifold model 
for 3D OSSI joint reconstruction and quantification
with more aggressive undersampling.

We form patch-tensors as in \autoref{chap:2tensor}, and impose low-rank constraints on the first and third unfoldings of the patch-tensors \cite{tensor2020}.
For the second unfolding of the patch-tensor that is not very low-rank along the fast time dimension, instead of enforcing low-rankness,
we fit the fast time signals to the manifold model and encourage the voxel-wise signal values to lie close to the manifold. 

The cost function of the proposed L+N model is:
\begin{align}
\hat{\X} &= \argmin{\X} f(\X)
\nonumber\\[0.1em]
f(\X) &=
\sum_{i = 1,\,3} \M \lambda_i\lVert\Pm(\X)_{(i)}\rVert_* 
+ \beta \sum_{n=1}^N \mathcal{R} \left(\X_{(2)}[:,n]\right)
+ \frac{1}{2}\lVert\mathcal{A}(\X)-\y \rVert_2^2
, \quad
\nonumber\\[0.6em]
\mathcal{R}(\bv) &=
\underset{m_0,T_2', f_0}{\textrm{min}}\ \lVert \bv-m_0 \BP (T_2', f_0; T_1, T_2)\rVert_2^2,
\label{ef1}
\end{align}
where
$\mathbf{X}\in\mathbb{C}^{x \times y \times z \times t}$
is a complex OSSI fMRI time block to be reconstructed.
$\mathcal{P}(\mathbf{\cdot})$
partitions and reshapes the input into $M$ low-rank patch-tensors
with $\Pm(\mathbf{X})\in\mathbb{C}^{s_p \times n_c \times t_s}$,
$m = 1,\ldots,M$.
$\Pm(\mathbf{X})_{(i)}$ is the mode-$i$ unfolding of $\Pm(\mathbf{X})$.
$\lambda_i$ is the regularization parameter for low-rankness of the mode-$i$ unfolding
. $N = xyzt_s$ for the near-manifold regularizer, and $\beta$ is the regularization parameter.
$\bv \in \Cp^{n_c}$ is a vector of fast time signal values for each voxel in $\X$,
$m_0\BP(T_2', f_0; T_1, T_2) \in \Cp^{n_c}$ denotes the manifold estimates.
$\mathcal{A}(\mathbf{\cdot})$ is a linear operator
consisting of coil sensitivities and the non-uniform Fourier transform including undersampling,
$\y$ represents sparsely sampled k-space measurements.

We optimize the cost function \eqref{ef1} using the alternating direction method of multipliers algorithm in \autoref{chap:2tensor}. The main difference is for updating the second unfolding that is regularized by the near-manifold model, instead of applying the singular value soft-thresholding operator in \eqref{Xk1}, we use the conjugate gradient method for the quadratic least-squares problem.

\begin{figure}[htb!]
\centering
    \includegraphics[width=0.97\textwidth]{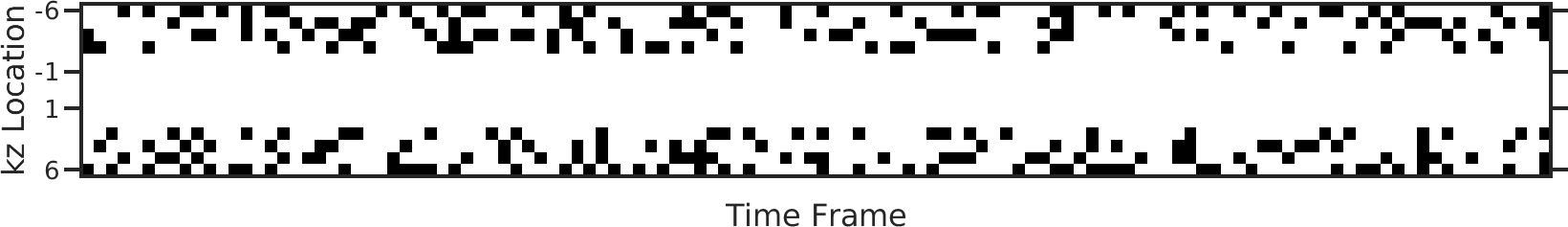}
    \caption{
Poisson-disk sampling of $k_z-t$ planes that keeps 80\% of the variable-density spirals. White color denotes sampled location, and black color denotes $k_z$ planes that are not acquired.
    }\label{4f1}
\end{figure}

\begin{figure}[htb!]
\centering
    \includegraphics[width=0.8\textwidth]{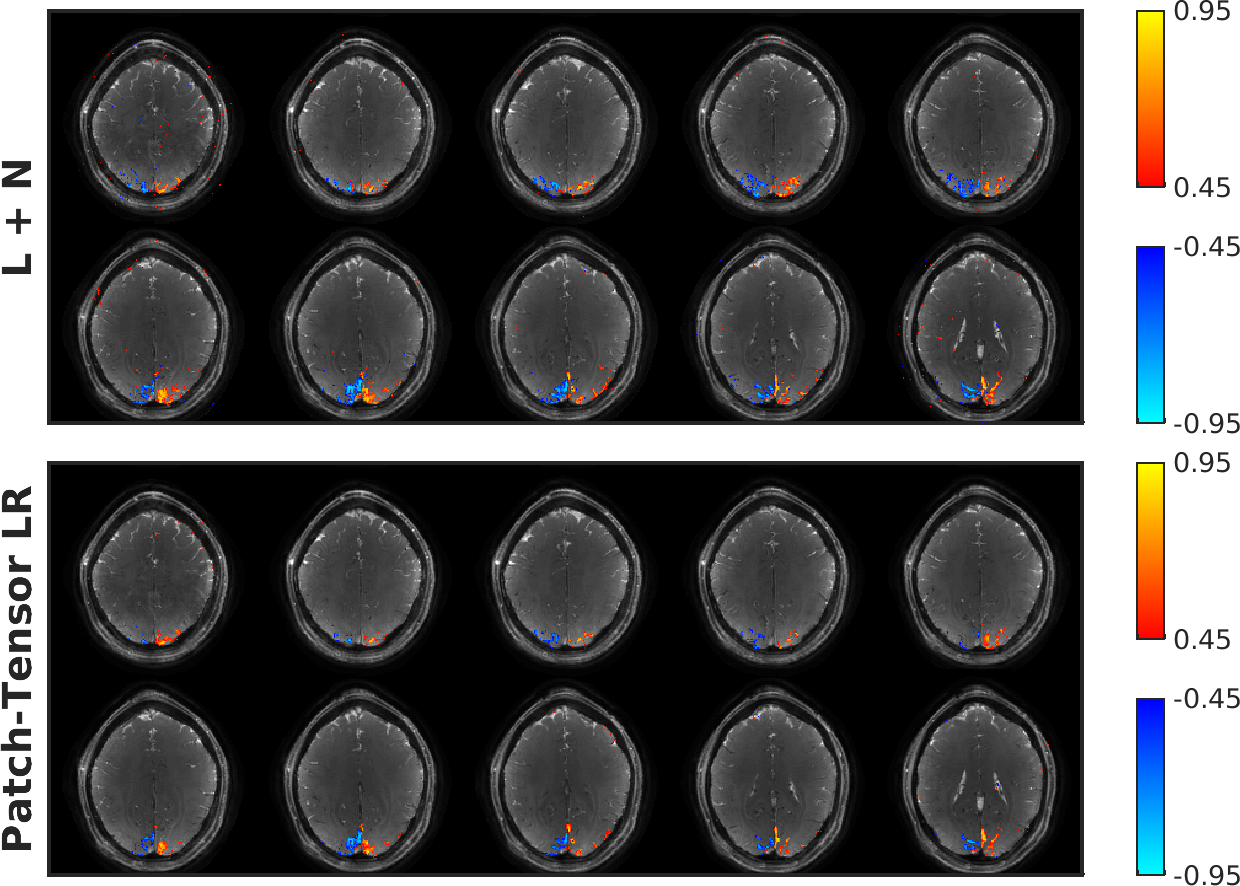}
    \caption{
3D OSSI activation map of the proposed model yields more activation
than the patch-tensor low-rank model in \autoref{chap:2tensor}.
    }\label{4f2}
\end{figure}

\begin{figure}[htb!]
\centering
    \includegraphics[width=0.8\textwidth]{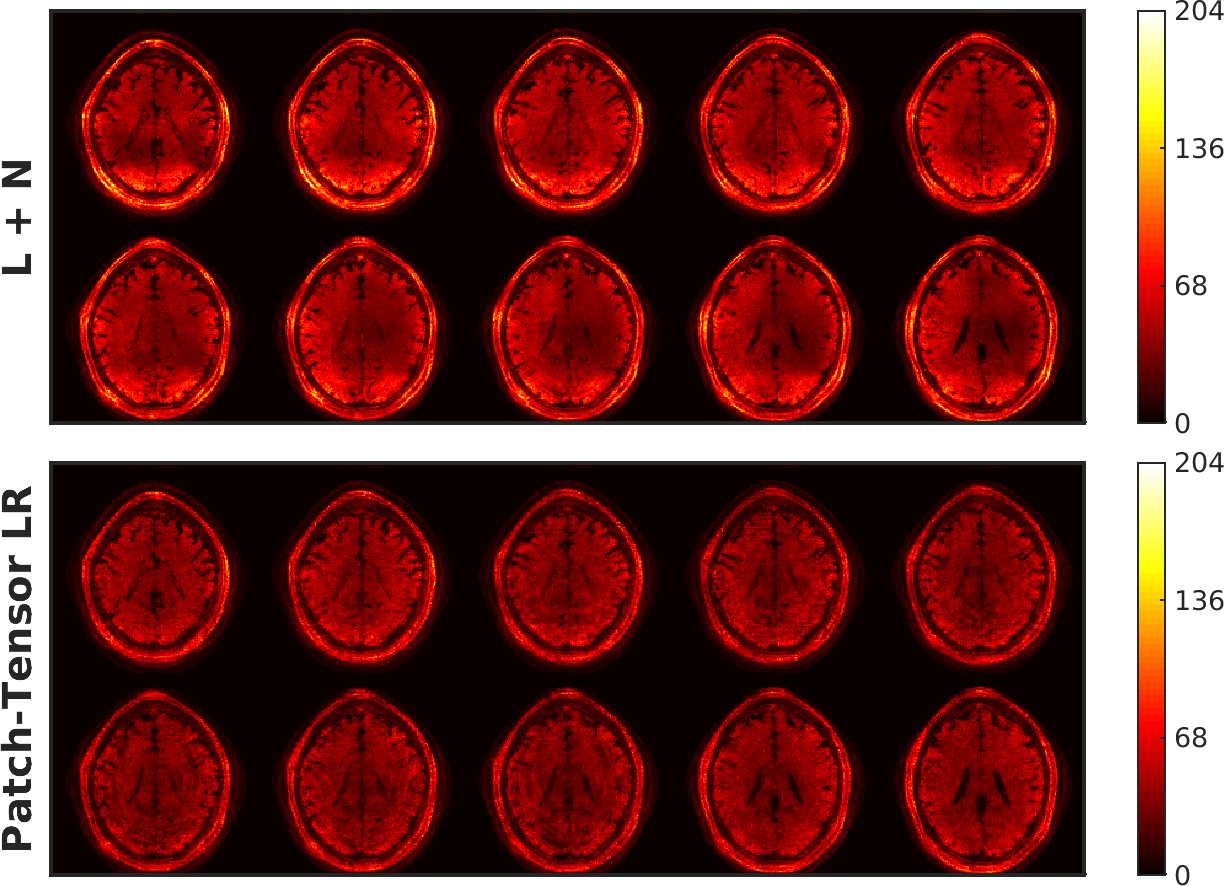}
    \caption{
3D OSSI temporal SNR map of the proposed model presents higher temporal SNR
than the tensor low-rank model in \autoref{chap:2tensor}.
    }\label{4f3}
\end{figure}

The sampling pattern we investigated initially is Poisson-disk undersampling of stack-of-spirals
with a factor of 12 acceleration.
\cref{4f1,4f2,4f3} show
the preliminary results
for Poisson-disk sampling of $k_z - t$ planes,
activation maps,
and temporal SNR maps using the proposed reconstruction.

\newpage

\section{Other Ideas and Approaches}

Our work presents innovations in high-dimensional modeling, physics constraint modeling, and MR image sequence modeling. In particular, the physics-based manifold model inspires us to design an unsupervised learning approach that models MR physics for OSSI signal generation to reconstruct OSSI images directly from physics parameters. In general, we hope to combine signal processing and physics insights with deep learning to advance different types of applications.

\subsection{Transformers Related}

Current Transformers use attention mechanisms that operate on matrices, which might be insufficiently general for high-dimensional data. 
To fully exploit high-dimensional information with attention mechanisms, we propose multi-dimensional attention.
The multi-dimensional attention could model unfoldings of the high-dimensional tensor data as shown in recent work \cite{babiloni2020tesa}. 
Furthermore, prior work \cite{wang2020linformer} exploits low-rankness in 2D attention maps for efficient Transformers, while no works have been done for higher dimensional attention maps. Similar to the patch-tensor low-rank model in \autoref{chap:2tensor}, we propose to impose tensor low-rankness on high-dimensional attention maps.

We can also design a multi-dimensional attention mechanism. Instead of implementing it through unfolded matrices and matrix multiplication, we propose to implement the tensor attention with tensor multiplication.
We could further incorporate sparsity constraints to encourage the high-dimensional attention maps to have sparse core tensors.  
Related works \cite{martins2016softmax,martins2020sparse} present sparse 2D attention maps.

Previous works such as inverting a CNN \cite{mahendran2016visualizing} for visualizing CNN representations could potentially be extended to Transformers and would be interesting to investigate. 
More importantly, with the deep image prior \cite{ulyanov2018deep} concept for unsupervised learning, we propose to use Transformers as the network architecture in deep image prior regularization for dynamic MRI sequence reconstruction. 

For Transformer network design, we propose to use the hierarchical attention network as in \cite{guo2022paying} to model multi-scale spatial-temporal dependency of the image sequences for both self-supervised and unsupervised tasks.







\subsection{More General Directions}

We have considered some other interesting research directions and approaches:

\begin{itemize}

\item We can design a union of sparse subspaces model \cite{elhamifar2013sparse} to serve as a regularizer for for dynamic MRI reconstruction.

\item We propose an MRI k-space inpainting model that exploits neighboring sample similarity using a local/adaptive SIREN network \cite{sitzmann2020implicit} for undersampled reconstruction.

\item Another future research direction
is to consider hierarchical and probabilistic network structures \cite{hinton2021represent,morin2005hierarchical,yang2016hierarchical,lakshminarayanan2017simple,giryes2016deep} to advance the learning and reasoning capacity of neural networks.

\end{itemize}



\bibliographystyle{plainnat}
\bibliography{ms}


\end{document}